\documentclass[12pt] {article}
\usepackage{psfrag}
\usepackage{graphicx}
\usepackage{latexsym,amsfonts}
\begin{document}
\setcounter{page}{1}
\def\theequation{\arabic{section}.\arabic{equation}}
\def\theequation{\thesection.\arabic{equation}}
\setcounter{section}{0}

\title{~\rightline{\normalsize {\rm IK-TUW-Preprint 0012401}}\\
[0.2in]On the equivalence between sine--Gordon model and Thirring
model in the chirally broken phase of the Thirring model}

\author{M. Faber\thanks{E--mail: faber@kph.tuwien.ac.at, Tel.:
+43--1--58801--14261, Fax: +43--1--5864203} ~~and~
A. N. Ivanov\thanks{E--mail: ivanov@kph.tuwien.ac.at, Tel.:
+43--1--58801--14261, Fax: +43--1--5864203}~\thanks{Permanent Address:
State Technical University, Department of Nuclear Physics, 195251
St. Petersburg, Russian Federation}}

\date{\today}

\maketitle
\vspace{-0.5in}
\begin{center}
{\it Atominstitut der \"Osterreichischen Universit\"aten,
Arbeitsbereich Kernphysik und Nukleare Astrophysik, Technische
Universit\"at Wien, \\ Wiedner Hauptstr. 8-10, A-1040 Wien,
\"Osterreich }
\end{center}

\begin{center}
\begin{abstract}
We investigate the equivalence between Thirring model and sine--Gordon
model in the chirally broken phase of the Thirring model. This is
unlike all other available approaches where the fermion fields of the
Thirring model were quantized in the chiral symmetric phase. In the
path integral approach we show that the bosonized version of the
massless Thirring model is described by a quantum field theory of a
massless scalar field and exactly solvable, and the massive Thirring
model bosonizes to the sine--Gordon model with a new relation between
coupling constants. We show that the non--perturbative vacuum of the
chirally broken phase in the massless Thirring model can be described
in complete analogy with the BCS ground state of superconductivity.
The Mermin--Wagner theorem and Coleman's statement concerning the
absence of Goldstone bosons in the 1+1--dimensional quantum field
theories are discussed.  We investigate the current algebra in the
massless Thirring model and give a new value of the Schwinger term. We
show that the topological current in the sine-Gordon model coincides
with the Noether current responsible for the conservation of the
fermion number in the Thirring model. This allows to identify the
topological charge in the sine-Gordon model with the fermion number.
\end{abstract}
\end{center}

\newpage

\section{Introduction}
\setcounter{equation}{0}

\hspace{0.2in} In 1+1--dimensional space--time there are two
non--trivial minimal quantum field theories which describe
non--perturbative phenomena: the sine--Gordon (SG) model \cite{[1]}
and the Thirring model \cite{[2]}. The SG model is a quantum field
theory of a single scalar field $\vartheta(x)$ self--coupled through
the dynamics determined by the Lagrangian \cite{[3]}\,\footnote{Below
we follow Coleman's notation \cite{[3]}.}
\begin{eqnarray}\label{label1.1}
{\cal L}(x) =
\frac{1}{2}\partial_{\mu}\vartheta(x)\partial^{\mu}\vartheta(x) +
\frac{\alpha}{\beta^2}\,(\cos\beta\vartheta(x) - 1),
\end{eqnarray}
where $\alpha$ and $\beta$ are real positive parameters \cite{[3]}. The
Lagrangian (\ref{label1.1}) is invariant under the transformations
\begin{eqnarray}\label{label1.2}
\vartheta(x) \to \vartheta^{\,\prime}(x) = \vartheta(x) + \frac{2\pi
n}{\beta},
\end{eqnarray}
where $n$ is an integer number running over $n = 0,\pm 1, \pm 2,\ldots$.

The most interesting property of the SG model is the existence of
classical, stable solutions of the equations of motion -- solitons and
anti--solitons \cite{[1]}. Solitons can annihilate with
anti--solitons. Many--soliton solutions obey Pauli's exclusion
principle. As pointed out by Skyrme \cite{[4]} this can be interpreted
as a fermion--like behaviour.

In turn, the Thirring model \cite{[2]} is a theory of a self--coupled
Dirac field $\psi(x)$ \cite{[2],[3]}
\begin{eqnarray}\label{label1.3}
{\cal L}(x) = \bar{\psi}(x)(i\gamma^{\mu}\partial_{\mu} - m)\psi(x) -
\frac{1}{2}\,g\,\bar{\psi}(x)\gamma^{\mu}\psi(x)\bar{\psi}(x)
\gamma_{\mu}\psi(x),
\end{eqnarray}
where $m$ is the mass of the fermion field and $g$ is a
dimensionless coupling constant. The field $\psi(x)$ is a spinor field
with two components $\psi_1(x)$ and $\psi_2(x)$. The
$\gamma$--matrices are defined in terms of the well--known $2\times 2$
Pauli matrices $\sigma_1$, $\sigma_2$ and $\sigma_3$
\begin{eqnarray}\label{label1.4}
\gamma^0 = \sigma_1\quad,\quad \gamma^1 = - i\sigma_2\quad,\quad
\gamma^5 = \gamma^0\gamma^1 = -i\sigma_1\sigma_2 = \sigma_3.
\end{eqnarray}
These $\gamma$--matrices obey the standard relations
\begin{eqnarray}\label{label1.5}
\gamma^{\mu}\gamma^{\nu}&+& \gamma^{\nu}\gamma^{\mu} = 2
g^{\mu\nu},\nonumber\\ 
\gamma^{\mu}\gamma^5&+& \gamma^5\gamma^{\mu} = 0.
\end{eqnarray}
We use the metric tensor $g^{\mu\nu}$ defined by $g^{00} = - g^{11} =
1$ and $g^{01} = g^{10} = 0$. The axial--vector product
$\gamma^{\mu}\gamma^5$ can be expressed in terms of $\gamma^{\nu}$
\begin{eqnarray}\label{label1.6}
\gamma^{\mu}\gamma^5 &=& - \epsilon^{\mu\nu}\gamma_{\nu},\
\end{eqnarray}
where $\epsilon^{\mu\nu}$ is the anti--symmetric tensor defined by
$\epsilon^{01} = - \epsilon^{10} = 1$. Further, we also use the
relation $\gamma^{\mu}\gamma^{\nu} = g^{\mu\nu} +
\epsilon^{\mu\nu}\gamma^5$.

The Lagrangian (\ref{label1.3}) is obviously invariant under $U_{\rm
V}(1)$ transformations
\begin{eqnarray}\label{label1.7}
\psi(x) \stackrel{\rm V}{\longrightarrow} \psi^{\prime}(x) &=&
e^{\textstyle i\alpha_{\rm V}}\psi(x).
\end{eqnarray}
For $m = 0$ the Lagrangian (\ref{label1.3}) is invariant under the
chiral group $U_{\rm V}(1)\times U_{\rm A}(1)$
\begin{eqnarray}\label{label1.8}
\psi(x) \stackrel{\rm V}{\longrightarrow} \psi^{\prime}(x) &=&
e^{\textstyle i\alpha_{\rm V}}\psi(x) ,\nonumber\\ \psi(x)
\stackrel{\rm A}{\longrightarrow} \psi^{\prime}(x) &=& e^{\textstyle
i\alpha_{\rm A}\gamma^5}\psi(x),
\end{eqnarray}
where $\alpha_{\rm V}$ and $\alpha_{\rm A}$ are real parameters
defining global rotations.

As has been shown in Refs.\cite{[5],[6]} the massless Thirring model
can be exactly solved in the sense that all correlation functions can
be calculated explicitly. The solution of the massless Thirring model
has been obtained in the traditional quantum field theoretic way by
Klaiber \cite{[5]} within the operator technique and within the path
integral approach by Furuya, Gamboa Saravi and Schaposnik \cite{[6]}
by using the technique of auxiliary vector fields. Within the path
integral approach and without the introduction of auxiliary vector
fields Fr\"ohlich and Marchetti \cite{[7]} have shown that the
evaluation of the Green functions in the massless Thirring model runs
parallel the evaluation of the Green functions in the quantum field
theory of a massless scalar field coupled to external sources via SG
model--like couplings.

The problem of the equivalence between the SG and the Thirring model
has a long history. The first discussion of this topic has been
started by Skyrme \cite{[4]} and continued by Coleman \cite{[3]} and
Mandelstam \cite{[8]}. Skyrme argued that the soliton modes of the SG
model possess the properties of fermion fields and couple through an
interaction of Thirring--model type. Coleman suggested a perturbative
approach to the understanding of an equivalence between the SG and the
Thirring model. He developed a perturbation theory with respect to
$\alpha$ and $m$ in order to compare the $n$--point Green functions in
the SG and the massive Thirring model in coordinate
representation. Under the assumption of the existence of these two
theories in the strict sense of constructive quantum field theory,
Coleman concluded that they should be equivalent if the coupling
constants $\beta$ and $g$ obey the relation \cite{[3]}
\begin{eqnarray}\label{label1.9}
\frac{4\pi}{\beta^2} = 1 + \frac{g}{\pi}
\end{eqnarray}
and the operators $\psi(x)$ and
$\vartheta(x)$ satisfy the Abelian bosonization rules \cite{[3]}
\begin{eqnarray}\label{label1.10}
Z\,m\,\bar{\psi}(x)\Bigg(\frac{1\mp \gamma^5}{2}\Bigg)\psi(x) =
-\frac{\alpha}{\beta^2}\,\,e^{\textstyle \pm i\beta \vartheta(x)},
\end{eqnarray}
where the constant $Z$ depends on the regularization \cite{[3]}. The
results obtained by Coleman are fully based on the solution of the
massless Thirring model given by Klaiber \cite{[5]} and recovered in
pure Euclidean formulation by Fr\"ohlich and Marchetti \cite{[7]}.

Unlike Coleman's analysis dealing with Green functions, i.e. matrix
elements of some products of field operators, Mandelstam has
undertaken an attempt of an explicit derivation of the operators being
functionals of the scalar field of the SG model and possessing the
properties of the fermionic field operators. Mandelstam identified
these operators with the interpolating operators of Thirring fields
and showed that these fermionic operators have a Lagrangian
of the Thirring model type.

Recently, another approach to the derivation of the equivalence
between the SG and the Thirring model was developed by Damgaard,
Nielsen and Sollacher \cite{[9]} and Thomassen \cite{[10]} within the
so--called {\it smooth bosonization approach} based on the path
integral method and using an enlarged set of field variables. In the
{\it smooth bosonization approach} this enlargement of field variables
appears via local chiral rotations \cite{[11]}, where the local chiral
phase is identified with a local pseudoscalar field. Its Lagrangian is
determined by the Jacobian of the fermion path integral measure
depending explicitly on a local chiral phase \cite{[12]}--\cite{[17]}.

The common point of all approaches to the solution of the massless
Thirring model \cite{[5]}--\cite{[7]} and to the derivation of the
equivalence between the SG and the massive Thirring model
\cite{[3],[9],[10]} is a quantization of the fermionic   system around
the trivial perturbative vacuum.

In order to justify our statement we suggest to follow the procedure
used by Nambu and Jona--Lasinio \cite{[18]}. Let us consider the
massless Thirring model defined by the Lagrangian (\ref{label1.3}) at
$m=0$. Then, following Nambu and Jona--Lasinio we supplement and
subtract the term $M\,\bar{\psi}(x)\psi(x)$, where $M$ is an arbitrary
parameter with the meaning of the dynamical mass of fermions.  That is
similar to the Hartee--Fock approximation where a two--body
interaction is approximated by a one--body term. The Lagrangian of the
massless Thirring model acquires the form
\begin{eqnarray}\label{label1.11}
{\cal L}(x) = \bar{\psi}(x)(i\gamma^{\mu}\partial_{\mu} - M)\psi(x) +
{\cal L}_{\rm int}(x)
\end{eqnarray}
with the interaction ${\cal L}_{\rm int}(x)$ given by
\begin{eqnarray}\label{label1.12} 
{\cal L}_{\rm int}(x)= M\,\bar{\psi}(x)\psi(x) -
\frac{1}{2}\,g\,\bar{\psi}(x)\gamma^{\mu}\psi(x)\bar{\psi}(x)
\gamma_{\mu}\psi(x).
\end{eqnarray}
Following the Nambu--Jona--Lasinio prescription one can show that the
dynamical mass $M$ satisfies the gap--equation
\begin{eqnarray}\label{label1.13} 
M = g\,\gamma^{\mu}(-i)S_F(0)\gamma_{\mu} =
g\gamma^{\mu}\int\frac{d^2p}{(2\pi)^2i}\,\frac{1}{M -
\hat{p}}\gamma_{\mu} = M\,\frac{g}{2\pi}\,{\ell n}\left(1 +
\frac{\Lambda^2}{M^2}\right),
\end{eqnarray}
where $\Lambda$ is an ultra--violet cut--off. Thus the gap--equation
reads\,\footnote{We would like to accentuate that the gap--equation is
calculated in the one--fermion loop approximation. As has been shown
in \cite{[19]}--\cite{[22]} the effective Lagrangian of a bosonized
version of a fermion system self--coupled via a four--fermion
interaction is defined by an functional determinant that can be
represented in terms of an infinite series of one--fermion loop
diagrams.}
\begin{eqnarray}\label{label1.14} 
M = M\,\frac{g}{2\pi}\,{\ell
n}\left(1 + \frac{\Lambda^2}{M^2}\right).
\end{eqnarray}
There are two solutions of this equation: $M=0$ and 
\begin{eqnarray}\label{label1.15} 
M = \frac{\Lambda}{\displaystyle \sqrt{e^{\textstyle 2\pi/g} - 1}}.
\end{eqnarray}
The $M=0$ solution is trivial and corresponds to a chiral symmetric
phase with a trivial perturbative vacuum. In turn, the $M\not= 0$
solution (\ref{label1.15}) is non--trivial and related to the chirally
broken phase with a non--perturbative vacuum. This chirally broken
phase is characterized by the appearance of dynamical fermions with a
dynamical mass $M$ and $\bar{\psi}\psi$ pairing \cite{[18]}. By
retracing Refs.\cite{[3],[5]}--\cite{[11]} it becomes obvious that all
results obtained there can be assigned to the $M = 0$ solution
characterizing the quantization of fermion fields around the trivial
perturbative vacuum.

In order to show that the chirally broken $M\not= 0$ phase of the
Thirring model is more preferable than the chiral symmetric $M = 0$
phase we have to calculate the energy density of the vacuum state
${\cal E}(M)$. This can be carried out only by using the exact
expression for the wave function of the non--perturbative vacuum. In
Section 6 we show that the wave function of the non--perturbative
vacuum in the massless Thirring model can be taken in the form of the
wave function of the ground state in the Bardeen--Cooper--Schrieffer
(BCS) theory of superconductivity \cite{[23]} (see also
\cite{[18],[22],[24]}). The energy density ${\cal E}(M)$
calculated in Section 6 has a minimum at $M\not= 0$ that satisfies the
gap--equation (\ref{label1.14}) and a maximum at $M = 0$. This
evidences that for the Thirring fermions the chirally broken phase is
energetically preferable with respect to the chiral symmetric phase.

It is well--known that the chirally broken phase is characterized by a
non--zero value of the fermion condensate, $\langle
0|\bar{\psi}(0)\psi(0)|0\rangle \not= 0$. In the massless Thirring
model the fermion condensate is defined by
\begin{eqnarray}\label{label1.16}
\langle 0|\bar{\psi}(x)\psi(x)|0\rangle_{\rm one-loop} = i\,{\rm
tr}\{S_F(0)\} = - \frac{M}{2\pi}\,{\ell n}\left(1 +
\frac{\Lambda^2}{M^2}\right) = -\,\frac{M}{g},
\end{eqnarray}
where we have taken into account the gap--equation
(\ref{label1.14}). 

Below we denote the fermion condensate (\ref{label1.16}) calculated in
the one--fermion loop approximation by $\langle\bar{\psi}\psi\rangle$,
$\langle 0|\bar{\psi}(x)\psi(x)|0\rangle_{\rm one-loop} =
\langle\bar{\psi}\psi\rangle = - M/g$.

Since the massless Thirring model possesses the same non--perturbative
properties as the Nambu--Jona--Lasinio model \cite{[18]}, we suggest
to recast the four--fermion interaction of the Thirring model into the
form given by Nambu and Jona--Lasinio.  After a Fierz transformation
\begin{eqnarray}\label{label1.17}
- \bar{\psi}(x)\gamma^{\mu}\psi(x)\bar{\psi}(x)\gamma_{\mu}\psi(x) =
  (\bar{\psi}(x)\psi(x))^2 + (\bar{\psi}(x)i\gamma^5\psi(x))^2
\end{eqnarray}
the Lagrangian (\ref{label1.3}) acquires the form
\begin{eqnarray}\label{label1.18}
{\cal L}(x) = \bar{\psi}(x)(i\gamma^{\mu}\partial_{\mu} - m)\psi(x) +
\frac{1}{2}\,g\,[(\bar{\psi}(x)\psi(x))^2 +
(\bar{\psi}(x)i\gamma^5\psi(x))^2].
\end{eqnarray}
In this form the Thirring model coincides with the
Nambu--Jona--Lasinio (NJL) model in 1+1--dimensional space--time. It
is well--known that the NJL model is a relativistic covariant
generalization of the BCS theory of superconductivity. The wave
function of the non--perturbative vacuum of the NJL model coincides
with the wave function of the ground state in the BCS theory
\cite{[23]}.

The main aim of this article is to solve the massless Thirring model
in the chirally broken phase and to derive the equivalence with the SG
model.  We also want to show explicitly that this is possible without
an enlargement of the number of degrees of freedom but via a reduction
of them. In fact, in the Thirring model the fermion  field has two
independent degrees of freedom. Since the SG model describes a scalar
field with only one degree of freedom, one of the two fermion degrees
of freedom should die out. How this goes dynamically in a
non--perturbative way should be the matter of our investigation.

On this way it is rather useful to follow the approach developed in
Refs.\cite{[19]}--\cite{[22]} for the derivation of effective Chiral
Lagrangians in the extended Nambu--Jona--Lasinio (ENJL) model with
chiral $U(3)\times U(3)$ symmetry \cite{[19]}--\cite{[21]} and the
effective Lagrangian in the Monopole Nambu--Jona--Lasinio model with
magnetic $U(1)$ symmetry \cite{[22]}.

This paper is organized as follows. In Section 2 we discuss Coleman's
derivation of the equivalence between the massive Thirring model and
the SG model. In Section 3 we bosonize the massless Thirring model. We
show that the bosonized version of the massless Thirring model is a
quantum field theory of a free massless scalar field. In Section 4 we
evaluate the generating functional of Green functions in the massless
Thirring model and show that any Green function in the massless
Thirring model can be expressed in terms of vacuum expectation values
of the operators $e^{\textstyle \pm i\beta\vartheta(x)}$, where
$\vartheta(x)$ is a massless scalar field with values $0 \le
\vartheta(x) \le 2\pi$. Using a trivial cut--off regularization of the
massless $\vartheta$--field in the infrared region we obtain results
that coincide with those derived by Klaiber, Coleman, Fr\"ohlich and
Marchetti, but give another relation between the coupling constant
$\beta$ and $g$ than that given by Coleman \cite{[3]}. The problem of
the vanishing of the fermion condensate averaged over the
$\vartheta$--field fluctuations is accentuated. The solution of this
problem is discussed in Section 8. In Section 5 we bosonize the
massive Thirring model. We show that the bosonized version of the
massive Thirring model is just the SG model. We express the parameters
of the SG model in terms of the parameters of the massive Thirring
model. In Section 6 we investigate the massless Thirring model in the
operator formulation. We analyse the normal ordering of the fermionic
operators and chiral symmetry breaking, the equations of motion for
fermion fields, the current algebra and the energy--momentum tensor.
We discuss the phenomenon of spontaneous breaking of chiral symmetry
in the massless Thirring model from the point of view of the BCS
theory of superconductivity. We use the exact expression for the wave
function of the non--perturbative vacuum and calculate the energy
density of this non--perturbative vacuum state. We show that the
energy density of the non--perturbative vacuum acquires a minimum
just, when the dynamical mass $M$ of fermions satisfies the
gap--equation (\ref{label1.14}).  Then, we show that the Schwinger
term calculated for the fermion system in the chirally broken phase
becomes depending on the coupling constant $g$.  In Section 7 we show
that the topological current of the SG model coincides with the
Noether current of the massive Thirring model related to the
invariance under global $U_{\rm V}(1)$ rotation.  As far as this
Noether current is responsible for the conservation of the fermion
number in the massive Thirring model, the topological charge of
soliton solutions of the SG model inherits the meaning of the fermion
number.  This proves Skyrme$^{\prime}$s statement \cite{[4]} that the
SG model solitons can be interpreted as massive fermions. In Section 8
we discuss the spontaneous breaking of chiral symmetry in the massless
Thirring model, the Mermin--Wagner theorem \cite{[25]} about the
vanishing of long--range order in two dimensional quantum field
theories and Coleman's statement concerning the absence of Goldstone
bosons in the 1+1--dimensional quantum field theory of a massless
scalar field. We show that in our approach the problem of the
vanishing of the fermion condensate averaged over the
$\vartheta$--field fluctuations can be solved by means of dimensional
and analytical regularization. We give the solution of the massless
Thirring model in the sense of an explicit evaluation of any
correlation function. In the Conclusion we discuss our results. In
Appendix A we calculate the Jacobian caused by local chiral rotations
and show that by using an appropriate regularization scheme this
Jacobian can be equal to unity. In Appendix B we demonstrate the
stability of the chirally broken phase under fluctuations of the
radial scalar field around the minimum of the effective potential
calculated in Section 3. We show that the radial scalar field
fluctuating around the minimum of the effective potential is decoupled
from the system. In Appendix C we give a classical solution of the
equations of motion of the massless Thirring model for the ansatz
discussed in Section 6. In Appendix D we give a detail description of
free massive and massless fermion fields in 1+1--dimensional
space--time.

\section{On Coleman's analysis of equivalence}
\setcounter{equation}{0}

\hspace{0.2in} In this section we would like to repeat Coleman's
derivation of the equivalence between the massive Thirring model and
the SG model within the path integral approach. Let $Z_{\rm SG}$ and
$Z_{\rm Th}$ be the partition functions of the SG and the massive
Thirring model defined by
\begin{eqnarray}\label{label2.1}
Z_{\rm SG} &=& \int {\cal D}\vartheta\,\exp i\int
d^2x\,\Big\{\frac{1}{2}\,\partial_{\mu}\vartheta(x)\partial^{\mu}\vartheta(x)
+ \frac{\alpha}{\beta^2}\,(\cos\beta\vartheta(x) - 1)\Big\},\nonumber\\
 Z_{\rm Th} &=& \int {\cal D}\psi{\cal D}\bar{\psi}\nonumber\\ &\times&\exp
i\int d^2x\,\Big\{\bar{\psi}(x)(i\gamma^{\mu}\partial_{\mu} -
m)\psi(x) -
\frac{1}{2}\,g\,\bar{\psi}(x)\gamma_{\mu}\psi(x)\bar{\psi}(x)
\gamma^{\mu}\psi(x)\Big\}.
\end{eqnarray}
Formally, in order to get convinced that the SG and the massive
Thirring model are equivalent it is sufficient to show that $Z_{\rm
SG} = Z_{\rm Th}$. Coleman suggested to prove this relation
perturbatively. For this aim he developed perturbation theories with
respect to $\alpha$ and $m$ \cite{[3]}. According to Coleman we have to
expand the partition functions with respect to the interaction terms
\cite{[3]}:
\begin{eqnarray}\label{label2.2}
{\cal L}^{\rm SG}_{\rm
int}(x)&=&\frac{\alpha}{\beta^2}\,\cos\beta\vartheta(x) = 
\frac{\alpha}{2\beta^2}\,(A_+(x) + A_-(x)),\nonumber\\ {\cal L}^{\rm
Th}_{\rm int}(x)&=& -\,m\,\bar{\psi}(x)\psi(x) = -\,m\,(\sigma_+(x) +
\sigma_-(x)),
\end{eqnarray}
where \cite{[3]}
\begin{eqnarray}\label{label2.3}
A_{\pm}(x) &=& e^{\textstyle \pm\,i\,\beta\,\vartheta(x)},\nonumber\\
\sigma_{\pm}(x)&=&\bar{\psi}(x)\Bigg(\frac{1 \pm
\gamma^5}{2}\Bigg)\psi(x).
\end{eqnarray}
In terms of the components $\psi_1(x)$ and $\psi_2(x)$ the fermion
densities $\sigma_{\pm}(x)$ are defined by $\sigma_+(x) =
\psi^{\dagger}_2(x)\psi_1(x)$ and $\sigma_-(x)=
\psi^{\dagger}_1(x)\psi_2(x)$.

The expansions for the partition functions in powers of $\alpha$ and $m$ read
\begin{eqnarray}\label{label2.4}
Z_{\rm SG} &=&
\sum^{\infty}_{n=0}\frac{i^n}{n!}\,\Bigg(\frac{\alpha}{2\beta^2}\Bigg)^n\int
{\cal D}\vartheta\,\exp i\int
d^2x\,\Big\{\frac{1}{2}\,\partial_{\mu}\vartheta(x)
\partial^{\mu}\vartheta(x)\Big\}\nonumber\\ &&\times \int\!\!\!\int\!\!
\ldots \!\!\int d^2x_1d^2x_2\ldots d^2x_n\,\prod^{n}_{k=1}(A_+(x_k) +
A_-(x_k)),\nonumber\\ Z_{\rm Th} &=&
\sum^{\infty}_{n=0}\frac{i^n}{n!}\,(-m)^n\nonumber\\ &&\times\int
{\cal D}\psi{\cal D}\bar{\psi}\,\exp i\int
d^2x\,\Big\{\bar{\psi}(x)i\gamma^{\mu}\partial_{\mu} \psi(x) -
\frac{1}{2}\,g\,\bar{\psi}(x)\gamma_{\mu}\psi(x)\bar{\psi}(x)
\gamma^{\mu}\psi(x)\Big\}\nonumber\\ &&\times \int\!\!\!\int\!\!
\ldots \!\!\int d^2x_1d^2x_2\ldots
d^2x_n\,\prod^{n}_{k=1}(\sigma_+(x_k) + \sigma_-(x_k)).
\end{eqnarray}
Every term of these expansions corresponds to a vacuum expectation
value of a massless free scalar field $\vartheta(x)$\,\footnote{Of
course, in reality the $\vartheta$--field is a massless pseudoscalar
field. As we show below (see also \cite{[3],[6]}--\cite{[16]}) it is
related to a chiral phase of a fermion field. Since we will not use
the properties of the $\vartheta$--field under parity transformations,
further on we call it for simplicity {\it a massless scalar field}.}
and a massless self--coupled fermion field $\psi(x)$.

A general term of $Z_{\rm SG}$ can be taken in the form \cite{[3]}
\begin{eqnarray}\label{label2.5}
&&\Big\langle 0\Big|{\rm T}\Big(\prod^{p}_{k=1}A_+(x_k)
\prod^{n}_{j=1}A_-(y_j)\Big)\Big|0\Big\rangle =\nonumber\\ &&=\int
{\cal D}\vartheta\,e^{\textstyle -i\,\frac{1}{2}\int
d^2x\,\vartheta(x)(\Box + \mu^2)\vartheta(x)}\prod^{p}_{k=1}A_+(x_k)
\prod^{n}_{j=1}A_-(y_j),
\end{eqnarray}
where $\mu$ is an infrared cut--off regularizing the free massless
$\vartheta$--field in the infrared region. The vacuum expectation
value (\ref{label2.5}) should be taken in the limit $\mu \to 0$
\cite{[3]}.

The evaluation of this Gaussian path integral is rather
straightforward. The result reads
\begin{eqnarray}\label{label2.6}
\hspace{-0.3in}&&\Big\langle 0\Big|{\rm T}\Big(\prod^{p}_{i=1}A_+(x_i)
\prod^{n}_{j=1}A_-(y_j)\Big)\Big|0\Big\rangle
=\exp\Big\{
\frac{1}{2}\,\beta^2\,(p+n)\,i\Delta(0)\Big\}\,\exp\Big\{ \beta^2\sum^{p}_{j
<k}i\Delta(x_j - x_k)\nonumber\\
\hspace{-0.3in}&& + \beta^2\sum^{n}_{j <k}i\Delta(y_j - y_k) -
\beta^2\sum^{p}_{j = 1}\sum^{n}_{k = 1}i\Delta(x_j - y_k)\Big\},
\end{eqnarray}
where the Green function $\Delta(x-y)$ is determined by \cite{[3]}
\begin{eqnarray}\label{label2.7}
\Delta(x-y) = i\langle 0|{\rm T}(\vartheta(x)\vartheta(y))|0\rangle
\end{eqnarray}
and obeys the equation \cite{[3]}
\begin{eqnarray}\label{label2.8}
(\Box + \mu^2)\,\Delta(x-y) = \delta^{(2)}(x - y).
\end{eqnarray}
In the limit $\mu \to 0$ the Green function $\Delta(x-y)$ is given by
the expression \cite{[3]}
\begin{eqnarray}\label{label2.9}
\Delta(x-y) = -\,\frac{i}{4\pi}\,{\ell n}[-\mu^2(x-y)^2 + i\,0].
\end{eqnarray}
Using the explicit form of the Green function (\ref{label2.9}) the vacuum expectation value
(\ref{label2.6}) transforms to
\begin{eqnarray}\label{label2.10}
\hspace{-0.3in}&&\Big\langle 0\Big|{\rm T}\Big(\prod^{p}_{k=1}A_+(x_k)
\prod^{n}_{j=1}A_-(y_j)\Big)\Big|0\Big\rangle =\nonumber\\
\hspace{-0.3in}&&=\exp\Big\{ \frac{1}{2}\,\beta^2\,(p+n)\,i\Delta(0)\Big\}\,
\frac{\displaystyle \prod^{p}_{j <k}[-\mu^2(x_j -
x_k)^2]^{\beta^2/4\pi}\prod^{n}_{j <k}[-\mu^2(y_j -
y_k)^2]^{\beta^2/4\pi}}{\prod^{p}_{j = 1}\prod^{n}_{k = 1}[-\mu^2(x_j -
y_k)^2]^{\beta^2/4\pi}},
\end{eqnarray}
in agreement with Coleman's result (see Eq.(4.11) of Ref.\cite{[3]}).

In the limit $\mu^2 \to 0$ this vacuum expectation value behaves like
\begin{eqnarray}\label{label2.11}
\Big\langle 0\Big|{\rm T}\Big(\prod^{p}_{k=1}A_+(x_k)
\prod^{n}_{j=1}A_-(y_j)\Big)\Big|0\Big\rangle \sim
(\mu^2\,)^{(p-n)^2\beta^2/8\pi}
\end{eqnarray}
and vanishes if $p\not= n$ \cite{[3]}. An analogous evaluation of the
vacuum expectation value (\ref{label2.5}) has been carried out by
Fr\"ohlich and Marchetti \cite{[7]}.

We would like to accentuate that the evaluation of the vacuum
expectation value (\ref{label2.10}) has been carried out with respect
to the trivial perturbative vacuum with $\langle \vartheta(x)\rangle =
0$ and with the trivial two--point Green function
(\ref{label2.9}). Therefore, none non--perturbative properties of the
SG model caused by the existence of non--trivial soliton states are
involved.

Now let us turn to the evaluation of the partition function $Z_{\rm
Th}$. From Eq.(\ref{label2.4}) one can see that every term of the
expansion in powers of $m$ is related to the vacuum expectation value
of a product of operators of massless self--coupled fermion  fields
$\psi(x)$ and $\bar{\psi}(x)$
\begin{eqnarray}\label{label2.12}
\Big\langle 0\Big|{\rm
T}\Big(\prod^{n}_{k=1}\sigma_-(x_k)\sigma_+(y_k)\Big)\Big|0\Big\rangle
= \Big\langle 0\Big|{\rm
T}\Big(\prod^n_{i=1}[\psi^{\dagger}_1(x_i)\psi_2(x_i)]
[\psi^{\dagger}_2(y_i)\psi_1(y_i]\Big)\Big|0\Big\rangle.
\end{eqnarray}
For the evaluation of these vacuum expectation values Coleman has
decided to follow as close as possible the results obtained by Klaiber
in his lectures \cite{[5]}. According to Klaiber's statement the massless
Thirring model can be reduced to a quantum field theory of a massless
free fermion  field $\Psi(x)$ by a corresponding canonical
transformation of the self--coupled fermion field $\psi(x) \to
\Psi(x)$ \cite{[5]}. The two--point Green function $S_{\rm F}(x-y)$ of the
free massless field $\Psi(x)$ is defined by \cite{[5]}
\begin{eqnarray}\label{label2.13}
S_{\rm F}(x-y) = i\langle 0|{\rm T}(\Psi(x)\bar{\Psi}(y))|0\rangle =
\frac{1}{2\pi}\frac{\hat{x} - \hat{y}}{(x-y)^2 - i\,0},
\end{eqnarray}
where $\hat{x} - \hat{y} = \gamma^{\mu}(x-y)_{\mu}$. We would like to
emphasize that the fermion condensate, determined in the usual way, is
equal to zero
\begin{eqnarray}\label{label2.14}
\lim_{y\to x}i\langle 0|{\rm T}(\bar{\Psi}(y)\Psi(x))|0\rangle = -
\lim_{y\to x}{\rm tr}[S_{\rm F}(x-y)] = 0.
\end{eqnarray}
This shows clearly that fermion  fields are quantized around the
trivial perturbative vacuum.

The result of the calculation of the vacuum expectation value
(\ref{label2.12}) was obtained by Klaiber \cite{[5]} and used by
Coleman \cite{[3]} in the form
\begin{eqnarray}\label{label2.15}
\Big\langle 0\Big|{\rm
T}\Big(\prod^{n}_{k=1}\sigma_-(x_k)\sigma_+(y_k)
\Big)\Big|0\Big\rangle \sim \frac{\displaystyle \prod^{p}_{j
<k}[-\bar{\mu}^2(x_j - x_k)^2]^{1 + b/\pi}\prod^{n}_{j
<k}[-\bar{\mu}^2(y_j - y_k)^2]^{1 + b/\pi}}{\displaystyle \prod^{n}_{j
= 1}\prod^{n}_{k = 1}[-\bar{\mu}^2(x_j - y_k)^2]^{1 + b/\pi}},
\end{eqnarray}
where $\bar{\mu}$ is an arbitrary scale and the parameter $b$ is given
by \cite{[3],[5]}
\begin{eqnarray}\label{label2.16}
1 + \frac{b}{\pi} = \frac{1}{\displaystyle 1 + \frac{g}{\pi}}.
\end{eqnarray}
The comparison of Eq.(\ref{label2.10}) for $p=n$ with
Eq.(\ref{label2.15}) led Coleman to the relation between the coupling
constants given by Eq.(\ref{label1.9}) at $\bar{\mu} \sim \mu$
\cite{[3]}.

The evaluation of the Green functions in the massless Thirring model
carried out by Klaiber within the operator technique was
then confirmed by Furuya, Gamboa Saravi and Schaposnik within the path
integral approach supplemented by the method of auxiliary vector
fields \cite{[6]}.

Thus, we have to emphasize that the fermion fields in Coleman's
derivation of the equivalence between the SG and the Thirring model
have been obviously quantized around the trivial perturbative vacuum
in the chiral symmetric phase. Therefore, it is not a surprise that
Coleman's relation between coupling constants differs from our
relation valid for fermion fields quantized around a non--trivial,
non--perturbative vacuum in the chirally broken phase.

\section{Bosonization of the massless Thirring model}
\setcounter{equation}{0}

\hspace{0.2in} Within our approach to the equivalence between the SG
and the Thirring model we suggest to start, first, with the massless
Thirring model and bosonize it by integrating over fermionic degrees
of freedom. We consider the partition function
\begin{eqnarray}\label{label3.1}
Z_{\rm Th} = \int {\cal D}\psi{\cal D}\bar{\psi}\,\exp i\int
d^2x\,\Big\{\bar{\psi}(x)i\gamma^{\mu}\partial_{\mu}\psi(x) -
\frac{1}{2}\,g\,\bar{\psi}(x)\gamma_{\mu}\psi(x)\bar{\psi}(x)
\gamma^{\mu}\psi(x)\Big\}.
\end{eqnarray}
After a Fierz transformation of the four--fermion interaction we get
\begin{eqnarray}\label{label3.2}
\hspace{-0.5in}&&Z_{\rm Th} = \nonumber\\
\hspace{-0.5in}&&= \int {\cal D}\psi{\cal D}\bar{\psi}\,\exp \,i\int
d^2x\,\Big\{\bar{\psi}(x)i\gamma^{\mu}\partial_{\mu}\psi(x) +
\frac{1}{2}\,g\,[(\bar{\psi}(x)\psi(x))^2 +
(\bar{\psi}(x)i\gamma^5\psi(x))^2]\Big\}.
\end{eqnarray}
Collective $\bar{\psi}\psi$ excitations, a local scalar field
$\sigma(x)$ and a pseudoscalar $\varphi(x)$, can be introduced in the
theory as usual \cite{[18]}--\cite{[22]}
\begin{eqnarray}\label{label3.3}
\hspace{-0.5in}&&Z_{\rm Th} = \int {\cal D}\psi{\cal D}
\bar{\psi}{\cal D}\sigma {\cal D}\varphi\nonumber\\
\hspace{-0.5in}&&\times\exp i\!\!\int
d^2x\,\Big\{\bar{\psi}(x)i\gamma^{\mu}\partial_{\mu}\psi(x) -
\bar{\psi}(x)(\sigma(x) + i\gamma^5\varphi(x))\psi(x) -
\frac{1}{2g}\,[\sigma^2(x) + \varphi^2(x)]\Big\}.
\end{eqnarray}
Integrating over fermionic  degrees of freedom we recast the integrand
into the form
\begin{eqnarray}\label{label3.4}
Z_{\rm Th} =\int {\cal D}\sigma {\cal D}\varphi\,{\rm
Det}(i\gamma^{\mu}\partial_{\mu} - \sigma - i\gamma^5\varphi)\exp
\,i\int d^2x\,\Big\{ - \frac{1}{2g}\,[\sigma^2(x) +
\varphi^2(x)]\Big\}.
\end{eqnarray}
This reduces the problem of the bosonization of the massless Thirring
model to the evaluation of the functional determinant
\begin{eqnarray}\label{label3.5}
{\rm Det}(i\gamma^{\mu}\partial_{\mu} - \sigma - i\gamma^5\varphi).
\end{eqnarray}
This determinant is related to the effective Lagrangian in the usual way
\begin{eqnarray}\label{label3.6}
&&{\rm Det}(i\gamma^{\mu}\partial_{\mu} - \sigma - i\gamma^5\varphi) =
\exp {\rm Tr}\,{\ell n}(i\gamma^{\mu}\partial_{\mu} - \sigma -
i\gamma^5\varphi) = \nonumber\\ &&=\exp \,i\int d^2x\,(-i)\,{\rm
tr}\,\langle x|{\ell n}(i\gamma^{\mu}\partial_{\mu} - \sigma -
i\gamma^5\varphi)|x\rangle = \exp \,i\int d^2x\,\tilde{{\cal L}}_{\rm
eff}(x),
\end{eqnarray}
where 
\begin{eqnarray}\label{label3.7}
\tilde{{\cal L}}_{\rm eff}(x) = (-i){\rm tr}\langle x|{\ell
n}(i\gamma^{\mu}\partial_{\mu} - \sigma - i\gamma^5\varphi)|x\rangle.
\end{eqnarray}
First, let us drop the contribution of gradients
$\partial_{\mu}\sigma$ and $\partial_{\mu}\varphi$ and evaluate the
effective potential $\tilde{V}[\sigma(x), \varphi(x)] = - \tilde{{\cal
L}}_{\rm eff}(x)|_{\partial_{\mu}\sigma = \partial_{\mu}\varphi = 0}$.

Dropping the contribution of the gradients $\partial_{\mu}\sigma$ and
$\partial_{\mu}\varphi$, the evaluation of the functional determinant
(\ref{label3.5}) runs in the following way
\begin{eqnarray}\label{label3.8}
\hspace{-0.3in}&&{\rm Det}(i\gamma^{\mu}\partial_{\mu} - \sigma -
i\gamma^5\varphi)|_{\partial_{\mu}\sigma = 
\partial_{\mu}\varphi = 0} = {\rm Det}(\Box + \Phi^{\dagger}\Phi) =
\nonumber\\
\hspace{-0.3in}&&= \exp\,i\int d^2x\,(-i){\rm tr}\langle x|{\ell
n}(\Box + \Phi^{\dagger}\Phi)|x\rangle = \exp\,i\int d^2x\int
\frac{d^2k}{(2\pi)^2i}{\ell n}(-k^2 + \Phi^{\dagger}(x)\Phi(x))
=\nonumber\\
\hspace{-0.3in}&&= \exp\,i\int d^2x\int \frac{d^2k_{\rm
E}}{(2\pi)^2}{\ell n}(k^2_{\rm E} + \Phi^{\dagger}(x)\Phi(x)),
\end{eqnarray}
where $\Phi(x) = \sigma(x) + i\varphi(x)$ and $k_{\rm E}$ is the
2--momentum in Euclidean momentum space obtained from the 2--momentum
$k$ in Minkowski momentum space by means of a Wick rotation $k_0 =
ik_2$ \cite{[29]}. The effective potential defined by the functional
determinant (\ref{label3.8}) amounts to
\begin{eqnarray}\label{label3.9}
\hspace{-0.5in}&&- \tilde{V}[\sigma(x), \varphi(x)] = \tilde{{\cal
L}}_{\rm eff}(x)|_{\partial_{\mu}\sigma = \partial_{\mu}\varphi = 0}
=\int \frac{d^2k_{\rm E}}{(2\pi)^2}{\ell n}(k^2_{\rm E} +
\Phi^{\dagger}(x)\Phi(x)) =\nonumber\\
\hspace{-0.5in}&&= \frac{1}{4\pi}\Big[(\Lambda^2 +
\Phi^{\dagger}(x)\Phi(x)){\ell n}(\Lambda^2 +
\Phi^{\dagger}(x)\Phi(x)) - \Phi^{\dagger}(x)\Phi(x){\ell
n}\Phi^{\dagger}(x)\Phi(x) - \Lambda^2\Big].
\end{eqnarray}
This result can be obtained differently by representing the effective
Lagrangian (\ref{label3.7}) in terms of one--fermion loop diagrams
\cite{[22]}. Denoting $\tilde{\Phi}=\sigma + i\gamma^5\varphi$ we get \cite{[22]}
\begin{eqnarray}\label{label3.10}
&&\tilde{{\cal L}}_{\rm eff}(x) = -i\,{\rm tr}\langle x|{\ell
n}(i\gamma^{\mu}\partial_{\mu} - \tilde{\Phi})|x\rangle = -i\,{\rm
tr}\langle x|{\ell n}(i\gamma^{\mu}\partial_{\mu})|x\rangle
\nonumber\\ &&+ \sum^{\infty}_{n=1}\frac{i}{n}{\rm tr}\Big\langle
x\Big|\Big(\frac{1}{i\gamma^{\mu}\partial_{\mu}}\tilde{\Phi}\Big)^n
\Big|x\Big\rangle
= -i\,{\rm tr}\langle x|{\ell n}(i\gamma^{\mu}\partial_{\mu})|x\rangle
+ \sum^{\infty}_{n=1}\tilde{{\cal L}}^{(n)}_{\rm eff}(x),
\end{eqnarray}
where the effective Lagrangian $\tilde{{\cal L}}^{(n)}_{\rm eff}(x)$
is defined by \cite{[22]}
\begin{eqnarray}\label{label3.11}
&&\tilde{{\cal L}}^{(n)}_{\rm eff}(x) = \int
\prod^{n-1}_{\ell}\frac{d^2x_{\ell}d^2k_{\ell}}{(2\pi)^2}\,e^{\textstyle
-ik_1\cdot x_1 -ik_2\cdot x_2 - \ldots -ik_n\cdot
x}\,\Big(-\frac{1}{n}\frac{1}{4\pi}\Big)\nonumber\\
&&\times\int\frac{d^2k}{\pi i}\,{\rm
tr}\Big\{\frac{1}{\hat{k}}\tilde{\Phi}(x_1)\frac{1}{\hat{k} +
\hat{k}_1}\tilde{\Phi}(x_2)\ldots\tilde{\Phi}(x_{n-1})\frac{1}{\hat{k}
+ \hat{k}_1 + \ldots + \hat{k}_{n-1}}\tilde{\Phi}(x)\Big\}
\end{eqnarray}
at $k_1 + k_2 + \ldots + k_n = 0$.

Dropping the momenta $k_i\,(i = 1,2,\ldots, n-1)$ giving the
contributions of the gradients $\partial_{\mu}\sigma$ and
$\partial_{\mu}\varphi$ in the effective Lagrangian we recast
$\tilde{{\cal L}}^{(n)}_{\rm eff}(x)$ into the form
\begin{eqnarray}\label{label3.12}
\tilde{{\cal L}}^{(n)}_{\rm eff}(x) =
-\frac{1}{n}\frac{1}{4\pi}\int\frac{d^2k}{\pi i}\,{\rm
tr}\Big\{\frac{1}{\hat{k}}\tilde{\Phi}(x)\frac{1}{\hat{k}}
\tilde{\Phi}(x)\ldots\tilde{\Phi}(x)\frac{1}{\hat{k}}\tilde{\Phi}(x)\Big\}.
\end{eqnarray}
A non--zero contribution comes only from even $n$, $n = 2 m\,(m =
1,2,\ldots)$
\begin{eqnarray}\label{label3.13}
\tilde{{\cal L}}^{(2m)}_{\rm eff}(x)&=&
-\frac{1}{m}\frac{1}{4\pi}(\Phi^{\dagger}(x)\Phi(x))^m\int\frac{d^2k}{\pi
i}\,\frac{1}{(k^2)^m} =\nonumber\\
&=&\frac{(-1)^{m+1}}{m}\frac{1}{4\pi}(\Phi^{\dagger}(x)\Phi(x))^m
\int^{\Lambda}_{\mu}\frac{dk^2_{\rm
E}}{(k^2_{\rm E})^m},
\end{eqnarray}
where $\Lambda$ and $\mu$ are the ultra--violet and infra--red
cut--offs. For $m=1$ we get
\begin{eqnarray}\label{label3.14}
\tilde{{\cal L}}^{(2)}_{\rm eff}(x) =
\Phi^{\dagger}(x)\Phi(x)\,\frac{1}{4\pi}\,{\ell
n}\frac{\Lambda^2}{\mu^2}.
\end{eqnarray}
In turn for $m \not= 1$ we obtain
\begin{eqnarray}\label{label3.15}
\tilde{{\cal L}}^{(2m)}_{\rm eff}(x) =
\frac{(-1)^m}{m(m-1)}(\Phi^{\dagger}(x)\Phi(x))^m\,\frac{1}{4\pi}\,
\Bigg[\Bigg(\frac{1}{\Lambda^2}\Bigg)^{m-1}
- \Bigg(\frac{1}{\mu^2}\Bigg)^{m-1}\Bigg].
\end{eqnarray}
The total effective Lagrangian is given by 
\begin{eqnarray}\label{label3.16}
&&\tilde{{\cal L}}_{\rm eff}(x) = -i\,{\rm tr}\langle x|{\ell
n}(i\gamma^{\mu}\partial_{\mu})|x\rangle +
\Phi^{\dagger}(x)\Phi(x)\,\frac{1}{4\pi}\,{\ell
n}\frac{\Lambda^2}{\mu^2}\nonumber\\ &&+
\frac{1}{4\pi}\sum^{\infty}_{n=1}\frac{(-1)^{n+1}}{n(n+1)}\,
(\Phi^{\dagger}(x)\Phi(x))^{n+1}\,\Bigg[\Bigg(\frac{1}{\Lambda^2}\Bigg)^{n}
- \Bigg(\frac{1}{\mu^2}\Bigg)^{n}\Bigg].
\end{eqnarray}
Summing up the infinite series we arrive at the expression
\begin{eqnarray}\label{label3.17}
\hspace{-0.7in}&&\tilde{{\cal L}}_{\rm eff}(x) = 
\frac{1}{4\pi}\Bigg[(\Lambda^2{\ell
n}\Lambda^2 - \Lambda^2 - \mu^2{\ell n}\mu^2 + \mu^2) +
\Phi^{\dagger}(x)\Phi(x)\,{\ell n}\frac{\Lambda^2}{\mu^2}\nonumber\\
\hspace{-0.7in}&&+ (\Lambda^2 + \Phi^{\dagger}(x)\Phi(x)){\ell n}\Bigg(1 +
\frac{\Phi^{\dagger}(x)\Phi(x)}{\Lambda^2}\Bigg) - (\mu^2 +
\Phi^{\dagger}(x)\Phi(x)){\ell n}\Bigg(1 +
\frac{\Phi^{\dagger}(x)\Phi(x)}{\mu^2}\Bigg)\Bigg],
\end{eqnarray}
where we have taken into account that (see (\ref{label3.8}))
\begin{eqnarray}\label{label3.18}
-i\,{\rm tr}\langle x|{\ell n}(i\gamma^{\mu}\partial_{\mu})|x\rangle =
\frac{1}{4\pi}\int^{\Lambda}_{\mu}dk^2_{\rm E}\,{\ell n}k^2_{\rm E} =
\frac{1}{4\pi}\,(\Lambda^2{\ell n}\Lambda^2 - \Lambda^2 - \mu^2{\ell
n}\mu^2 + \mu^2).
\end{eqnarray}
Eq.(\ref{label3.17}) can be simplified
\begin{eqnarray}\label{label3.19}
\hspace{-0.5in}&&\tilde{{\cal L}}_{\rm eff}(x) =
\frac{1}{4\pi}\,\Big[(\mu^2 - \Lambda^2)\nonumber\\
\hspace{-0.5in}&& + (\Lambda^2 + \Phi^{\dagger}(x)\Phi(x)){\ell
n}(\Lambda^2 + \Phi^{\dagger}(x)\Phi(x)) - (\mu^2 +
\Phi^{\dagger}(x)\Phi(x)){\ell n}(\mu^2 +
\Phi^{\dagger}(x)\Phi(x))\Big].
\end{eqnarray}
Setting $\mu = 0$ we arrive at the effective potential (\ref{label3.9}).

The total effective potential we obtain by summing up
Eq.(\ref{label3.9}) and the quadratic term of Eq.(\ref{label3.3})
which has the form $(1/2g)\,\Phi^{\dagger}(x)\Phi(x)$:
\begin{eqnarray}\label{label3.20}
&&V[\Phi^{\dagger}(x)\Phi(x)] = \tilde{V}[\Phi^{\dagger}(x)\Phi(x)] +
\frac{1}{2g}\,\Phi^{\dagger}(x)\Phi(x) = \frac{1}{4\pi}\Big[
\Phi^{\dagger}(x)\Phi(x){\ell n}\Phi^{\dagger}(x)\Phi(x)\nonumber\\
&&- (\Lambda^2 + \Phi^{\dagger}(x)\Phi(x)){\ell n}(\Lambda^2 +
\Phi^{\dagger}(x)\Phi(x)) + \frac{2\pi}{g}\,\Phi^{\dagger}(x)\Phi(x) +
\Lambda^2\Big].
\end{eqnarray}
In polar representation $\Phi(x) = \rho(x)\,e^{\textstyle
i\,\vartheta(x)}$ corresponding to $\sigma(x) = \rho(x)\,\cos
\vartheta(x)$ and $\varphi(x) = \rho(x)\,\sin \vartheta(x)$ the
effective potential depends only on the $\rho$--field and reads
\begin{eqnarray}\label{label3.21}
V[\rho(x)] = \frac{1}{4\pi}\Bigg[ \rho^2(x){\ell
n}\,\frac{\rho^2(x)}{\Lambda^2} - (\Lambda^2 + \rho^2(x)){\ell
n}\Bigg(1+ \frac{\rho^2(x)}{\Lambda^2}\Bigg)
+\frac{2\pi}{g}\,\rho^2(x) \Bigg],
\end{eqnarray}
where we have dropped the unimportant divergent contribution
$(\Lambda^2 - \Lambda^2\,{\ell n}\,\Lambda^2)/4\pi$. This shifts the
effective potential to $V[0] = 0$.

It is well--known that a quantum system has to be quantized around the
minima of the effective potential. They are defined by\,\footnote{The
vacuum average $\langle 0|V[\rho(x)]|0 \rangle$ of the effective
potential we carry out in the tree--approximation for the
$\rho$--field \cite{[19]}--\cite{[22]}. This yields $\langle
0|V[\rho(x)]|0 \rangle_{\rm tree} = V[\langle \rho(x) \rangle]$.}
\begin{eqnarray}\label{label3.22}
\frac{\delta V[\bar{\rho}(x)]}{\delta \bar{\rho}(x)} =
\frac{1}{2\pi}\,\bar{\rho}(x)\,\Bigg[- {\ell n}\Bigg(1 +
\frac{\Lambda^2}{\bar{\rho}^{\,2}(x)}\Bigg) + \frac{2\pi}{g} \Bigg] =
0,
\end{eqnarray}
where $\bar{\rho}(x)$ is the vacuum expectation value of the
$\rho$--field, $\bar{\rho}(x) = \langle \rho(x) \rangle$. The equation
(\ref{label3.22}) has a trivial solution $\bar{\rho}(x) = 0$ which
corresponds to a maximum of the potential and a non--trivial one
\begin{eqnarray}\label{label3.23}
\bar{\rho}(x) = \frac{\Lambda}{\displaystyle \sqrt{e^{\textstyle
2\pi/g} - 1}}.
\end{eqnarray}
The only constraint on the existence of the non--trivial solution is
$g > 0$. This condition is trivial, since according to the analysis by
Nambu and Jona--Lasinio \cite{[18]} bound collective $\bar{\psi}\psi$
excitations can appear in a theory with the Lagrangian
(\ref{label1.18}) only in the case of attraction between fermions,
i.e. for positive $g$.

\begin{figure}[h]
\psfrag{xl}{$\displaystyle{\frac{\rho}{\Lambda}}$}
\psfrag{ylabel}{$\displaystyle{\frac{4\pi}{\Lambda^2}} \; V\left(\rho\right)$}
\centerline{\scalebox{1.0}{\includegraphics{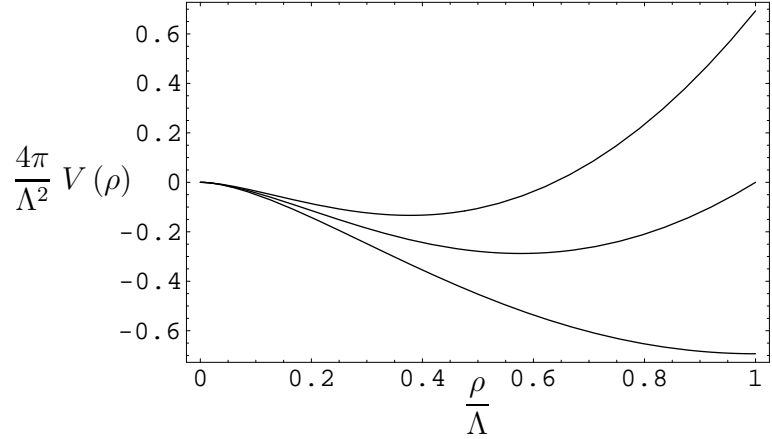}}}
\caption{The effective potential $V\left(\rho\right)$ of
Eq.(\ref{label3.21}) as a function of $\rho/\Lambda$ for $2\pi/g =
{\ell n}\,2^k$ with $k=1,2$ and 3.}
\label{fig1}
\end{figure}
The effective potential $V\left(\rho\right)$ of Eq.(\ref{label3.21})
as a function of $\rho/\Lambda$ is depicted in Fig.\ref{fig1} for
$2\pi/g = {\ell n}\,2^k$ with $k=1,2$ and 3. One can clearly see the
maximum at $\bar{\rho} = 0$ and the minimum at $\bar{\rho}^2/\Lambda^2
= 1/(2^k - 1)$ corresponding to a non--trivial solution of the
gap--equation (\ref{label3.22}).

From the second derivative one can see that the effective potential
(\ref{label3.21}) has a minimum only for the non--trivial solution
of $\bar{\rho}(x)$ defined by Eq.(\ref{label3.23}). We denote
this non--trivial solution $\bar{\rho}(x) = \rho_0$.

One can show that $\rho_0$ coincides with the dynamical mass $M$ given
by Eq.(\ref{label1.15}). To show this we derive the equations of
motion
\begin{eqnarray}\label{label3.24}
\bar{\psi}(x)\psi(x) &=& -\,\frac{\sigma(x)}{g},\nonumber\\
\bar{\psi}(x)i\gamma^5\psi(x) &=& -\,\frac{\varphi(x)}{g},
\end{eqnarray}
from the linearized Lagrangian defining the
partition function $Z_{\rm Th}$ in Eq.(\ref{label3.3}).

The vacuum average $\langle 0|\bar{\psi}(x)\psi(x)|0\rangle$ of the
first equation of motion in the one--fermion loop approximation for
the $\psi$--field and in the tree--approximation of the
$\sigma$--field gives
\begin{eqnarray}\label{label3.25}
\langle 0|\bar{\psi}(x)\psi(x)|0\rangle_{\rm one-loop} =
-\,\frac{\langle 0|\sigma(x)|0 \rangle_{\rm tree}}{g} =
-\,\frac{\rho_0}{g},
\end{eqnarray}
where $\langle 0|\sigma(x)|0 \rangle_{\rm tree} =
\langle\rho(x)\rangle \langle 0|\cos\vartheta(x)|0 \rangle_{\rm tree}
= \rho_0\,\cos \langle 0|\vartheta(x)|0 \rangle = \rho_0.$ 

Matching the r.h.s. of Eq.(\ref{label3.25}) with Eq.(\ref{label1.16})
for the fermion condensate one obtains $\rho_0 = M$. This demonstrates
the complete agreement between the fermionic and bosonic description
of the massless Thirring model. This result runs parallel the dynamics
of the evolution of the fermion system in NJL models describing well
both low--energy interactions of hadrons \cite{[19]}--\cite{[21]} and
confinement \cite{[22]}. Below we would use $M$ instead of $\rho_0$.

Expanding the effective potential around the minimum $\rho(x) = \rho_0
+ \tilde{\rho}(x)$ we get
\begin{eqnarray}\label{label3.26}
V[\tilde{\rho}(x)] &=&V[\rho_0] + \frac{1}{2\pi}\,\Big(1 -
e^{\textstyle -2\pi/g}\Big)\,\tilde{\rho}^{\,2}(x)\nonumber\\ &+&
\frac{1}{6\pi}\,e^{\textstyle -2\pi/g}\,\Big(1 - e^{\textstyle
-2\pi/g}\,\Big)^{3/2}\Big(1 - 2\,e^{\textstyle
-2\pi/g}\,\Big)\,\frac{\tilde{\rho}^{\,3}(x)}{\Lambda} +
O\Big(\frac{1}{\Lambda^2}\Big).
\end{eqnarray}
Keeping only terms surviving in the $\Lambda \to \infty$ limit we
arrive at the expression
\begin{eqnarray}\label{label3.27}
V[\tilde{\rho}(x)] =\frac{1}{2\pi}\,\Big(1 - e^{\textstyle
-2\pi/g}\,\Big)\,\tilde{\rho}^{\,2}(x),
\end{eqnarray}
where we have dropped the trivial infinite constant $V[M]$.

It is clear from dimensional considerations that the
gradient terms of the $\tilde{\rho}$--field
$\partial_{\mu}\tilde{\rho}(x)$ appear in the effective Lagrangian
only in the ratio $\partial_{\mu}\tilde{\rho}(x)/\Lambda$. Thereby,
they vanish in the limit $\Lambda \to \infty$.

\begin{figure}[h]
\centerline{\scalebox{1.0}{\includegraphics{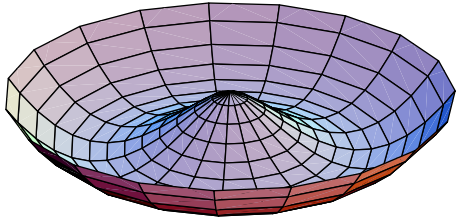}}}
\caption{The effective potential $V\left(\rho\right)$ of 
Eq.(\ref{label3.21}) as a function of $\rho$ and $\vartheta$ 
for $2\pi/g = 2{\ell n}\,2$.}
\label{fig2}
\end{figure}
Hence, the effective potential defined by Eq.(\ref{label3.27}) implies
that the fluctuations of the $\rho$--field around the minimum
(\ref{label3.21}) of the effective potential are described by a free
scalar field $\tilde{\rho}(x)$ decoupled from the phase field
$\vartheta(x)$.\footnote{The decoupling of the $\tilde{\rho}$--field
is demonstrated in more detail in Appendix B.} The $\rho$ and
$\vartheta$ dependence of the effective potential $V\left(\rho\right)$
of Eq.(\ref{label3.21}) is shown in Fig. \ref{fig2} for $2\pi/g =
{\ell n}\,2^2$. For further investigation of the dynamics of the
$\vartheta$--field one can integrate out the degrees of freedom
related to the $\tilde{\rho}$--field and employ the representation
\begin{eqnarray}\label{label3.28}
\Phi(x) = \sigma(x) + i\varphi(x) =M\,e^{\textstyle i\,\vartheta(x)}.
\end{eqnarray}
Hence, a bosonized version of the massless Thirring model obtained
from the chirally broken phase of the fermion  system is defined by
only one degree of freedom, a scalar field $\vartheta(x)$.

In the approximation presented by Eq.(\ref{label3.28}) the partition
function (\ref{label3.4}) reduces to the form
\begin{eqnarray}\label{label3.29}
Z_{\rm Th} =\int {\cal D}\vartheta\,{\rm
Det}\Big(i\gamma^{\mu}\partial_{\mu} - M\,e^{\textstyle i\,\gamma^5
\vartheta}\,\Big),
\end{eqnarray}
where we have dropped a trivial infinite constant. The functional
determinant can be transformed as follows
\begin{eqnarray}\label{label3.30}
{\rm Det}\Big(i\gamma^{\mu}\partial_{\mu} - M\,e^{\textstyle
i\,\gamma^5 \vartheta}\,\Big) &=&{\rm Det}\Big(e^{\textstyle
i\,\gamma^5 \vartheta/2}\,(i\gamma^{\mu}\partial_{\mu} +
\gamma^{\mu}A_{\mu} - M)\,e^{\textstyle i\,\gamma^5
\vartheta/2}\,\Big) =\nonumber\\ &=&J[\vartheta]\,{\rm
Det}(i\gamma^{\mu}\partial_{\mu} + \gamma^{\mu}A_{\mu} - M),
\end{eqnarray}
where we have denoted 
\begin{eqnarray}\label{label3.31}
A_{\mu}(x) =
\frac{1}{2}\,\varepsilon_{\mu\nu}\,\partial^{\nu}\vartheta(x).
\end{eqnarray}
The Jacobian $J[\vartheta]$ induced by a local chiral rotation can be
calculated in the usual way \cite{[12]}--\cite{[17]}. In the Appendix we show that by
using an appropriate regularization scheme this Jacobian can be
obtained to be equal to unity
\begin{eqnarray}\label{label3.32}
J[\vartheta] =1.
\end{eqnarray}
The partition function (\ref{label3.29}) reads then
\begin{eqnarray}\label{label3.33}
Z_{\rm Th} = \int {\cal D}\vartheta\,{\rm
Det}(i\gamma^{\mu}\partial_{\mu} + \gamma^{\mu}A_{\mu} - M) =
\int {\cal D}\vartheta\,\exp\,i\,\int d^2x\,{\cal L}_{\rm eff}(x).
\end{eqnarray}
The simplest way to calculate the effective Lagrangian ${\cal L}_{\rm
eff}(x)$ is to represent it in the form of one--fermion loop diagrams
\cite{[19]}--\cite{[22]}. Since $M$ is proportional to $\Lambda$, it is clear
from dimensional considerations that the main contribution should come
from the diagram with two vertices. The contribution of the diagram
with $n > 2$ vertices falls as $O(1/\Lambda^{n-2})$ at $\Lambda \to
\infty$. That is why the effective Lagrangian ${\cal L}_{\rm eff}(x)$
is determined by
\begin{eqnarray}\label{label3.34}
\hspace{-0.3in}{\cal L}_{\rm eff}(x)&=& -i\,\langle x|{\rm tr}{\ell
n}(i\gamma^{\mu}\partial_{\mu} - M)|x \rangle -
\frac{1}{8\pi}\int\frac{d^2x_1d^2k_1}{(2\pi)^2}\,e^{\textstyle
-ik_1\cdot(x_1 - x)}A_{\mu}(x)A_{\nu}(x_1)\nonumber\\
\hspace{-0.3in}&&\times\int \frac{d^2k}{\pi
i}\,{\rm tr}\Bigg\{\frac{1}{M -
\hat{k}}\gamma^{\mu}\frac{1}{M - \hat{k} -
\hat{k}_1}\gamma^{\nu}\Bigg\}.
\end{eqnarray}
Omitting a trivial infinite constant and keeping only the leading
contribution at $\Lambda \to \infty$ we get
\begin{eqnarray}\label{label3.35}
{\cal L}_{\rm eff}(x) = \frac{1}{16\pi}\,\Big(1 - e^{\textstyle
-2\pi/g}\,\Big)\,\partial_{\mu}\vartheta(x)\,\partial^{\mu}\vartheta(x),
\end{eqnarray}
where we have used the relation $\varepsilon_{\mu\alpha}
\,\varepsilon^{\nu\alpha} = - g^{\nu}_{\mu}$.

This result testifies that the bosonized version of the
massless Thirring model obtained from the chirally broken phase of the
fermion  system is a quantum field theory of a free massless scalar
field $\vartheta(x)$.

\section{Generating functional of Green functions in the massless 
Thirring model. Bosonization rules} 
\setcounter{equation}{0}

\hspace{0.2in} Now we are able to turn to the problem of an explicit
evaluation of arbitrary correlation functions in the massless Thirring
model.  For this aim we consider the generating functional of Green
functions defined by
\begin{eqnarray}\label{label4.1}
\hspace{-0.5in}&&Z_{\rm Th}[J,\bar{J}] = \int {\cal D}\psi{\cal
D}\bar{\psi}\,\exp i\int
d^2x\,\Big\{\bar{\psi}(x)i\gamma^{\mu}\partial_{\mu}\psi(x) -
\frac{1}{2}\,g\,\bar{\psi}(x)\gamma_{\mu}\psi(x)\bar{\psi}(x)
\gamma^{\mu}\psi(x)\nonumber\\
\hspace{-0.5in}&& + \bar{\psi}(x)J(x) + \bar{J}(x)\psi(x)\Big\}=\nonumber\\
\hspace{-0.5in}&&= \int {\cal D}\psi{\cal D}\bar{\psi}\,\exp \,i\int
d^2x\,\Big\{\bar{\psi}(x)i\gamma^{\mu}\partial_{\mu}\psi(x) +
\frac{1}{2}\,g\,[(\bar{\psi}(x)\psi(x))^2 +
(\bar{\psi}(x)i\gamma^5\psi(x))^2]\nonumber\\
\hspace{-0.5in}&& + \bar{\psi}(x)J(x) + \bar{J}(x)\psi(x)\Big\}=\nonumber\\
\hspace{-0.5in}&&= \int {\cal D}\psi{\cal D}\bar{\psi}{\cal D}\sigma
{\cal D}\varphi\,\exp \,i\int
d^2x\,\Big\{\bar{\psi}(x)i\gamma^{\mu}\partial_{\mu}\psi(x) -
\bar{\psi}(x)(\sigma(x) + i\gamma^5\varphi(x))\psi(x) \nonumber\\
\hspace{-0.5in}&& + \bar{\psi}(x)J(x) + \bar{J}(x)\psi(x) -
\frac{1}{2g}\,[\sigma^2(x) + \varphi^2(x)]\Big\},
\end{eqnarray}
where $\bar{J}(x)$ and $J(x)$ are external sources of the Thirring
fields $\psi(x)$ and $\bar{\psi}(x)$. Therewith, the
external source $J(x)$, a column matrix with components $J_1(x)$ and
$J_2(x)$, is responsible for the production of the
$\psi^{\dagger}_2(x)$ and $\psi^{\dagger}_1(x)$ fields, whereas the
external source $\bar{J}(x)$, a row matrix with components
$J^{\dagger}_2(x)$ and $J^{\dagger}_1(x)$, produces the fields
$\psi_1(x)$ and $\psi_2(x)$.

Integrating over the fermion fields we arrive at
\begin{eqnarray}\label{label4.2}
Z_{\rm Th}[J,\bar{J}] &=&\int {\cal D}\sigma {\cal D}\varphi\,{\rm
Det}(i\gamma^{\mu}\partial_{\mu} - \sigma - i\gamma^5\varphi)\exp
\,i\int d^2x\,\Big\{ - \frac{1}{2g}\,[\sigma^2(x) +
\varphi^2(x)]\Big\}\nonumber\\ &&\times\,\exp \,i\int d^2x\,\Big\{-\,
\bar{J}(x)\,\frac{1}{\displaystyle i\gamma^{\mu}\partial_{\mu} -
\sigma(x) - i\gamma^5\varphi(x)}\,J(x)\Big\}.
\end{eqnarray}
Skipping intermediate steps expounded in detail in Section 3 we get
\begin{eqnarray}\label{label4.3}
Z_{\rm Th}[J,\bar{J}] &=&\int {\cal D}\vartheta\,\exp \,i\int
d^2x\,\Big\{\frac{1}{16\pi}\,\Big(1 - e^{\textstyle
-2\pi/g}\,\Big)\,\partial_{\mu}\vartheta(x)\,\partial^{\mu}\vartheta(x)
\nonumber\\ && - \bar{J}(x)\,\frac{1}{\displaystyle
i\gamma^{\mu}\partial_{\mu} - M\,e^{\textstyle
i\gamma^5\vartheta(x)}}\,J(x)\Big\}.
\end{eqnarray}
Keeping only leading terms in the $1/M$ expansion (or equivalently in $1/\Lambda$) we obtain
\begin{eqnarray}\label{label4.4}
Z_{\rm Th}[J,\bar{J}] &=&\int {\cal D}\vartheta\,\exp \,i\int
d^2x\,\Big\{\frac{1}{16\pi}\,\Big(1 - e^{\textstyle
-2\pi/g}\,\Big)\,\partial_{\mu}\vartheta(x)\,\partial^{\mu}\vartheta(x)
\nonumber\\ && + \frac{1}{M}\bar{J}(x)\Big(\frac{1 -
\gamma^5}{2}\Big)J(x)\,e^{\textstyle i\vartheta(x)} +
\frac{1}{M}\bar{J}(x)\Big(\frac{1 +
\gamma^5}{2}\Big)J(x)\,e^{\textstyle -
i\vartheta(x)}\Big\}=\nonumber\\ &=&\int {\cal D}\vartheta\,\exp
\,i\int d^2x\,\Big\{\frac{1}{2}\,\frac{1}{8\pi}\,\Big(1 -
e^{\textstyle
-2\pi/g}\,\Big)\,\partial_{\mu}\vartheta(x)\,
\partial^{\mu}\vartheta(x)\nonumber\\
&&+ \frac{1}{M}\,J^{\dagger}_1(x)J_2(x)\,e^{\textstyle
i\vartheta(x)} +
\frac{1}{M}\,J^{\dagger}_2(x)J_1(x)\,e^{\textstyle -
i\vartheta(x)}\Big\}.
\end{eqnarray}
By normalizing the $\vartheta$ field, $\vartheta(x) \to
\beta\,\vartheta(x)$, with $\beta$ given by the condition
\begin{eqnarray}\label{label4.5}
\frac{8\pi}{\beta^2} = 1 - e^{\textstyle -2\pi/g}
\end{eqnarray}
resembling Coleman's relation \cite{[3]} and defining correctly the
kinetic term of the renormalized field $\vartheta(x)$, we arrive at
\begin{eqnarray}\label{label4.6}
\hspace{-0.5in}Z_{\rm Th}[J,\bar{J}] &=& \int {\cal D}\vartheta\,\exp
\,i\int d^2x\,\Big\{\frac{1}{2}\,\partial_{\mu}\vartheta(x)\,
\partial^{\mu}\vartheta(x)\nonumber\\
\hspace{-0.5in}&&+
\frac{1}{M}\,J^{\dagger}_1(x)J_2(x)\,\,e^{\textstyle i\beta
\vartheta(x)} +
\frac{1}{M}\,J^{\dagger}_2(x)J_1(x)\,e^{\textstyle - i\beta
\vartheta(x)}\Big\}.
\end{eqnarray}
The vacuum expectation value of fermion fields considered by Coleman
\cite{[3]}
\begin{eqnarray}\label{label4.7}
\Big\langle 0\Big|{\rm
T}\Big(\prod^{n}_{k=1}\sigma_+(x_k)\sigma_-(y_k)\Big)\Big|0\Big\rangle
= \Big\langle
0\Big|{\rm T}\Big(\prod^n_{k=1}[\psi^{\dagger}_2(x_k)\psi_1(x_k)]
[\psi^{\dagger}_1(y_k)\psi_2(y_k)]\Big)\Big|0\Big\rangle
\end{eqnarray}
can be represented in the form of functional derivatives with respect
to external sources $J^{\dagger}_1(x)$, $J^{\dagger}_2(x)$ and
$J_1(x)$, $J_2(x)$
\begin{eqnarray}\label{label4.8}
&&\Big\langle 0\Big|{\rm
T}\Big(\prod^{n}_{k=1}\sigma_+(x_k)\sigma_-(y_k)\Big)\Big|0\Big\rangle
= \Big\langle 0\Big|{\rm
T}\Big(\prod^n_{k=1}[\psi^{\dagger}_2(x_k)\psi_1(x_k)]
[\psi^{\dagger}_1(y_k)\psi_2(y_k)]\Big)\Big|0\Big\rangle=\nonumber\\
&&=\prod^n_{k=1}\frac{\delta}{i\delta J_1(x_k)}\frac{\delta}{i\delta
J^{\dagger}_2(x_k)}\frac{\delta}{i\delta J_2(y_k)}\frac{\delta}{i\delta
J^{\dagger}_1(y_k)}Z_{\rm Th}[J,\bar{J}]\Big|_{\textstyle J_1 = J_2
=J^{\dagger}_1 = J^{\dagger}_2 = 0}.
\end{eqnarray}
Using the generating functional $Z_{\rm Th}[J,\bar{J}]$ in the form
(\ref{label4.6}) the r.h.s. of Eq.(\ref{label4.8}) can be written as
\begin{eqnarray}\label{label4.9}
&&\Big\langle 0\Big|{\rm
T}\Big(\prod^{n}_{k=1}\sigma_+(x_k)\sigma_-(y_k)\Big)\Big|0\Big\rangle
= \Big\langle 0\Big|{\rm
T}\Big(\prod^n_{k=1}[\psi^{\dagger}_2(x_k)\psi_1(x_k)]
[\psi^{\dagger}_1(y_k)\psi_2(y_k)]\Big)\Big|0\Big\rangle = \nonumber\\
&&= (-1)^n\,\left(\frac{\delta^2(0)}{M}\right)^{2n}\,\Big\langle
0\Big|{\rm
T}\Big(\prod^n_{i=1}[A_-(x_i)A_+(y_i)]\Big)\Big|0\Big\rangle,
\end{eqnarray}
where $\delta^2(0) = \int d^2p/(2\pi)^2 =i\int d^2p_{\rm E}/(2\pi)^2 =
i \bar{\Lambda}^2/4\pi$. The cut--off $\bar{\Lambda}$ is invented to
regularize divergences coming from the closed loops of the
$\vartheta$--field. The two--point Green function of the
$\vartheta$--field (\ref{label2.9}) regularized at $x=y$ is defined by 
\begin{eqnarray}\label{label4.10}
i\Delta(0)= -\,\frac{1}{4\pi}\,{\ell
n}\Bigg(\frac{\bar{\Lambda}^2}{\mu^2}\Bigg).
\end{eqnarray}
Thereby, the relation (\ref{label4.9}) can be rewritten as follows
\begin{eqnarray}\label{label4.11}
&&\Big\langle 0\Big|{\rm
T}\Big(\prod^{n}_{k=1}\sigma_+(x_k)\sigma_-(y_k)\Big)\Big|0\Big\rangle
= \Big\langle 0\Big|{\rm
T}\Big(\prod^n_{k=1}[\psi^{\dagger}_2(x_k)\psi_1(x_k)]
[\psi^{\dagger}_1(y_k)\psi_2(y_k)]\Big)\Big|0\Big\rangle = \nonumber\\
&&= \left(\frac{\bar{\Lambda}^2}{4\pi M}\right)^{2n}\,\Big\langle
0\Big|{\rm
T}\Big(\prod^n_{i=1}[A_-(x_i)A_+(y_i)]\Big)\Big|0\Big\rangle,
\end{eqnarray}
Relation (\ref{label4.11}) demonstrates the equivalence between vacuum
expectation values in the massless Thirring model and vacuum
expectation values in a quantum field theory of a massless scalar
field $\vartheta(x)$ coupled to external sources via exponential
couplings $A_+(x) = e^{\textstyle i\beta \vartheta(x)}$ and $A_-(x) =
e^{\textstyle - i \beta \vartheta(x)}$ (see Eq.(\ref{label2.3})).

In order to fix the value of $\bar{\Lambda}$ in terms of $M$ we
suggest to evaluate the vacuum expectation value of the operator
\begin{eqnarray}\label{label4.12}
(\bar{\psi}(x)\psi(x))^2 + (\bar{\psi}(x)i\gamma^5\psi(x))^2 =
4\,\sigma_+(x)\,\sigma_-(x).
\end{eqnarray}
In Section 6 (see Eq.(\ref{labelH.29})) we show that this operator is
an integral of motion and is equal to $M^2/g^2$.  The evaluation of
the vacuum expectation value of the operator (\ref{label4.12}) can be
carried out with the help of (\ref{label4.11}). The result reads
\begin{eqnarray}\label{label4.13}
\hspace{-0.3in}&&\langle 0|[(\bar{\psi}(x)\psi(x))^2 +
(\bar{\psi}(x)i\gamma^5\psi(x))^2]|0\rangle =\nonumber\\
\hspace{-0.3in}&&= \langle 0|[4\,\sigma_+(x)\,\sigma_-(x)]|0 \rangle =
\left(\frac{\bar{\Lambda}^2}{2\pi M}\right)^2\langle
0|[A_-(x)A_+(x)]|0 \rangle =\left(\frac{\bar{\Lambda}^2}{2\pi
M}\right)^2.
\end{eqnarray}
Equating the r.h.s. of (\ref{label4.13}) to $M^2/g^2$ we obtain the
cut--off $\bar{\Lambda}$ in terms of $M$ and $g$
\begin{eqnarray}\label{label4.14}
\bar{\Lambda} = \sqrt{\frac{2\pi}{g}}\,M.
\end{eqnarray}
In the strong coupling limit $g\to \infty$ we get $\bar{\Lambda} \to
\Lambda$, whereas in the weak coupling limit $g \to 0$ the cut--off
$\bar{\Lambda}$ vanishes. The former corresponds to the absence of the
$\vartheta$--field fluctuations in a free massless fermion field
theory.

Using the relation (\ref{label4.14}) we recast the r.h.s. of
(\ref{label4.11}) into the form
\begin{eqnarray}\label{label4.15}
&&\Big\langle 0\Big|{\rm
T}\Big(\prod^{n}_{k=1}\sigma_+(x_k)\sigma_-(y_k)\Big)\Big|0\Big\rangle
= \Big\langle 0\Big|{\rm
T}\Big(\prod^n_{k=1}[\psi^{\dagger}_2(x_k)\psi_1(x_k)]
[\psi^{\dagger}_1(y_k)\psi_2(y_k)]\Big)\Big|0\Big\rangle = \nonumber\\
&&= \frac{\langle\bar{\psi}\psi\rangle^{2n}}{2^{2n}}\,\Big\langle
0\Big|{\rm
T}\Big(\prod^n_{i=1}[A_-(x_i)A_+(y_i)]\Big)\Big|0\Big\rangle,
\end{eqnarray}
where we have used that $\langle\bar{\psi}\psi\rangle = - M/g$
(\ref{label1.16}).  

Using Eq.(\ref{label2.10}) the r.h.s. of Eq.(\ref{label4.15}) can
be calculated explicitly and reads
\begin{eqnarray}\label{label4.16}
\hspace{-0.5in}&&\Big\langle 0\Big|{\rm
T}\Big(\prod^{n}_{k=1}\sigma_+(x_k)\sigma_-(y_k)\Big)\Big|0\Big\rangle
=  \nonumber\\
\hspace{-0.5in}&&=
\frac{\langle\bar{\psi}\psi\rangle^{2n}}{2^{2n}}\,e^{\textstyle n
\beta^2 i\Delta(0)}\;\frac{\displaystyle \prod^{n}_{j <k}[-\mu^2(x_j -
x_k)^2]^{\beta^2/4\pi}[-\mu^2(y_j -
y_k)^2]^{\beta^2/4\pi}}{\prod^{n}_{j = 1}\prod^{n}_{k = 1}[-\mu^2(x_j
- y_k)^2]^{\beta^2/4\pi}},
\end{eqnarray}
where $i\Delta(0)$ is defined by (\ref{label4.10}).  Formula
(\ref{label4.16}) reproduces, in principle, Klaiber's equations
\cite{[5]} used further by Coleman \cite{[3]} but with a relation
between the coupling constants $\beta$ and $g$ (\ref{label4.5})
different to that suggested by Coleman (\ref{label1.9}) \cite{[3]}.
The new relation (\ref{label4.5}) is caused by the fact that in our
approach unlike in that of Coleman the fermion system is in the
chirally broken phase.

Relation (\ref{label4.15}) between the
vacuum expectation values can be represented in operator form by the
Abelian bosonization rules
\begin{eqnarray}\label{label4.17}
Z\,\bar{\psi}(x)\Bigg(\frac{1\mp \gamma^5}{2}\Bigg)\psi(x) =
\frac{1}{2}\,\langle \bar{\psi}\psi\rangle \,e^{\textstyle \pm i \beta
\vartheta(x)}.
\end{eqnarray}
They can be derived straightforwardly from the equations of motion
(\ref{label3.24}) for $\sigma(x)$ and $\varphi(x)$ connected by
Eq.(\ref{label3.28}), where $M/g = -\langle \bar{\psi}\psi\rangle$,
\begin{eqnarray}\label{label4.18}
\bar{\psi}(x)\Bigg(\frac{1\mp \gamma^5}{2}\Bigg)\psi(x) =
\frac{1}{2}\,\langle \bar{\psi}\psi\rangle \,e^{\textstyle \pm i \beta
\vartheta(x)}
\end{eqnarray}
with a subsequent renormalization of the fermion field $\psi(x) \to
Z^{1/2}\psi(x)$, where $Z$ is a renormalization constant. The
parameter $Z$ is invented to remove divergences appearing in the
evaluation of the vacuum expectation values of $A_-(x)$ and
$A_+(y)$. If such divergences do not appear the parameter $Z$ should
be set unity, $Z=1$. For example, in one--loop approximation for the
fermion field and tree--approximation for the $\vartheta$--field one
obtains $Z = 1$.

Relation (\ref{label4.17}) is analogous to the Abelian bosonization
rules derived by Coleman (\ref{label1.10}) in the massive Thirring
model. For the massless Thirring model due to the employment of the
chiral symmetric phase with a chiral symmetric vacuum giving $\langle
\bar{\psi}\psi\rangle = 0$ Coleman$^{\prime}$s procedure fails in
deriving a relation like (\ref{label4.17}).

In Section 6 we show that the Abelian bosonization rules
(\ref{label4.17}) are consistent with the equations of motion for
fermionic fields evolving out of the chirally broken phase.

Using the Abelian bosonization rules (\ref{label4.17}) we are able to
evaluate the vacuum expectation value of the
$\bar{\psi}(x)\psi(x)$--operator
\begin{eqnarray}\label{label4.19}
\hspace{-0.3in}\langle 0|\bar{\psi}(x)\psi(x)|0 \rangle &=& \langle
0|[\sigma_+(x) + \sigma_-(x)]|0 \rangle = \frac{1}{2}\,Z^{-1}\langle
\bar{\psi}\psi\rangle\langle 0|[A_-(x) + A_+(x)]|0 \rangle =
\nonumber\\
\hspace{-0.3in}&=& Z^{-1}\langle \bar{\psi}\psi\rangle\langle
0|\cos\beta\vartheta(x)|0 \rangle = 0,
\end{eqnarray}
where we have used (\ref{label2.10}) and (\ref{label2.11}).

We would like to emphasize that the vanishing of the vacuum
expectation value (\ref{label4.19}) is caused by the infrared
behaviour of the $\vartheta$--field. This is related to the $\mu \to
0$ limit which takes into account long--range fluctuations. In this
region the $\vartheta$--field is ill--defined \cite{[26],[27]} that
leads to the randomization of the $\vartheta$--field in the infrared
region \cite{[28]}. Due to this $\cos\beta\vartheta(x)$ is averaged to
zero \cite{[28]}. This result agrees with the Mermin--Wagner theorem
\cite{[25]} pointing out the absence of long--range order in
two--dimensional models. However, since the randomization of the
$\vartheta$--field in the infrared region is fully a 1+1--dimensional
problem, one can avoid the vanishing of $\langle
0|\bar{\psi}(x)\psi(x)|0 \rangle$ by means of dimensional
regularization. In more details we discuss this problem in Section
8. There we give also an exact solution for the massless Thirring
model in the sense of the evaluation of any correlation function.

\section{Bosonization of the massive Thirring model}
\setcounter{equation}{0}

\hspace{0.2in} The massive Thirring model differs from the massless
model by the term $-m\,\bar{\psi}(x)\psi(x)$ in the Lagrangian, where
$m$ is the fermion mass.

Skipping intermediate steps which we have carried out explicitly in
Section 2 we arrive at the partition function of the massive Thirring
model given in terms of the path integral over fermion fields and over
fields of collective excitations
\begin{eqnarray}\label{label5.1}
Z_{\rm Th} &=& \int {\cal D}\psi{\cal D}\bar{\psi}{\cal D}\sigma {\cal
D}\varphi\,\exp \,i\int
d^2x\,\Big\{\bar{\psi}(x)i\gamma^{\mu}\partial_{\mu}\psi(x)
-m\,\bar{\psi}(x)\psi(x) \nonumber\\ &&- \bar{\psi}(x)(\sigma(x) +
i\gamma^5\varphi(x))\psi(x) - \frac{1}{2g}\,[\sigma^2(x) +
\varphi^2(x)]\Big\}.
\end{eqnarray}
By a shift of the $\sigma$--field, $m + \sigma \to \sigma$, we obtain
\begin{eqnarray}\label{label5.2}
Z_{\rm Th} &=& \int {\cal D}\psi{\cal D}\bar{\psi}{\cal D}\sigma {\cal
D}\varphi\,\exp \,i\int
d^2x\,\Big\{\bar{\psi}(x)i\gamma^{\mu}\partial_{\mu}\psi(x) -
\bar{\psi}(x)(\sigma(x) + i\gamma^5\varphi(x))\psi(x)\nonumber\\ &&-
\frac{1}{2g}\,[\sigma^2(x) + \varphi^2(x)] +
\frac{m}{g}\,\sigma(x)\Big\},
\end{eqnarray}
where we have dropped an infinite constant proportional to $m^2$.

Integrating over fermionic  degrees of freedom we recast the integrand
into the form
\begin{eqnarray}\label{label5.3}
Z_{\rm Th} &=&\int {\cal D}\sigma {\cal D}\varphi\,{\rm
Det}(i\gamma^{\mu}\partial_{\mu} - \sigma -
i\gamma^5\varphi)\nonumber\\ &&\times\,\exp \,i\int d^2x\,\Big\{ -
\frac{1}{2g}\,[\sigma^2(x) + \varphi^2(x)] +
\frac{m}{g}\,\sigma(x)\Big\}.
\end{eqnarray}
Since the functional determinant coincides completely with the
determinant calculated in Section 3, we can immediately write down the
total effective potential 
\begin{eqnarray}\label{label5.4}
\hspace{-0.3in}&&V[\Phi^{\dagger}(x),\Phi(x)] =
\tilde{V}[\Phi^{\dagger}(x)\Phi(x)] +
\frac{1}{2g}\,\Phi^{\dagger}(x)\Phi(x) - \frac{m}{g}\,\sigma(x) =
\frac{1}{4\pi}\Big[ \Phi^{\dagger}(x)\Phi(x){\ell
n}\Phi^{\dagger}(x)\Phi(x)\nonumber\\
\hspace{-0.3in}&&- (\Lambda^2 + \Phi^{\dagger}(x)\Phi(x)){\ell
n}(\Lambda^2 + \Phi^{\dagger}(x)\Phi(x)) +
\frac{2\pi}{g}\,\Phi^{\dagger}(x)\Phi(x) - \frac{4\pi
m}{g}\,\sigma(x) + \Lambda^2\Big].
\end{eqnarray}
In polar representation the effective potential
(\ref{label5.4}) up to an infinite constant takes the form
\begin{eqnarray}\label{label5.5}
V[\rho(x),\vartheta(x)] &=& \frac{1}{4\pi}\Bigg[ \rho^2(x){\ell
n}\,\frac{\rho^2(x)}{\Lambda^2} - (\Lambda^2 + \rho^2(x)){\ell
n}\Bigg(1 + \frac{\rho^2(x)}{\Lambda^2}\Bigg)
+\frac{2\pi}{g}\,\rho^2(x)\nonumber\\ && - \frac{4\pi m}{g}\,\rho(x) -
\frac{4\pi m}{g}\,\rho(x) \,(\cos\vartheta(x) - 1)\Bigg].
\end{eqnarray}
\begin{figure}[h]
\centerline{\scalebox{1.0}{\includegraphics{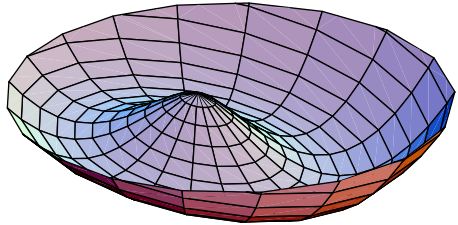}}}
\caption{The effective potential $V\left(\rho\right)$ of
Eq.(\ref{label5.5}) as a function of $\rho$ and $\vartheta$ for
$2\pi/g = {\ell n}\,2^2$ and $4\pi m/g = 0.2$ in units of
$\Lambda$.}
\label{fig3}
\end{figure}
A graphical representation of this potential as a function of $\rho$
and $\vartheta$ is shown in Fig.~\ref{fig3} for $2\pi/g = {\ell
n}\,2^2$ and $4\pi m/g = 0.2$ in units of $\Lambda$.

For the calculation of the minimum of the effective potential
(\ref{label5.5}) we have to calculate the vacuum expectation value
\begin{eqnarray}\label{label5.6}
\hspace{-0.3in}&&V[\bar{\rho}(x),0] = \nonumber\\ 
\hspace{-0.3in}&&= \frac{1}{4\pi}\Bigg[
\bar{\rho}^2(x){\ell n}\,\frac{\bar{\rho}^2(x)}{\Lambda^2} - (\Lambda^2 +
\bar{\rho}^2(x)){\ell n}\Bigg(1 + \frac{\bar{\rho}^2(x)}{\Lambda^2}\Bigg) +
\frac{2\pi}{g}\,\bar{\rho}^2(x) - \frac{4\pi m}{g}\,\bar{\rho}(x)
\Bigg],
\end{eqnarray}
where we have used $\langle 0|\vartheta(x)|0 \rangle = 0$ and $\langle
0|\cos\vartheta(x) - 1 |0\rangle = \cos\langle 0|\vartheta(x)|0
\rangle - 1 = 0$ that corresponds the tree--approximation for the
$\vartheta$--field. The first derivative of the effective potential
(\ref{label5.6}) with respect to $\bar{\rho}(x)$ is given by
\begin{eqnarray}\label{label5.7}
\frac{\delta V[\bar{\rho}(x),0]}{\delta \bar{\rho}(x)} =
\frac{1}{2\pi}\,\bar{\rho}(x)\,\Bigg[- {\ell n}\Bigg(1 +
\frac{\Lambda^2}{\bar{\rho}^{\,2}(x)}\Bigg) + \frac{2\pi}{g} -
\frac{2\pi m}{g}\,\frac{1}{\bar{\rho}(x)}\Bigg] = 0.
\end{eqnarray}
The r.h.s of Eq.(\ref{label5.7}) can be rewritten in a more
convenient form
\begin{eqnarray}\label{label5.8}
\bar{\rho}(x) = m + \bar{\rho}(x)\,\frac{g}{2\pi}\,{\ell n}\Bigg(1 +
\frac{\Lambda^2}{\bar{\rho}^{\,2}(x)}\Bigg).
\end{eqnarray}
This result agrees well with the gap--equation
(\ref{label1.14}) modified for $m\not= 0$
\begin{eqnarray}\label{label5.9} 
M = m + M\,\frac{g}{2\pi}\,{\ell n}\left(1 +
\frac{\Lambda^2}{M^2}\right)
\end{eqnarray}
with $\bar{\rho}(x) = M$.  By using (\ref{label1.16}) relation
(\ref{label5.9}) reads
\begin{eqnarray}\label{label5.10} 
M - m = -g\,\langle \bar{\psi}\psi\rangle.
\end{eqnarray}
The solution of Eq.(\ref{label5.8}) is equal to
\begin{eqnarray}\label{label5.11}
\bar{\rho}(x) = M = \frac{\Lambda}{\displaystyle
\sqrt{e^{\textstyle 2\pi/g} - 1}} +
\frac{\pi}{g}\,\frac{m}{\displaystyle 1 - e^{\textstyle - 2\pi/g}} +
O\Big(\frac{m^2}{\Lambda}\Big).
\end{eqnarray}
Since $\Lambda \gg m$, our statement concerning the decoupling of the
$\tilde{\rho}$--field is also valid for the bosonization of the
massive Thirring model. This implies that the bosonized version of the
massive Thirring model as well as the massless one is described by one
degree of freedom, the scalar field $\vartheta(x)$.

The partition function of the bosonized version of the massive
Thirring model defined in the vicinity of the minimum of the effective
potential (\ref{label5.5}) acquires the form
\begin{eqnarray}\label{label5.12}
Z_{\rm Th} &=&\int {\cal D}\vartheta\,{\rm
Det}\Big(i\gamma^{\mu}\partial_{\mu} - M\,e^{\textstyle
i\,\gamma^5\,\vartheta}\,\Big)\,\exp\,i\int d^2x\,\frac{m
M}{g}\,(\cos \vartheta(x) - 1) =\nonumber\\ &=&\int {\cal
D}\vartheta\,{\rm Det}(i\gamma^{\mu}\partial_{\mu} +
\gamma^{\mu}A_{\mu} - M)\,\exp\,i\int d^2x\,\frac{m
M}{g}\,(\cos \vartheta(x) - 1) = \nonumber\\ &=&\int {\cal
D}\vartheta\,\exp\,i\,\int d^2x\,{\cal L}_{\rm eff}(x),
\end{eqnarray}
where $A_{\mu}(x) =
(1/2)\varepsilon_{\mu\nu}\partial^{\nu}\vartheta(x)$ and the effective
Lagrangian ${\cal L}_{\rm eff}(x)$ is determined by
\begin{eqnarray}\label{label5.13}
\hspace{-0.3in}{\cal L}_{\rm eff}(x)&=& -i\,\langle x|{\rm tr}{\ell
n}(i\gamma^{\mu}\partial_{\mu} - M)|x \rangle -
\frac{1}{8\pi}\int\frac{d^2x_1d^2k_1}{(2\pi)^2}\,e^{\textstyle
-ik_1\cdot(x_1 - x)}A_{\mu}(x)A_{\nu}(x_1)\nonumber\\
\hspace{-0.3in}&&\times \int \frac{d^2k}{\pi i}\,
{\rm tr}\Bigg\{\frac{1}{M - \hat{k}}\gamma^{\mu}
\frac{1}{M - \hat{k} - \hat{k}_1}
\gamma^{\nu}\Bigg\} + \frac{m M}{g}\,(\cos \vartheta(x) - 1)
\end{eqnarray}
in complete analogy with the massless case (\ref{label3.34}).

Omitting a trivial infinite constant and the terms proportional to
inverse powers of $\Lambda$ leads to
\begin{eqnarray}\label{label5.14}
\hspace{-0.3in}{\cal L}_{\rm eff}(x) = \frac{1}{16\pi}\,\Big(1 -
e^{\textstyle
-2\pi/g}\,\Big)\,\partial_{\mu}\vartheta(x)\,\partial^{\mu}\vartheta(x)
+ \frac{m M}{g}\,(\cos \vartheta(x) - 1).
\end{eqnarray}
In order to get the correct kinetic term of the $\vartheta$--field, we
renormalize the $\vartheta$--field, $\vartheta(x) \to
\beta\,\vartheta(x)$, where the renormalization constant $\beta$ obeys
relation (\ref{label4.5}). Introducing a parameter $\alpha$
\begin{eqnarray}\label{label5.15}
\alpha =\beta^2\,\frac{m M}{g} =
-\,m\,\beta^2\,\langle\bar{\psi}\psi\rangle + \frac{m^2}{g}\,\beta^2,
\end{eqnarray}
where we have used Eq.(\ref{label5.10}), we transform the effective
Lagrangian (\ref{label5.14}) to the standard form of the Lagrangian of
the SG model \cite{[3]}
\begin{eqnarray}\label{label5.16}
{\cal L}_{\rm eff}(x) =
\frac{1}{2}\,\partial_{\mu}\vartheta(x)\,\partial^{\mu}\vartheta(x) +
\frac{\alpha}{\beta^2}\,(\cos \beta\vartheta(x) - 1).
\end{eqnarray}
This testifies the complete equivalence of the bosonized version of
the massive Thirring model and the SG model
\begin{eqnarray}\label{label5.17}
Z_{\rm Th} = Z_{\rm SG},
\end{eqnarray}
with the relation (\ref{label4.5}) between the coupling constants
$\beta$ and $g$.

The generating functional of the Green functions $Z_{\rm
Th}[J,\bar{J}]$ in the massive Thirring model can be derived in
analogy to Eq.(\ref{label4.6}), the generating functional of the
Green functions in the massless Thirring model, and reads
\begin{eqnarray}\label{label5.18}
\hspace{-0.5in}Z_{\rm Th}[J,\bar{J}] &=& \int {\cal D}\vartheta\,\exp
\,i\int d^2x\,\Big\{\frac{1}{2}\,\partial_{\mu}\vartheta(x)\,
\partial^{\mu}\vartheta(x) + \frac{\alpha}{\beta^2}\,(\cos
\beta\vartheta(x) - 1)\nonumber\\
\hspace{-0.5in}&&+
\frac{1}{M}\,J^{\dagger}_1(x)J_2(x)\,\,e^{\textstyle i\beta
\vartheta(x)} +
\frac{1}{M}\,J^{\dagger}_2(x)J_1(x)\,e^{\textstyle - i\beta
\vartheta(x)}\Big\}.
\end{eqnarray}
The Abelian bosonization rules analogous to Eq.(\ref{label4.17}) in
the massless Thirring model can be derived from the equations of
motion
\begin{eqnarray}\label{label5.19}
\bar{\psi}(x)\psi(x)&=& -\frac{\sigma(x) - m}{g},\nonumber\\
\bar{\psi}(x)i\gamma^5\psi(x)&=& -\frac{\varphi(x)}{g}.
\end{eqnarray}
Setting $\sigma(x) = M\,\cos\beta\vartheta(x)$ and $\varphi(x) =
M\,\sin\beta\vartheta(x)$ we get
\begin{eqnarray}\label{label5.20}
m\,\bar{\psi}(x)\Bigg(\frac{1\mp \gamma^5}{2}\Bigg)\psi(x) =
-\frac{\alpha}{2\beta^2}\,e^{\textstyle \pm i \beta \vartheta(x)} +
\frac{m^2}{2g}.
\end{eqnarray}
Renormalizing the fermion field $\psi(x) \to Z^{1/2}\,\psi(x)$ we
arrive at the relation
\begin{eqnarray}\label{label5.21}
Z\,m\,\bar{\psi}(x)\Bigg(\frac{1\mp \gamma^5}{2}\Bigg)\psi(x) =
-\frac{\alpha}{2\beta^2}\,e^{\textstyle \pm i \beta \vartheta(x)} +
\frac{m^2}{2g}.
\end{eqnarray}
For the evaluation of the vacuum expectation value in one--loop
approximation for the fermion field and in tree--approximation for the
scalar field the parameter $Z$ amounts to $Z = 1$.

The operator relation (\ref{label5.21}) can be considered as a
generalization of the Abelian bosonization rules (\ref{label1.10}) derived by Coleman. The term proportional to $m^2$ can be dropped at
leading order in the $m$ expansion \cite{[3]}.

We would like to accentuate that in our case the coupling constant
$\beta^2$ is always greater than $8\pi$, $\beta^2 > 8\pi$. This is in
disagreement with Coleman's statement pointing out that the
equivalence between the massive Thirring model and the SG model can
exist only if $\beta^2 < 8\pi$ \cite{[3]}. Such a disagreement can be
explained by different starting phases of the fermion system evolving
to the bosonic phase. In fact, in Coleman's approach the fermion
system has been considered in the chiral symmetric phase, whereas in
our case the fermion  system is in the phase of spontaneously broken
chiral symmetry. We would like to remind that in the Thirring model
with an attractive four--fermion interaction the chirally broken phase
is preferable.

\section{\hspace{-0.2in} Operator formalism for the massless 
Thirring model} 
\setcounter{equation}{0}

\hspace{0.2in} In this section we investigate the massless Thirring
model in the operator formalism. We analyse the normal ordering of
fermion operators and chiral symmetry breaking, the equations of
motion for fermion fields, the current algebra and the connection of
the energy--momentum tensor $\theta_{\mu\nu}(x,t)$ \cite{[30]} with
its Sugawara form \cite{[31], [32]}. We show that the Schwinger
term \cite{[33]} calculated for the fermion system in the chirally
broken phase depends on the coupling constant $g$ and reduces to the
result obtained by Sommerfield \cite{[34]} in the limit $g\to 0$. We
demonstrate that Sommerfield$^{\prime}$s value for the Schwinger term
corresponds to a trivial vacuum of the fermion system.  

We discuss the phenomenon of spontaneous breaking of chiral symmetry
in the massless Thirring from the point of view of the BCS theory of
superconductivity. We use the BCS expression for the wave function of
the non--perturbative vacuum and calculate the energy density of this
non--perturbative vacuum state. We show that the energy density of the
non--perturbative vacuum acquires a minimum just, when the dynamical
mass $M$ of fermions satisfies the gap--equation
(\ref{label1.14}).

\subsection{Normal ordering and chiral symmetry breaking}

\hspace{0.2in} The Lagrangian of the massless Thirring that we use
below reads
\begin{eqnarray}\label{labelH.1}
{\cal L}(x,t) =:\bar{\psi}(x,t)i\gamma^{\mu}\partial_{\mu}\psi(x,t): -
\frac{1}{2}\,g :\bar{\psi}(x,t)\gamma_{\mu}\psi(x,t)
\bar{\psi}(x,t)\gamma^{\mu}\psi(x,t):,
\end{eqnarray}
where $:\ldots:$ denotes normal ordering. A vector current
$j_{\mu}(x,t)$ and the divergence $\partial^{\mu}j_{\mu}(x,t)$ can be
derived in the usual way by a local gauge transformation $U_{\rm
V}(1)$ with a parameter $\alpha_{\rm V}(x,t)$:
\begin{eqnarray}\label{labelH.2}
\psi(x,t) &\to& e^{\textstyle i\alpha_{\rm
V}(x,t)}\psi(x,t),\nonumber\\ 
\bar{\psi}(x,t)&\to&
\bar{\psi}(x,t)\,e^{\textstyle - i\alpha_{\rm V}(x,t)}.
\end{eqnarray}
This changes the Lagrangian (\ref{labelH.1}) as follows
\begin{eqnarray}\label{labelH.3}
{\cal L}(x,t) \to {\cal L}[\alpha_{\rm V}(x,t)] = {\cal L}(x,t) -
:\bar{\psi}(x,t)\gamma_{\mu}\psi(x,t):\partial^{\mu}\alpha_{\rm V}(x,t).
\end{eqnarray}
Therefore, the vector current $j_{\mu}(x,t)$ and its divergence
$\partial^{\mu}j_{\mu}(x,t)$ are equal to
\begin{eqnarray}\label{labelH.4}
j_{\mu}(x,t) &=& - \frac{\delta {\cal
L}[\alpha_{\rm V}(x,t)]}{\delta\partial^{\mu}\alpha_{\rm V}(x,t)} =
:\bar{\psi}(x,t)\gamma_{\mu}\psi(x,t):,\nonumber\\
\partial^{\mu}j_{\mu}(x,t)&=& - \frac{\delta {\cal
L}[\alpha_{\rm V}(x,t)]}{\delta\alpha_{\rm V}(x,t)} = 0.
\end{eqnarray}
For the subsequent analysis we need the interaction term in the
Lagrangian (\ref{labelH.1}) in the form of a product of currents
$j_{\mu}(x,t)j^{\mu}(x,t)$. In order to understand the replacement
\begin{eqnarray*}
:\bar{\psi}(x,t)\gamma_{\mu}\psi(x,t)
\bar{\psi}(x,t)\gamma^{\mu}\psi(x,t): \to j_{\mu}(x,t)j^{\mu}(x,t)
\end{eqnarray*}
we suggest to start with the product $j_{\mu}(x,t)j^{\mu}(x,t)$ and
Wick$^{\prime}$s theorem to reduce this product to the form of the
interaction term in (\ref{labelH.1}). It is useful to employ
Schwinger$^{\prime}$s method of separation \cite{[33]}. Denoting
$(x,t) \to x$ we obtain
\begin{eqnarray}\label{labelH.5}
\hspace{-0.5in}&&j_{\mu}(x)j^{\mu}(x) =
:\bar{\psi}(x)\gamma_{\mu}\psi(x)::\bar{\psi}(x)\gamma^{\mu}\psi(x): =
\nonumber\\ \hspace{-0.5in}&&=\lim_{\varepsilon \to
0}:\bar{\psi}\Big(x +
\frac{1}{2}\,\varepsilon\Big)\gamma_{\mu}\psi\Big(x +
\frac{1}{2}\,\varepsilon\Big)::\bar{\psi}\Big(x -
\frac{1}{2}\,\varepsilon\Big)\gamma^{\mu}\psi\Big(x -
\frac{1}{2}\,\varepsilon\Big):=\nonumber\\
\hspace{-0.5in}&&= \lim_{\varepsilon \to
0}\Big[:\bar{\psi}\Big(x +
\frac{1}{2}\,\varepsilon\Big)\gamma_{\mu}\psi\Big(x +
\frac{1}{2}\,\varepsilon\Big)\bar{\psi}\Big(x -
\frac{1}{2}\,\varepsilon\Big)\gamma^{\mu}\psi\Big(x -
\frac{1}{2}\,\varepsilon\Big):\nonumber\\\hspace{-0.5in} && + :\bar{\psi}\Big(x +
\frac{1}{2}\,\varepsilon\Big)\gamma_{\mu}\Big\langle 0\Big|\psi\Big(x
+ \frac{1}{2}\,\varepsilon\Big)\bar{\psi}\Big(x -
\frac{1}{2}\,\varepsilon\Big)\Big|0 \Big\rangle \gamma^{\mu}\psi\Big(x
- \frac{1}{2}\,\varepsilon\Big):\nonumber\\ 
\hspace{-0.5in} && + :\bar{\psi}\Big(x -
\frac{1}{2}\,\varepsilon\Big)\gamma_{\mu}\Big\langle 0\Big|\psi\Big(x -
\frac{1}{2}\,\varepsilon\Big)\bar{\psi}\Big(x +
\frac{1}{2}\,\varepsilon\Big)\Big|0 \Big\rangle \gamma^{\mu}\psi\Big(x +
\frac{1}{2}\,\varepsilon\Big):\nonumber\\
\hspace{-0.5in}&& - {\rm
tr}\,\Big\{\gamma_{\mu}\Big\langle 0\Big|\psi\Big(x -
\frac{1}{2}\,\varepsilon\Big)\bar{\psi}\Big(x +
\frac{1}{2}\,\varepsilon\Big)\Big|0 \Big\rangle
\gamma^{\mu}\Big\langle 0\Big|\psi\Big(x +
\frac{1}{2}\,\varepsilon\Big)\bar{\psi}\Big(x -
\frac{1}{2}\,\varepsilon\Big)\Big|0 \Big\rangle \Big\}\Big],
\end{eqnarray}
where $\varepsilon = (\varepsilon^0, \varepsilon^1)$ is an
infinitesimal 2--vector.

Now we would like to discuss the contributions caused by the vacuum
expectation values in Eq.(\ref{labelH.5}). For the free massless
fermion field we get
\begin{eqnarray}\label{labelH.6}
\Big\langle 0\Big|\psi\Big(x \pm
\frac{1}{2}\,\varepsilon\Big)\bar{\psi}\Big(x \mp
\frac{1}{2}\,\varepsilon\Big)\Big|0 \Big\rangle &=&
\int\limits^{\infty}_{-\infty}\frac{dp}{4\pi}\,\frac{\gamma^0|p| -
\gamma^1p}{|p|}\,e^{\textstyle \mp\,i\,(|p|\varepsilon^0 -
p\varepsilon^1)}=\nonumber\\
&=&\hat{\varepsilon}\,\Bigg[\delta(\varepsilon^2)
\mp\,\frac{i}{2\pi}\,\frac{1}{\varepsilon^2}\Bigg].
\end{eqnarray}
Due to the identity $\gamma_{\mu}\gamma^{\alpha}\gamma^{\mu} = 0$ for
$\alpha = 0,1$ these vacuum expectation values taken
between $\gamma$--matrices $\gamma_{\mu}\ldots \gamma^{\mu}$ vanish
\begin{eqnarray}\label{labelH.7}
\gamma_{\mu}\Big\langle 0\Big|\psi\Big(x \pm 
\frac{1}{2}\,\varepsilon\Big)\bar{\psi}\Big(x \mp
\frac{1}{2}\,\varepsilon,\Big)\Big|0 \Big\rangle \gamma^{\mu} = 0.
\end{eqnarray}

This result persists for the interacting massless fermion field. In
fact, the vacuum expectation values calculated for the trivial vacuum
should have the following general form
\begin{eqnarray}\label{labelH.8}
\Big\langle 0\Big|\psi\Big(x \pm
\frac{1}{2}\,\varepsilon\Big)\bar{\psi}\Big(x \mp
\frac{1}{2}\,\varepsilon\Big)\Big|0 \Big\rangle =
\gamma^{\alpha}\,\Phi_{\alpha}(x,\varepsilon),
\end{eqnarray}
where $\Phi_{\alpha}(x,\varepsilon)$ is an arbitrary function, and
vanishes again between $\gamma$--matrices $\gamma_{\mu}\ldots
\gamma^{\mu}$.

\noindent Substituting (\ref{labelH.7}) in (\ref{labelH.5}) we obtain
\begin{eqnarray}\label{labelH.9}
\hspace{-0.5in}&&j_{\mu}(x)j^{\mu}(x) =
:\bar{\psi}(x)\gamma_{\mu}\psi(x)::\bar{\psi}(x)
\gamma_{\mu}\psi(x):
= \nonumber\\ \hspace{-0.5in}&&=\lim_{\varepsilon \to 0}
:\bar{\psi}\Big(x +
\frac{1}{2}\,\varepsilon\Big)\gamma_{\mu}\psi\Big(x +
\frac{1}{2}\,\varepsilon\Big)::\bar{\psi}\Big(x -
\frac{1}{2}\,\varepsilon\Big)\gamma_{\mu}\psi\Big(x -
\frac{1}{2}\,\varepsilon\Big):=\nonumber\\ 
\hspace{-0.5in}&&= \lim_{\varepsilon \to 0}:\bar{\psi}\Big(x +
\frac{1}{2}\,\varepsilon\Big)\gamma_{\mu}\psi\Big(x +
\frac{1}{2}\,\varepsilon\Big)\bar{\psi}\Big(x -
\frac{1}{2}\,\varepsilon\Big)\gamma_{\mu}\psi\Big(x -
\frac{1}{2}\,\varepsilon\Big):.
\end{eqnarray}
Therefore, in the trivial vacuum we get the relation
\begin{eqnarray}\label{labelH.10}
j_{\mu}(x,t)j^{\mu}(x,t) &=&
:\bar{\psi}(x,t)\gamma_{\mu}\psi(x,t)::\bar{\psi}(x,t)
\gamma_{\mu}\psi(x,t):=\nonumber\\ &=&:\bar{\psi}(x,t)\gamma_{\mu}
\psi(x,t)\bar{\psi}(x,t)\gamma_{\mu}\psi(x,t):,
\end{eqnarray}
where we have taken the limit $\varepsilon \to 0$ and come back to the
notation $x \to (x,t)$.

Due to the relation (\ref{labelH.10}), for a system of massless
fermions self--coupled in the chiral symmetric phase with a trivial
chirally invariant vacuum, the Lagrangian (\ref{labelH.1}) acquires
the form
\begin{eqnarray}\label{labelH.11}
{\cal L}(x,t) &=&:\bar{\psi}(x,t)i\gamma^{\mu}\partial_{\mu}\psi(x,t):
- \frac{1}{2}\,g :\bar{\psi}(x,t)\gamma_{\mu}\psi(x,t):
:\bar{\psi}(x,t)\gamma^{\mu}\psi(x,t):=\nonumber\\
&=&:\bar{\psi}(x,t)i\gamma^{\mu}\partial_{\mu}\psi(x,t): -
\frac{1}{2}\,g\,j_{\mu}(x,t)j^{\mu}(x,t).
\end{eqnarray}
Now let us show that the fermion system described by the Lagrangian
(\ref{labelH.11}) is unstable under chiral symmetry breaking. For
this aim we rewrite the Lagrangian in an equivalent form
\begin{eqnarray}\label{labelH.12}
{\cal L}(x,t) &=& :\bar{\psi}(x,t)(i\gamma^{\mu}\partial_{\mu} -
M)\psi(x,t): + M:\bar{\psi}(x,t)\psi(x,t):\nonumber\\ && -
\frac{1}{2}\,g\,:\bar{\psi}(x,t)\gamma_{\mu}\psi(x,t):
:\bar{\psi}(x,t)\gamma^{\mu}\psi(x,t):
\end{eqnarray}
and normal order the interaction term at the scale $M$
\begin{eqnarray}\label{labelH.13}
&&:\bar{\psi}(x,t)\gamma_{\mu}\psi(x,t):
:\bar{\psi}(x,t)\gamma^{\mu}\psi(x,t): =
:\bar{\psi}(x,t)\gamma_{\mu}\psi(x,t)\bar{\psi}(x,t)\gamma^{\mu}\psi(x,t):\nonumber\\
&&+ 2:\bar{\psi}(x,t)\gamma_{\mu}\langle
0|\psi(x,t)\bar{\psi}(x,t)|0\rangle \gamma^{\mu}\psi(x,t):\nonumber\\
&& - {\rm tr}\,\{\gamma_{\mu}\langle
0|\psi(x,t)\bar{\psi}(x,t)|0\rangle \gamma^{\mu}\langle
0|\psi(x,t)\bar{\psi}(x,t)|0\rangle\}.
\end{eqnarray}
The vacuum expectation value in the r.h.s. of (\ref{labelH.13})
calculated in the one--fermion loop approximation for massive fermions
of mass $M$ reads
\begin{eqnarray}\label{labelH.14}
&&\langle 0|\psi(x,t)\bar{\psi}(x,t)|0\rangle \to \Big\langle
0\Big|\psi\Big(x \pm \frac{1}{2}\,\varepsilon\Big)\bar{\psi}\Big(x \mp
\frac{1}{2}\,\varepsilon\Big)\Big|0\Big\rangle =\nonumber\\
&&=\int\limits^{\infty}_{-\infty}\frac{dp}{4\pi}\,\frac{\gamma^0E_p -
\gamma^1p + M}{E_p}\,e^{\textstyle \mp\,i\,(E_p\varepsilon^0 -
p\varepsilon^1)} =\nonumber\\
&&=\pm \frac{\hat{\varepsilon}}{\sqrt{\varepsilon^2}}\frac{M}{4\pi}\int\limits^{\infty}_{-\infty}d\varphi\,\cosh\varphi\,e^{\textstyle
- i M\sqrt{\varepsilon^2}\cosh\varphi} +
\frac{M}{4\pi}\int\limits^{\infty}_{-\infty}d\varphi\,e^{\textstyle -
i M\sqrt{\varepsilon^2}\cosh\varphi}=\nonumber\\
&&=\pm\frac{\hat{\varepsilon}}{\sqrt{\varepsilon^2}}
\frac{M}{2\pi}K_1(iM\sqrt{\varepsilon^2}) + \frac{M}{2\pi}K_0(iM\sqrt{\varepsilon^2}),
\end{eqnarray}
where $E_p = \sqrt{p^2 + M^2}$ and $K_1(z)$ and $K_0(z)$ are
McDonald$^{\prime}$s functions. In the r.h.s. of (\ref{labelH.13}) the
contribution of the first term proportional to $\hat{\varepsilon}$
vanishes due to the identities
$\gamma_{\mu}\hat{\varepsilon}\gamma^{\mu} = 0$ and ${\rm
tr}\{\hat{\varepsilon}\} = 0$. A non--zero contribution comes only
from the second term that coincides with the causal Green function of
the scalar field with a mass M and can be regularized in the limit
$\varepsilon \to 0$ by the cut--off $\Lambda$:
\begin{eqnarray}\label{labelH.15}
&&\frac{M}{2\pi}K_0(iM\sqrt{\varepsilon^2})=
M\int\frac{d^2p}{(2\pi)^2i}\,\frac{e^{\textstyle \mp i p\cdot
\varepsilon}}{M^2 - p^2 -i0} \stackrel{\varepsilon \to
0}{\longrightarrow}\int\frac{d^2p}{(2\pi)^2i}\,\frac{M}{M^2 - p^2 -i0}
= \nonumber\\ 
&&=\frac{M}{4\pi}\,{\ell n}\Bigg(1 + \frac{\Lambda^2}{M^2}\Bigg).
\end{eqnarray}
Substituting (\ref{labelH.13}) with the vacuum expectation value
(\ref{labelH.14}) in (\ref{labelH.12}) we obtain
\begin{eqnarray}\label{labelH.16}
\hspace{-0.3in}{\cal L}(x,t) &=&:\bar{\psi}(x,t)
(i\gamma^{\mu}\partial_{\mu} - M)\psi(x,t): -
\frac{1}{2}\,g:\bar{\psi}(x,t)\gamma_{\mu}\psi(x,t)\bar{\psi}(x,t)
\gamma^{\mu}\psi(x,t):\nonumber\\ \hspace{-0.3in}&&+ \Bigg[M -
g\,\frac{M}{2\pi}\,{\ell n}\Bigg(1 +
\frac{\Lambda^2}{M^2}\Bigg)\Bigg]:\bar{\psi}(x,t)\psi(x,t):,
\end{eqnarray}
where we have omitted an insignificant constant. Self--consistency of
the approach demands the relation
\begin{eqnarray*}
M - g\,\frac{M}{2\pi}\,{\ell n}\Bigg(1 + \frac{\Lambda^2}{M^2}\Bigg) =
0,
\end{eqnarray*}
that is our gap--equation (\ref{label1.14}). This results in the
Lagrangian
\begin{eqnarray}\label{labelH.17}
\hspace{-0.3in}{\cal L}(x,t) =
:\bar{\psi}(x,t)(i\gamma^{\mu}\partial_{\mu} - M)\psi(x,t): -
\frac{1}{2}\,g:\bar{\psi}(x,t)\gamma_{\mu}\psi(x,t)\bar{\psi}(x,t)
\gamma^{\mu}\psi(x,t):.
\end{eqnarray}
For $M\not= 0$ the Lagrangian (\ref{labelH.16}) describes a system of
fermions with mass $M$ in the chirally broken phase. We conclude that
for an attractive two-body interaction the vacuum expectation values
in Eq.(\ref{labelH.5}) lead to an instability of the perturbative
vacuum. In the next subsection we show that a stable non--perturbative
vacuum can be determined within the BCS formalism.

\subsection{The massless Thirring model in the formalism of 
the BCS--theory of superconductivity. Chiral symmetry breaking}

\hspace{0.2in} We discuss the phenomenon of spontaneous breaking of
chiral symmetry in the massless Thirring model from the point of view
of the Bardeen--Cooper--Schrieffer (BCS) theory of superconductivity
\cite{[23]}. We show that the energy density of the non--perturbative
vacuum acquires a minimum, when the dynamical mass $M$ of fermions
satisfies the gap--equation (\ref{label1.14}).

The Hamiltonian of the massless Thirring model is equal to
\begin{eqnarray}\label{labelE.1}
{\cal H}(x,t) = - :\bar{\psi}(x,t)i\gamma^1\frac{\partial}{\partial
x}\psi(x,t): +
\frac{1}{2}\,g:\bar{\psi}(x,t)\gamma_{\mu}\psi(x,t)\bar{\psi}(x,t)
\gamma^{\mu}\psi(x,t):.
\end{eqnarray}
In terms of the components of the $\psi$--field, a column matrix
$\psi(x,t) = (\psi_1(x,t), \psi_2(x,t))$, the Hamiltonian
(\ref{labelE.1}) reads
\begin{eqnarray}\label{labelE.2}
{\cal H}(x,t) &=& - :\psi^{\dagger}_1(x,t)i\frac{\partial }{\partial
x}\psi_1(x,t): + :\psi^{\dagger}_2(x,t)i\frac{\partial }{\partial
x}\psi_2(x,t):\nonumber\\
&&+ 2\,g:\psi^{\dagger}_1(x,t)\psi_1(x,t)
\psi^{\dagger}_2(x,t)\psi_2(x,t):.
\end{eqnarray}
For the further evaluation it is convenient to embed the fermion
system into a finite volume $L$. For  periodical conditions
$\psi(0,t) = \psi(L,t)$ the expansion of $\psi(x,t)$ into
plane--waves reads (see Appendix D):
\begin{eqnarray}\label{labelE.3}
\psi(x,t) = \sum_{p^1}\, \frac{1}{\sqrt{ 2p^0
L}}\,\Big[u(p^0,p^1)a(p^1)\,e^{\textstyle -ip^0 t + ip^1 x} +
v(p^0,p^1)b^{\dagger}(p^1)\,e^{\textstyle ip^0 t - ip^1 x}\Big].
\end{eqnarray}
The creation and annihilation operators are dimensionless and obey the
anticommutation relations
\begin{eqnarray}\label{labelE.4}
&&\{a(p^1), a^{\dagger}(q^1)\} = \{b(p^1), b^{\dagger}(q^1)\} =
\delta_{\textstyle p^1 q^1},\nonumber\\
&&\{a(p^1), a(q^1)\} = \{a^{\dagger}(p^1), a^{\dagger}(q^1)\} = 
\{b(p^1), b(q^1)\} =
\{b^{\dagger}(p^1), b^{\dagger}(q^1)\} = 0.
\end{eqnarray}
They are related to the annihilation operators of fermions $A(p^1)$ and
antifermions $B(p^1)$ with mass $M$ by the
Bogoliubov transformation \cite{[18],[24]}
\begin{eqnarray}\label{labelE.40}
A(p^1) &=& u_{\textstyle p^1}\,a(p^1) - v_{\textstyle
p^1}\,b^{\dagger}(-p^1),\nonumber\\ 
B(p^1)&=&u_{\textstyle p^1}\,b(p^1) -
v_{\textstyle p^1}\,a^{\dagger}(-p^1).
\end{eqnarray}
The coefficient functions $u_{\textstyle
p^1}$ and $v_{\textstyle p^1}$ are equal to \cite{[18],[22]} and
\cite{[24]}:
\begin{eqnarray}\label{labelE.41}
u_{\textstyle p^1} = \sqrt{\frac{1}{2}\Bigg( 1 +
\frac{|p^1|}{\sqrt{(p^1)^2 + M^2}}\Bigg)}\quad,\quad v_{\textstyle
p^1} = \varepsilon(p^1)\,\sqrt{\frac{1}{2}\Bigg( 1 -
\frac{|p^1|}{\sqrt{(p^1)^2 + M^2}}\Bigg)},
\end{eqnarray}
where $\varepsilon(p^1)$ is a sign function, and obey the normalization condition 
\begin{eqnarray}\label{labelE.42}
 u^2_{\textstyle p^1} + v^2_{\textstyle p^1} = 1.
\end{eqnarray}
The wave function of the non--perturbative vacuum $|\Omega\rangle$ we
take in the BCS form \cite{[23]}:
\begin{eqnarray}\label{labelE.5}
|\Omega\rangle = \angle = prod_{\textstyle k^1}[u_{\textstyle k^1} +
 v_{\textstyle k^1}\, a^{\dagger}(k^1)b^{\dagger}(-k^1)]|0\rangle,
\end{eqnarray}
where $|0\rangle$ is a perturbative, chiral symmetric vacuum. The wave
function $|\Omega\rangle$ satisfies the equations
\begin{eqnarray}\label{labelE.43}
A(p^1)|\Omega\rangle = B(p^1)|\Omega\rangle = 0
\end{eqnarray}
and is invariant under parity transformation
\begin{eqnarray}\label{labelE.6}
{\cal P}\,\psi(x,t)\,{\cal P}^{\dagger} &=&
\gamma^0\psi(-x,t)\,\Longrightarrow \nonumber\\\Longrightarrow \,
{\cal P}\,\psi_1(x,t)\,{\cal P}^{\dagger} &=& \psi_2(-x,t)\quad,\quad
{\cal P}\,\psi_2(x,t)\,{\cal P}^{\dagger} = \psi_1(-x,t),\nonumber\\
{\cal P}\,a^{\dagger}(k^1)\,{\cal P}^{\dagger}&=& +\,
a^{\dagger}(-k^1)\quad,\quad{\cal P}\,b^{\dagger}(k^1)\,{\cal
P}^{\dagger} = -\, b^{\dagger}(-k^1),
\end{eqnarray}
where we have dropped insignificant phase factors.

Due to the relation (\ref{labelE.42}) the wave function of the
non--perturbative vacuum is normalized to unity $\langle
\Omega|\Omega\rangle = 1$. The coefficient functions $u_{\textstyle
k^1}$ and $v_{\textstyle k^1}$ depend explicitly on the dynamical $M$
which we treat as a variational parameter.

The energy of the ground state is equal to \cite{[23]}
\begin{eqnarray}\label{labelE.8}
E(M) &=& \int^{\infty}_{-\infty}dx\,\langle \Omega|{\cal
H}(x,t)|\Omega \rangle = \nonumber\\ &=&4\sum_{\textstyle p^1>
0}p^1\,v^2_{\textstyle p^1} - \frac{8g}{L}\Big[\Big(\sum_{\textstyle
p^1> 0}v_{\textstyle p^1}u_{\textstyle p^1}\Big)^2
+\frac{1}{2}\sum_{\textstyle p^1> 0}v^4_{\textstyle p^1}\Big].
\end{eqnarray}
The energy density we define by
\begin{eqnarray}\label{labelE.9}
{\cal E}(M) = \lim_{\textstyle L\to \infty}\frac{E(M)}{L} =
4\int^{\infty}_0\frac{dp^1}{2\pi}\,p^1\,v^2(p^1) -
8g\Bigg[\int^{\infty}_0\frac{dp^1}{2\pi}\,v(p^1)u(p^1)\Bigg]^2.
\end{eqnarray}
Substituting (\ref{labelE.41}) in (\ref{labelE.9}) we express the
energy density as a function of the variational parameter $M$:
\begin{eqnarray}\label{labelE.11}
{\cal E}(M) = 2\int^{\infty}_0\frac{dp^1}{2\pi}\,p^1\Bigg(1
-\frac{p^1}{\sqrt{(p^1)^2 + M^2}}\Bigg) -
2g\Bigg[\int^{\infty}_0\frac{dp^1}{2\pi}\,\frac{M}{\sqrt{(p^1)^2 +
M^2}}\Bigg]^2.
\end{eqnarray}
The minimum of the energy density is defined by 
\begin{eqnarray}\label{labelE.12}
\frac{d{\cal E}(M)}{d M} = \Bigg[M -
2g\int^{\infty}_0\frac{dp^1}{2\pi}\,\frac{M}{\sqrt{(p^1)^2 +
M^2}}\Bigg]\int^{\infty}_0\frac{dp^1}{\pi}\, \frac{(p^1)^2}{((p^1)^2 +
M^2)^{3/2}}= 0.
\end{eqnarray}
This yields the BCS--like gap--equation
\begin{eqnarray}\label{labelE.13}
M =
2g\int^{\infty}_0\frac{dp^1}{2\pi}\,\frac{M}{\sqrt{(p^1)^2 +
M^2}}.
\end{eqnarray}
Calculating the second derivative of ${\cal E}(M)$ with respect to $M$
one can show that the BCS gap--equation describes the minimum of the
energy density (\ref{labelE.11}) at $M\not= 0$. Using the obvious
relation
\begin{eqnarray}\label{labelE.14}
\int^{\infty}_0\frac{dp^1}{2\pi}\,\frac{M}{\sqrt{(p^1)^2 + M^2}} =
\int\frac{d^2p}{(2\pi)^2i}\,\frac{M}{M^2 - p^2 -i0} = \frac{M}{4\pi}\,
{\ell n}\Bigg(1 + \frac{\Lambda^2}{M^2}\Bigg)
\end{eqnarray}
we obtain the gap--equation (\ref{labelE.13}) in the form
({\ref{label1.14}).

Using the relations between momentum integrals (\ref{labelE.14}) and
\begin{eqnarray}\label{labelE.15}
&&2\int^{\infty}_0\frac{dp^1}{2\pi}\,p^1\Bigg(1
-\frac{p^1}{\sqrt{(p^1)^2 + M^2}}\Bigg) =\nonumber\\
&&= - \int\frac{d^2p}{(2\pi)^2i}\,{\ell n}(M^2 - p^2 - i 0) +
\int\frac{d^2p}{(2\pi)^2i}\,\frac{2M^2}{M^2 - p^2 - i 0} +
C,
\end{eqnarray}
where $C$ is a infinite constant independent of $M$, and the
gap--equation ({\ref{label1.14}) the energy density ${\cal E}(M)$ can
be transformed into the form
\begin{eqnarray}\label{labelE.16}
{\cal E}(M) = \frac{1}{4\pi}\,\Bigg[M^2\,{\ell n}\frac{M^2}{\Lambda^2}
- (\Lambda^2 + M^2)\,{\ell n}\Bigg(1 + \frac{M^2}{\Lambda^2}\Bigg) +
\frac{2\pi}{g}\,M^2\Bigg]  + C',
\end{eqnarray}
where $C'$ is an infinite constant independent on $M$.

Using the normalization ${\cal E}(0) = 0$ we obtain
\begin{eqnarray}\label{labelE.17}
{\cal E}(M) = \frac{1}{4\pi}\,\Bigg[M^2\,{\ell n}\frac{M^2}{\Lambda^2}
- (\Lambda^2 + M^2)\,{\ell n}\Bigg(1 + \frac{M^2}{\Lambda^2}\Bigg) +
\frac{2\pi}{g}\,M^2\Bigg].
\end{eqnarray}
This evidences the complete agreement of the energy density of the
BCS--like ground state with the effective potential $V[M]$ given by
(\ref{label3.21}) at $\bar{\rho}(x) = M$:
\begin{eqnarray}\label{labelE.18}
{\cal E}(M) = V[M].
\end{eqnarray}
Thus, we have shown explicitly that the chirally broken phase in the
massless Thirring model is a superconducting phase of the BCS type
with the BCS non--perturbative superconducting vacuum.

Since the energy density ${\cal E}(M)$ has a maximum at $M = 0$, it is
obvious that for Thirring fermions the chirally broken phase is
energetically preferred with respect to the chiral symmetric phase.

One can show that fixing $M=const$ and tending $\Lambda \to \infty$
the effective potential and the energy density tend to a finite limit
${\cal E}(M) = V[M] \to - M^2/4\pi$. This means that the energy
spectrum of the ground state of the massless Thirring model is
restricted from below for $\Lambda \to \infty$.

Now let investigate chiral properties of the wave function of the
ground state (\ref{labelE.5}) under chiral rotations of the massless
Thirring fermion fields
\begin{eqnarray}\label{labelE.19}
\psi(x,t) \to {\psi\,}'(x,t) &=& e^{\textstyle i\gamma^5\alpha_{\rm
A}}\psi(x,t),\nonumber\\ \bar{\psi}(x,t) \to {\bar{\psi}\,}'(x,t) &=&
\bar{\psi}(x,t)\,e^{\textstyle i\gamma^5\alpha_{\rm A}}.
\end{eqnarray}
The operators of annihilation and creation transform under chiral
rotations (\ref{labelE.19}) as follows \cite{[18]},\cite{[24]}:
\begin{eqnarray}\label{labelE.20}
\begin{array}{lcl}
a(k^1) \to {a\,}'(k^1) = e^{\textstyle + i\alpha_{\rm
A}\varepsilon(k^1)}a(k^1)\hspace{0.2in},& b(k^1) \to {b\,}'(k^1) = e^{\textstyle -
i\alpha_{\rm A}\varepsilon(k^1)}b(k^1), \\ a^{\dagger}(k^1) \to
{a^{\dagger}}'(k^1) = e^{\textstyle - i\alpha_{\rm
A}\varepsilon(k^1)}a^{\dagger}(k^1)\hspace{0.05in},& b^{\dagger}(k^1) \to
{b^{\dagger}}'(k^1) = e^{\textstyle + i\alpha_{\rm
A}\varepsilon(k^1)}b^{\dagger}(k^1).
\end{array}
\end{eqnarray}
The wave function of the non--perturbative vacuum of the massless
Thirring model (\ref{labelE.5}) is not invariant under chiral
rotations (\ref{labelE.19}) and (\ref{labelE.20})
\cite{[18]},\cite{[24]}:
\begin{eqnarray}\label{labelE.21}
|\Omega\rangle \to |\Omega; \alpha_{\rm A}\rangle = \prod_{\textstyle
 k^1}\Big[u_{\textstyle k^1} + v_{\textstyle
 k^1}\,e^{\textstyle -2i\alpha_{\rm
 A}\varepsilon(k^1)}\,
 a^{\dagger}(k^1)b^{\dagger}(-k^1)\Big]|0\rangle,
\end{eqnarray}
One can show that the energy density ${\cal E}(M)$ does not depend on
the phase of the function $v_{\textstyle k^1}$.

The scalar product $\langle \alpha\,'_{\rm A};\Omega|\Omega;
\alpha_{\rm A}\rangle$ of the wave function for $\alpha\,'_{\rm
A}\not= \alpha_{\rm A}$ is equal to \cite{[18], [24]}:
\begin{eqnarray}\label{labelE.22}
\langle \alpha\,'_{\rm A};\Omega|\Omega; \alpha_{\rm A}\rangle =
\exp\Bigg\{\frac{L}{2\pi}\int^{\infty}_0dk^1\,{\ell n}\Bigg[1 -
\sin^2(\alpha\,'_{\rm A} - \alpha_{\rm A})\frac{M^2}{M^2 +
(k^1)^2}\Bigg]\Bigg\}.
\end{eqnarray}
In the limit $L\to \infty$ the wave functions $|\Omega; \alpha\,'_{\rm
A}\rangle$ and $|\Omega; \alpha_{\rm A}\rangle$ are orthogonal for
$\alpha\,'_{\rm A}\not= \alpha_{\rm A}$ \cite{[18], [24]}.

The wave function (\ref{labelE.5}) is invariant under parity
transformation: ${\cal P}|\Omega\rangle =|\Omega\rangle$. The
operator
\begin{eqnarray}\label{labelE.23}
{\cal O}_+ =2
\sum_{p^1}\varepsilon(p^1)b^{\dagger}(-p^1)a^{\dagger}(p^1)
\end{eqnarray}
is invariant under parity transformations (\ref{labelE.6}): ${\cal
P}{\cal O}_+{\cal P}^{\dagger} = {\cal O}_+$. Its vacuum expectation
value of the operator ${\cal O}_+$ per unit volume can be identified
with the order parameter for the massless Thirring model in the BCS
formalism
\begin{eqnarray}\label{labelE.24}
\hspace{-0.3in}&&\langle {\cal O}_+\rangle= \frac{1}{L}\,\langle \Omega| {\cal
O}_+|\Omega \rangle = - \frac{2}{L}\sum_{p^1}u_{\textstyle
p^1}v_{\textstyle p^1} =-
\int^{\infty}_{-\infty}\frac{dp^1}{2\pi}\frac{M}{\sqrt{M^2 + (p^1)^2}}
= -\frac{M}{2\pi}{\ell n}\Bigg(1 +
\frac{\Lambda^2}{M^2}\Bigg),\nonumber\\
\hspace{-0.3in}&&
\end{eqnarray}
where we have used (\ref{labelE.14}). The expectation value $\langle
{\cal O}_+\rangle$ is the fermion condensate
\begin{eqnarray}\label{labelE.25}
\langle \Omega|:\bar{\psi}(0)\psi(0):|\Omega\rangle =
\langle {\cal O}_+\rangle = -\frac{M}{2\pi}{\ell n}\Bigg(1 +
\frac{\Lambda^2}{M^2}\Bigg),
\end{eqnarray}
where we have used $\langle {\cal O}_+\rangle = \langle {\cal
O}^{\dagger}_+\rangle$.  

Thus, we have shown that the ground state of the massless Thirring
model possesses all properties of a BCS--type superconducting vacuum,
and our results obtained by means of the path integral approach are
fully reproducible within the BCS formalism.

\subsection{Equations of motion and Integral of motion}

\hspace{0.2in} Now let us turn to the analysis of the equations of
motion. Using the Lagrangian (\ref{labelH.11}) we derive the equations
of motion
\begin{eqnarray}\label{labelH.20}
i\gamma^{\mu}\partial_{\mu}\psi(x,t) &=&
g\,j^{\mu}(x,t)\gamma_{\mu}\psi(x,t) ,\nonumber\\
-i\partial_{\mu}\bar{\psi}(x,t)\gamma^{\mu} &=&
g\,\bar{\psi}(x,t)\gamma_{\mu}j^{\mu}(x,t).
\end{eqnarray}
Due to the peculiarity of 1+1--dimensional quantum field theories of
fermion fields \cite{[32]} these equations are equivalent to
\begin{eqnarray}\label{labelH.21}
i\partial_{\mu}\psi(x,t) &=& a\,j_{\mu}(x,t)\psi(x,t) +
b\,\varepsilon_{\mu\nu}j^{\nu}(x,t)\gamma^5\psi(x,t),\nonumber\\
-i\partial_{\mu}\bar{\psi}(x,t)
&=&a\,\bar{\psi}(x,t)j_{\mu}(x,t) +
b\,\bar{\psi}(x,t)\gamma^5j^{\nu}(x,t)\varepsilon_{\nu\mu},
\end{eqnarray}
where the parameters $a$ and $b$ are constrained by the condition $a +
b = g$\,\footnote{Below we show that $a-b = 1/c$ where $c$ is the
Schwinger term \cite{[33]}.}. Multiplying the equations 
(\ref{labelH.21}) by $\gamma^{\mu}$ and summing over $\mu = 0,1$ we
end up with the equations of motion (\ref{labelH.20}).

From the equations of motion (\ref{labelH.21}) we can get a very
important information about the evolution of fermions in the massless
Thirring model. For this we write (\ref{labelH.21}) in the form
\begin{eqnarray}\label{labelH.22}
- \partial_{\mu}[\bar{\psi}(x,t)\psi(x,t)] &=& 2\,b\,
\varepsilon_{\mu\nu}j^{\nu}(x,t)
[\bar{\psi}(x,t)i\gamma^5\psi(x,t)],\nonumber\\
\partial_{\mu}[\bar{\psi}(x,t)i\gamma^5\psi(x,t)] &=& 2\,b\,
\varepsilon_{\mu\nu}j^{\nu}(x,t)[\bar{\psi}(x,t)\psi(x,t)],
\end{eqnarray}
exclude  $2\,b\,\varepsilon_{\mu\nu}j^{\nu}(x,t)$  and  arrive at  the
equation
\begin{eqnarray}\label{labelH.23}
\partial_{\alpha}([\bar{\psi}(x,t)\psi(x,t)]^2 +
[\bar{\psi}(x,t)i\gamma^5\psi(x,t)]^2) = 0.
\end{eqnarray}
This means that the quantity
\begin{eqnarray}\label{labelH.24}
[\bar{\psi}(x,t)\psi(x,t)]^2 +
[\bar{\psi}(x,t)i\gamma^5\psi(x,t)]^2 = const
\end{eqnarray}
is an integral of motion. Using the equations of motion
(\ref{label3.24}) and going to the polar representation $\sigma(x,t) =
\rho(x,t)\,\cos \beta \vartheta(x,t)$ and $\varphi(x,t) =
\rho(x,t)\,\sin \beta \vartheta(x,t)$ we get
\begin{eqnarray}\label{labelH.25}
[\bar{\psi}(x,t)\psi(x,t)]^2 + [\bar{\psi}(x,t)i\gamma^5\psi(x,t)]^2
= \frac{\rho^2(x,t)}{g^2} = const.
\end{eqnarray}
Substituting Eq.(\ref{labelH.25}) in Eq.(\ref{labelH.23}) we obtain the equation of motion for the $\rho$--field
\begin{eqnarray}\label{labelH.26}
\partial_{\alpha}\rho(x,t) = 0.
\end{eqnarray}
This gives $\rho(x,t) = \rho(0)$. The value of $\rho(0)$ can be fixed
by noticing that for $\partial_{\alpha}\rho(x,t) = 0$ the effective
Lagrangian of the $\rho$--field is defined by the effective potential
(\ref{label3.21}), ${\cal L}_{\rm eff}[\rho(x,t)] = -
V[\rho(x,t)]$. In this case the equation of motion for the
$\rho$--field reads
\begin{eqnarray}\label{labelH.27}
\frac{\delta {\cal L}_{\rm eff}[\rho(x,t)]}{\delta \rho(x,t)} = -
\frac{\delta V[\rho(x,t)]}{\delta \rho(x,t)} = 0
\end{eqnarray}
and coincides with the extremum condition (\ref{label3.22}) with the
solution $\rho(x,t) = \rho_0 = M$. This implies that the solution of
(\ref{labelH.26}) which is the equation of motion for the
$\rho$--field should be $\rho(x,t) = \rho(0) = M$.

Since the terms $[\bar{\psi}(x,t)\psi(x,t)]^2$ and
$[\bar{\psi}(x,t)i\gamma^5\psi(x,t)]^2$ are positively defined, we
predict $\rho(x,t) \not= 0$. This testifies the stability of the
chirally broken phase during the evolution of the fermion system
described by the massless Thirring model evolving from the symmetry
broken phase.

Setting $\rho(x,t) = M + \tilde{\rho}(x,t)$, where the
$\tilde{\rho}$--field describes fluctuations of the $\rho$--field
around the minimum of the effective potential, the equation of motion
(\ref{labelH.26}) reduces to the form
\begin{eqnarray}\label{labelH.28}
\partial_{\alpha}\tilde{\rho}(x,t) = 0.
\end{eqnarray}
In Appendix B we show that the $\tilde{\rho}$--field decouples from
the system. The direct consequence of this decoupling is
$\tilde{\rho}(x,t) = 0$ as solution of Eq.(\ref{labelH.28}).

An example of classical solutions of the equations of motion
(\ref{labelH.20}) and (\ref{labelH.21}) for fermions evolving in the
chirally broken phase and obeying the integral of motion 
\begin{eqnarray}\label{labelH.29}
[\bar{\psi}(x,t)\psi(x,t)]^2 + [\bar{\psi}(x,t)i\gamma^5\psi(x,t)]^2
= \frac{M^2}{g^2} 
\end{eqnarray}
is given by the ansatz 
\begin{eqnarray}\label{labelH.30}
\psi(x,t) = \sqrt{-\frac{M}{2g}}{\displaystyle
\left(\begin{array}{c}e^{\textstyle +\omega/2}\,e^{\textstyle
- i\xi(x,t)} \\
e^{\textstyle - \omega/2}\,e^{\textstyle
+ i\eta(x,t)}
\end{array}\right)},
\end{eqnarray}
where $\omega$ is a arbitrary real parameter and $\xi(x,t) +\eta(x,t)
= \beta\vartheta(x,t)$.  In terms of (\ref{labelH.30}) the scalar
$\bar{\psi}(x,t)\psi(x,t$ )and pseudoscalar
$\bar{\psi}(x,t)i\gamma^5\psi(x,t)$ densities read
\begin{eqnarray}\label{labelH.31}
\bar{\psi}(x,t)\psi(x,t) &=&
-\frac{M}{g}\,\cos\beta\vartheta(x,t),\nonumber\\
\bar{\psi}(x,t)i\gamma^5\psi(x,t) &=& -
\frac{M}{g}\,\sin\beta\vartheta(x,t).
\end{eqnarray}
The functions $\xi(x,t)$ and $\eta(x,t)$ are found in Appendix C.

\subsection{Current algebra and the Schwinger term}

\hspace{0.2in} The canonical conjugate momentum of the $\psi$--field
is defined by
\begin{eqnarray}\label{labelG.1}
\pi(x,t) = \frac{\delta {\cal L}(x,t)}{\delta \partial_0\psi(x,t)} =
i\psi^{\dagger}(x,t).
\end{eqnarray}
The canonical anti--commutation relations read therefore
\begin{eqnarray}\label{labelG.2}
\{\psi(x,t),\pi(y,t)\} = i\,\delta(x - y)\;\to \;
\{\psi(x,t),\psi^{\dagger}(y,t)\} = \delta(x - y).
\end{eqnarray}
Using the canonical anti--commutation relations (\ref{labelG.2}) one
can see that the equal--time commutation relations $[j_{\mu}(x,t),
j_{\nu}(y,t)]$ vanish for each choice of $\mu$ and $\nu$
\cite{[30]}. However, according to Schwinger \cite{[29]} the
equal--time commutation relations $[j_{\mu}(x,t), j_{\nu}(y,t)]$
should read 
\begin{eqnarray}\label{labelG.3}
&&[j_0(x,t), j_0(y,t)] = 0,\nonumber\\
&&[j_0(x,t), j_1(y,t)] = i\,c\,\frac{\partial}{\partial
x}\delta(x-y),\nonumber\\
&&[j_1(x,t), j_1(y,t)] = 0,
\end{eqnarray}
where $c$ is the Schwinger term \cite{[33]}.

In the massless Thirring model the Schwinger term $c$ has been
calculated by Sommerfield \cite{[34]}, with the result $c =
1/\pi$. Now let us show that this result is due to the triviality of
the vacuum in the chiral symmetric phase of the massless Thirring
model. We will get another value for $c$ in the spontaneously broken
phase.

For this aim we evaluate the vacuum expectation value of the
equal--time commutation relation $[j_0(x,t), j_1(y,t)]$. In the
one--fermion loop approximation for free fermions with mass $M$,
sufficient as has been shown above for the description of the dynamics
of a fermion system in the chirally broken phase, we get
\begin{eqnarray}\label{labelG.4}
&&\langle 0|[j_0(x,t), j_1(y,t)]|0\rangle =  \langle
0|[:\bar{\psi}(x,t)\gamma_0\psi(x,t):,
:\bar{\psi}(y,t)\gamma_1\psi(y,t):]|0\rangle =\nonumber\\ 
&&= - \int\limits^{\infty}_{-\infty}\frac{dk}{2\pi}\,e^{\textstyle
ik(x-y)}\int\limits^{\infty}_{-\infty}\frac{dq}{4\pi}\,
\Bigg[\frac{k-q}{\displaystyle \sqrt{(k-q)^2 + M^2}} +
\frac{k+q}{\displaystyle \sqrt{(k+q)^2 + M^2}}\Bigg].
\end{eqnarray}
For the integration over $q$
\begin{eqnarray}\label{labelG.5}
\hspace{-0.5in}&&\int\limits^{\infty}_{-\infty}\frac{dq}{4\pi}\,
\Bigg[\frac{k-q}{\displaystyle \sqrt{(k-q)^2 + M^2}} +
\frac{k+q}{\displaystyle \sqrt{(k+q)^2 + M^2}}\Bigg] =\nonumber\\
\hspace{-0.5in}&&=\int\limits^{\infty}_{-\infty}\frac{dq}{4\pi}\,
\Bigg[\frac{q+k}{\displaystyle \sqrt{(q+k)^2 + M^2}} -
\frac{q-k}{\displaystyle \sqrt{(q-k)^2 + M^2}}\Bigg]=\nonumber\\
\hspace{-0.5in}&&=\int\limits^{\infty}_{-\infty}\frac{dq}{4\pi}
\int\limits^1_{-1}d\alpha\,\frac{d}{d\alpha}
\Bigg[\frac{q+k\alpha}{\displaystyle \sqrt{(q+k\alpha)^2 + M^2}}\Bigg]
=\int\limits^{\infty}_{-\infty}\frac{dq}{4\pi}
\int\limits^1_{-1}d\alpha\, \frac{k M^2}{\displaystyle ((q+k\alpha)^2
+ M^2)^{3/2}}=\nonumber\\
\hspace{-0.5in}&&=\int\limits^{\infty}_{-\infty}\frac{dq}{4\pi}
\int\limits^1_{-1}d\alpha\,\frac{k M^2}{\displaystyle (q^2 +
M^2)^{3/2}} = k\int\limits^{\infty}_{-\infty}\frac{dq}{2\pi}
\,\frac{M^2}{\displaystyle (q^2 + M^2)^{3/2}} =
k\int\frac{d^2q}{\pi^2i}\,\frac{M^2}{(M^2 - q^2 - i 0)^2}=\nonumber\\
\hspace{-0.5in}&&= k\,\frac{1}{\pi}\,\frac{\Lambda^2}{M^2 + \Lambda^2},
\end{eqnarray}
we used the relation
\begin{eqnarray}\label{labelG.6}
\int\limits^{\infty}_{-\infty}\frac{dq}{2\pi}\,\frac{M^2}{\displaystyle
(q^2 + M^2)^{3/2}} = \int\frac{d^2q}{\pi^2i}\,\frac{M^2}{(M^2 - q^2 -
i 0)^2}.
\end{eqnarray}
The r.h.s. of Eq.(\ref{labelG.6}) represented in  relativistic
invariant form is regularized according to our approach to the
evaluation of the effective Lagrangians (\ref{label3.35}) and
(\ref{label5.14}).

Substituting Eq.(\ref{labelG.5}) in Eq.(\ref{labelG.4}) we obtain
\begin{eqnarray}\label{labelG.7}
&&\langle 0|[j_0(x,t), j_1(y,t)]|0\rangle = \langle
0|[:\bar{\psi}(x,t)\gamma_0\psi(x,t):,
:\bar{\psi}(y,t)\gamma_1\psi(y,t):]|0\rangle =\nonumber\\ &&=
i\,\frac{1}{\pi}\,\frac{\Lambda^2}{M^2 +
\Lambda^2}\,\frac{\partial}{\partial x}\delta(x-y).
\end{eqnarray}
Following Schwinger \cite{[33]} and Sommerfield \cite{[34]} we write
down the equal--time commutation relation
\begin{eqnarray}\label{labelG.8}
[j_0(x,t), j_1(y,t)] = i\,\frac{1}{\pi}\,\frac{\Lambda^2}{M^2 +
\Lambda^2}\,\frac{\partial}{\partial x}\delta(x-y).
\end{eqnarray}
When matching Eq.(\ref{labelG.8}) with Eq.(\ref{labelG.3}) we derive
the Schwinger term
\begin{eqnarray}\label{labelG.9}
c = \frac{1}{\pi}\,\frac{\Lambda^2}{M^2 + \Lambda^2}.
\end{eqnarray}
For the chiral symmetric phase with $M = 0$ the Schwinger term is equal
to that obtained by Sommerfield $c = 1/\pi$ \cite{[34]}, whereas in
the chirally broken phase, when $M$ is defined by (\ref{label1.15}),
we get the new value
\begin{eqnarray}\label{labelG.10}
c = \frac{1}{\pi}\,\Big(1 - e^{\textstyle -2\pi/g}\Big).
\end{eqnarray}
This agrees with the dependence of $\beta^2$ on $g$ given by
Eq.(\ref{label4.5}).

\subsection{Energy--momentum tensor and its Sugawara form}

\hspace{0.2in} The energy--momentum tensor $\theta_{\mu\nu}(x,t)$ of
the fermion system described by the Lagrangian ${\cal L}(x,t)$ is
defined by
\begin{eqnarray}\label{labelF.1}
\theta_{\mu\nu}(x,t) &=&
\frac{1}{2}:\partial_{\mu}\bar{\psi}(x,t)\,\frac{\delta {\cal
L}(x,t)}{\delta \partial^{\nu}\bar{\psi}(,t)}: + 
\frac{1}{2}:\frac{\delta
{\cal L}(x,t)}{\delta \partial^{\mu}\psi(x,t)}
\partial_{\nu}\psi(x,t): +
(\mu \leftrightarrow \nu)\nonumber\\
&& - {\cal L}(x,t)\,g_{\mu\nu}.
\end{eqnarray}
For the massless Thirring model with the Lagrangian (\ref{labelH.1})
the energy--momentum tensor $\theta_{\mu\nu}(x,t)$ reads
\begin{eqnarray}\label{labelF.2}
\hspace{-0.5in}\theta_{\mu\nu}(x,t) &=&
\frac{1}{4}:\bar{\psi}(x,t)i\gamma_{\mu}\partial_{\nu}\psi(x,t): +
\frac{1}{4}:\bar{\psi}(x,t)i\gamma_{\nu}
\partial_{\mu}\psi(x,t):\nonumber\\
\hspace{-0.5in}&-&
\frac{1}{4}:\partial_{\mu}\bar{\psi}(x,t)i\gamma_{\nu}\psi(x,t): -
\frac{1}{4}:\partial_{\nu}\bar{\psi}(x,t)i\gamma_{\mu}\psi(x,t):
\nonumber\\
\hspace{-0.5in}&-&g_{\mu\nu}:\Big[\bar{\psi}(x,t)i\gamma^{\alpha}
\partial_{\alpha}\psi(x,t) -
\frac{1}{2}\,g\,\bar{\psi}(x,t)\gamma_{\alpha}\psi(x,t)
\bar{\psi}(x,t)\gamma^{\alpha}\psi(x,t)]:.
\end{eqnarray}
As has been pointed out by Callan, Dashen and Sharp \cite{[30]} the
components of the energy momentum tensor $\theta_{\mu\nu}(x,t)$ in the
massless Thirring model obey the same equal--time commutation
relations as the components of the energy--momentum tensor
$\theta^{S}_{\mu\nu}(x,t)$ defined in terms of the vector current
$j_{\mu}(x,t)$ only
\begin{eqnarray}\label{labelF.3}
\theta^{S}_{\mu\nu}(x,t) = \frac{1}{2c}[j_{\mu}(x,t)j_{\nu}(x,t) +
j_{\nu}(x,t)j_{\mu}(x,t) -
g_{\mu\nu}\,j_{\alpha}(x,t)j^{\alpha}(x,t)],
\end{eqnarray}
where $c$ is the Schwinger term. A quantum field theory with currents
as quantum variables and an energy--momentum tensor of the kind
(\ref{labelF.3}) has been considered by Sugawara \cite{[31]}.

A direct reduction of the energy--momentum tensor (\ref{labelF.2}) to
Sugawara$^{\prime}$s form (\ref{labelF.3}) can be carried out using
the equations of motion (\ref{labelH.21}) as shown by Sommerfield
\cite{[32]}. Substituting (\ref{labelH.21}) in (\ref{labelF.2}) with
the Lagrangian defined by (\ref{labelH.11}) we arrive at the
expression:
\begin{eqnarray}\label{labelF.4}
\theta_{\mu\nu}(x,t) =
\frac{1}{2}\,(a-b)\,[j_{\mu}(x,t)j_{\nu}(x,t) +
j_{\nu}(x,t)j_{\mu}(x,t) -
g_{\mu\nu}\,j_{\alpha}(x,t)j^{\alpha}(x,t)],
\end{eqnarray}
When matching (\ref{labelF.4}) with (\ref{labelF.3}) we infer that
$a-b = 1/c$. In the chirally broken phase the Schwinger term is
defined by (\ref{labelG.10}).

\section{The fermion number as a topological charge of the SG model}
\setcounter{equation}{0}

\hspace{0.2in} The topological properties of the SG model are
characterized by the topological current ${\cal J}^{\mu}(x)$
\cite{[35]}:
\begin{eqnarray}\label{label7.1}
{\cal J}^{\mu}(x,t) =
\frac{\beta}{2\pi}\,\varepsilon^{\mu\nu}\,\partial_{\nu}\vartheta(x,t).
\end{eqnarray}
The spatial integral of its time--component, the topological charge,
\begin{eqnarray}\label{label7.2}
q = \int\limits^{\infty}_{-\infty}dx\,{\cal J}^0(x,t)=
\frac{\beta}{2\pi}\int\limits^{\infty}_{-\infty}dx\,\frac{\partial
}{\partial x}\vartheta(x,t) = \frac{\beta}{2\pi}\,[\vartheta(\infty) -
\vartheta(-\infty)].
\end{eqnarray}
is conserved irrespective of the equations of motion and is integer
valued \cite{[35]}. The field $e^{\textstyle
i\beta\vartheta(x,t)}$, where $\vartheta(x,t)$ is a solution of the
equation of motion \cite{[1]}
\begin{eqnarray}\label{label7.3}
\Box\vartheta(x,t) + \frac{\alpha}{\beta}\,\sin\beta\vartheta(x,t) = 0,
\end{eqnarray}
maps at any time $t$ the real axis $\mathbb{R}^{\,1}$ onto the circle
$\mathbb{S}^1$ with a winding number equal to the topological charge
$q$ \cite{[35]}.

For a solitary wave moving with a velocity $u$, the one--soliton
solution of the SG model, \cite{[1]}
\begin{eqnarray}\label{label7.4}
\vartheta(x,t) =
\frac{4}{\beta}\,\arctan\Bigg[\exp\Bigg(\sqrt{\alpha}\,
\frac{x-ut}{\sqrt{1-u^2}}\Bigg)\Bigg]
\end{eqnarray}
the topological charge $q$ is equal to unity
\begin{eqnarray}\label{label7.5}
q = \frac{2}{\pi}\,\arctan\Bigg[\exp\Bigg(\sqrt{\alpha}\,
\frac{x-ut}{\sqrt{1-u^2}}\Bigg)\Bigg] \Bigg|^{\infty}_{-\infty} = 1.
\end{eqnarray}
In turn, for the antisoliton solution, $\bar{\vartheta}(x,t)$, given by \cite{[1]}
\begin{eqnarray}\label{label7.6}
\bar{\vartheta}(x,t) =
\frac{4}{\beta}\,\arctan\Bigg[\exp\Bigg(-\sqrt{\alpha}\,
\frac{x-ut}{\sqrt{1-u^2}}\Bigg)\Bigg]
\end{eqnarray}
the topological charge $\bar{q}$ amounts to $\bar{q} = -1$.

We argue that in our approach to the bosonization of the massive
Thirring model the topological current (\ref{label7.1}) coincides with
the Noether current related to the global $U_{\rm V}(1)$ symmetry of
the massive Thirring model. Since this Noether current is responsible
for the conservation of the fermion number in the massive Thirring
model, this allows to identify the topological charge with the fermion
number.

For the derivation of the Noether current we write the effective
Lagrangian of the bosonized version of the massive Thirring model
(\ref{label5.12}) in the form
\begin{eqnarray}\label{label7.7}
\hspace{-0.3in}{\cal L}_{\rm eff}(x)&=& -i\,\langle x|{\rm tr}{\ell
n}(i\gamma^{\mu}\partial_{\mu} + \gamma^{\mu}A_{\mu} - M)|x
\rangle + \ldots,
\end{eqnarray}
where $A_{\mu}(x) =
\beta\,\varepsilon_{\mu\nu}\,\partial^{\nu}\vartheta(x)/2$ with
$\beta$ defined by Eq.(\ref{label4.5}) and included in the definition
of the $A_{\mu}$--field to get the correct kinetic term for the
$\vartheta$--field.

Under infinitesimal local $U_{\rm V}(1)$ rotations with a parameter
$\alpha_{\rm V}(x)$ the vector field $A_{\mu}(x)$ transforms
as\,\footnote{Transformations of fermion fields under local $U_{\rm
V}(1)$ rotations with a parameter $\alpha_{\rm V}(x)$ are defined by
Eq.(\ref{labelH.2}).}
\begin{eqnarray}\label{label7.8}
A_{\mu}(x) \to A^{\prime}_{\mu}(x) = A_{\mu}(x) +
\partial_{\mu}\alpha_{\rm V}(x),
\end{eqnarray}
and the effective Lagrangian (\ref{label7.7}) changes as follows
\begin{eqnarray}\label{label7.9}
\hspace{-0.3in}{\cal L}_{\rm eff}(x) \to {\cal L}_{\rm
eff}[\alpha_{\rm V}(x)] = {\cal L}_{\rm eff}(x) + i\,\Bigg\langle
x\Bigg|{\rm tr}\,\Bigg\{\frac{1}{M - i\gamma^{\nu}\partial_{\nu} -
\gamma^{\nu}A_{\nu}}\,\gamma^{\mu}\Bigg\}\Bigg
|x\Bigg\rangle\,\partial_{\mu}\alpha_{\rm V}(x).
\end{eqnarray}
The Noether current is defined by \cite{[22]}
\begin{eqnarray}\label{label7.10}
J^{\mu}(x) = - \frac{\delta {\cal L}_{\rm eff}[\alpha_{\rm
V}(x)]}{\delta \partial_{\mu}\alpha_{\rm V}(x)} = - i\,\Bigg\langle
x\Bigg|{\rm tr}\,\Bigg\{\frac{1}{M - i\gamma^{\nu}\partial_{\nu} -
\gamma^{\nu}A_{\nu}}\,\gamma^{\mu}\Bigg\}\Bigg |x\Bigg\rangle.
\end{eqnarray}
The explicit calculation of the matrix element in the r.h.s. of
Eq.(\ref{label7.10}) gives
\begin{eqnarray}\label{label7.11}
J^{\mu}(x) = \frac{1}{2\pi}\,\Big(1 - e^{\textstyle
-2\pi/g}\Big)A^{\mu}(x) =
\frac{2}{\beta}\,\varepsilon^{\mu\nu}\,\partial_{\nu}\vartheta(x),
\end{eqnarray}
where we have used relation (\ref{label4.5}).

Thus, the topological current ${\cal J}^{\mu}(x)$ of
Eq.(\ref{label7.1}) is proportional to the Noether current
(\ref{label7.11}) with $\beta^2/4\pi$ as coefficient of
proportionality.

Since the topological current coincides with the Noether current
related to the $U_{\rm V}(1)$ symmetry of the massive Thirring model,
which is responsible for the conservation of the fermion number, the
topological charge $q$ has the meaning of the fermion number.

This leads to the conclusion that the solitons of the SG model can be
identified with fermions, as conjectured by Skyrme \cite{[4]}.

It is interesting to note that the mass of the soliton \cite{[1]}
$M_{\rm sol} = 8\sqrt{\alpha}/\beta^2$ can be represented in a form
resembling the Gell-- Mann--Oakes--Renner low--energy theorem for the
mass spectrum of low--lying pseudoscalar mesons \cite{[36]}
\begin{eqnarray}\label{label7.12}
M^2_{\rm sol} = \frac{64\alpha}{\beta^4} = -
\frac{64}{\beta^2}\,m\,\langle \bar{\psi}\psi\rangle + O(m^2),
\end{eqnarray}
where we have used Eqs.(\ref{label4.5}) and (\ref{label5.15}).

\section{Chiral symmetry breaking in the massless Thirring model 
and the Mermin--Wagner theorem}
\setcounter{equation}{0}

\hspace{0.2in} In this Section we would like to show that our approach
to the bosonization of the massless Thirring model does not contradict
to the Mermin--Wagner theorem \cite{[25]}. According to this theorem
there is no spontaneously broken continuous symmetry in 2--dimensional
quantum field theories. The essence of the Mermin--Wagner theorem can
be illustrated by the classical Heisenberg model with a continuous
$O(n)$ symmetry, where dynamical variables are unit vectors
$\vec{S}_i$ on a sphere \cite{[27]}. Following Mermin and Wagner
\cite{[25]} one can show \cite{[27]} that there is no spontaneous
magnetization for $n < 3$. The applicability of the Mermin--Wagner
theorem to 1+1--dimensional quantum field theories has been pointed
out in \cite{[26],[37],[38]} (see also \cite{[27],[28]}). From a
dynamical point of view the Mermin--Wagner theorem states  the
absence of long--range order in 1+1--dimensional quantum field
theories.

In this connection Coleman \cite{[26]} argued that in a
1+1--dimensional quantum field theory of a massless scalar field there
are no Goldstone bosons.  They accompany, according to Goldstone's
theorem \cite{[39]}, the spontaneous breaking of a continuous
symmetry. In order to prove this statement Coleman considered a
quantum field theory of a massless free scalar field $\varphi(x,t)$
with the Lagrangian
\begin{eqnarray}\label{label8.1}
{\cal L}(x,t) =
\frac{1}{2}\,\partial_{\mu}\varphi(x,t)\partial^{\mu}\varphi(x,t).
\end{eqnarray}
The equation of motion of the $\varphi$--field reads
\begin{eqnarray}\label{label8.2}
\Box \varphi(x,t) = 0.
\end{eqnarray}
The Lagrangian (\ref{label8.1}) is invariant under the field
translations \cite{[40]}
\begin{eqnarray}\label{label8.3}
 \varphi(x,t) \to \varphi'(x,t) = \varphi(x,t) - 2\,\alpha_{\rm A},
\end{eqnarray}
where $\alpha_{\rm A}$ is an arbitrary parameter. The conserved
current associated with these field translations is equal to
\cite{[40]}
\begin{eqnarray}\label{label8.4}
j_{\mu}(x,t) = \partial_{\mu}\varphi(x,t).
\end{eqnarray}
The total ``charge'' is defined by the time--component of
$j_{\mu}(x,t)$ \cite{[40]}
\begin{eqnarray}\label{label8.5}
Q(t) = \lim_{L\to \infty}Q_L(t) =\lim_{L\to \infty}
\int^{ L/2}_{- L/2}dx\,\frac{\partial}{\partial
t}\varphi(x,t),
\end{eqnarray}
where $L$ is the volume occupied by the system.

It is well--known that the spontaneous breaking of a continuous
symmetry occurs when the ground state of the system is not invariant
under the symmetry group \cite{[40]}. The ground state of the system
described by the Lagrangian (\ref{label8.1}) is not invariant under
field translations (\ref{label8.3}) \cite{[40]}. Thereby, the
field--translation symmetry should be spontaneously broken and a
Goldstone boson should appear \cite{[40]}.

The absence of Goldstone bosons in the quantum field theory described
by the Lagrangian (\ref{label8.1}) Coleman argued by stating the
impossibility to construct a massless scalar field operator (see also
\cite{[27]}). This statement has been supported by the analysis of the
two--point Wightman function \cite{[26], [27]}:
\begin{eqnarray}\label{label8.6}
\langle 0|\varphi(x,t)\varphi(0)|0\rangle = 
\int\frac{d^2k}{2\pi}\,\theta(k^0)\,\delta(k^2)\, e^{\textstyle - ik^0t
+ ik^1x}=\frac{1}{2\pi}\int^{\infty}_0\frac{dk^1}{k^1}\,
\cos(k^1x)\,e^{\textstyle -ik^1t},
\end{eqnarray}
which is defined by {\it a meaningless infrared divergent integral. No
subtraction procedure can be devised to circumvent this difficulty
without spoiling the fundamental properties of field theory, for
instance, positivity of the Hilbert space metric. A massless scalar
field theory is undefined in a two--dimensional world due to severe
infrared divergences} \cite{[27]}. This corresponds to the destruction
of  long--range order pointed out by Mermin and Wagner \cite{[27]}.

However, in spite of the widely accepted Coleman's statement about the
absence of Goldstone bosons in a 1+1--dimensional quantum field theory
described by the Lagrangian (\ref{label8.1}) we argue, nevertheless,
that in this theory the field--translation symmetry is spontaneously
broken in the sense of the non--invariance of the ground state under
transformations (\ref{label8.3}) and Goldstone bosons appear.

Indeed, as has been pointed out by Itzykson and Zuber that in the case
of the quantum field theory described by the Lagrangian
(\ref{label8.1}) the Goldstone boson is the quantum of the
$\varphi$--field itself \cite{[40]}. Then, the non--invariance of the
ground state of the system can be demonstrated by acting the operator
$\exp\{-2i\alpha_{\rm A}Q(0)\}$ on the vacuum wave function
$|0\rangle$, i.e. $|\alpha_{\rm A}\rangle = \exp\{-2i\alpha_{\rm
A}Q(0)\}|0\rangle$ \cite{[40]}.

For the calculation of $|\alpha_{\rm A}\rangle$ we follow Itzykson and
Zuber \cite{[40]} and use the Fourier decomposition of the massless
scalar field $\varphi(x,t)$:
\begin{eqnarray}\label{label8.7}
\varphi(x,t) =
\int^{\infty}_{-\infty}\frac{dk^1}{2\pi}\,\frac{1}{2k^0}\,
\Big[a(k^1)\,e^{\textstyle
-ik^0t + ik^1x} + a^{\dagger}(k^1)\,e^{\textstyle ik^0t - ik^1x}\Big],
\end{eqnarray}
where $k^0 = |k^1|$, then $a(k^1)$ and $a^{\dagger}(k^1)$ are
annihilation and creation operators obeying the standard commutation
relation 
\begin{eqnarray}\label{label8.8}
[a(k^1), a^{\dagger}(q^1)] = (2\pi)\,2k^0\,\delta(k^1 - q^1).
\end{eqnarray}
From (\ref{label8.5}) we obtain the total ``charge'' operator $Q(0)$
 \cite{[40]}
\begin{eqnarray}\label{label8.9}
Q(0) = -\frac{i}{2}[a(0) - a^{\dagger}(0)].
\end{eqnarray}
Then, we get the wave function $|\alpha_{\rm A}\rangle$
\begin{eqnarray}\label{label8.10}
|\alpha_{\rm A}\rangle = e^{\textstyle -2i\alpha_{\rm A}Q(0)}|0\rangle
  = e^{\textstyle - \alpha_{\rm A}[a(0) - a^{\dagger}(0)]}|0\rangle.
\end{eqnarray}
For the subsequent mathematical operations with the wave functions
$|\alpha_{\rm A}\rangle$ it is convenient to use the regularization
procedure suggested by Itzykson and Zuber. We define the regularized
operator $Q(0)_{\rm R}$ as follows \cite{[40]}
\begin{eqnarray}\label{label8.11}
Q(0)_{\rm R} = \lim_{L \to
\infty}\int^{\infty}_{-\infty}dx\,\frac{\partial}{\partial
t}\varphi(x,t)\Bigg|_{t=0}e^{\textstyle - x^2/L^2}.
\end{eqnarray}
The regularized wave function $|\alpha_{\rm A}\rangle_{\rm R}$ reads
then
\begin{eqnarray}\label{label8.12}
|\alpha_{\rm A}\rangle_{\rm R} &=& e^{\textstyle -2i\alpha_{\rm
  A}Q(0)_{\rm R}}|0\rangle = \nonumber\\ &=&\lim_{L \to
  \infty}\exp\Big\{ -\frac{\alpha_{\rm
  A}L}{2\sqrt{\pi}}\int^{\infty}_{-\infty}dk^1[a(k^1) -
  a^{\dagger}(k^1)]\,e^{\textstyle - L^2(k^1)^2/4}\Big\}|0\rangle.
\end{eqnarray}
The energy operator of the massless scalar field described by the
Lagrangian (\ref{label8.1}) is equal to
\begin{eqnarray}\label{label8.13}
\hat{H}(t) &=& \int^{\infty}_{-\infty}dx\,{\cal H}(x,t) =
\frac{1}{2}\int^{\infty}_{-\infty}dx\,:\Bigg[\Bigg(\frac{\partial
\varphi(x,t)}{\partial t}\Bigg)^2 + \Bigg(\frac{\partial
\varphi(x,t)}{\partial x}\Bigg)^2\Bigg]: =\nonumber\\ &=&
\frac{1}{2}\int^{\infty}_{-\infty}\frac{dk^1}{2\pi}\,a^{\dagger}(k^1)a(k^1).
\end{eqnarray}
One can easily show that the wave functions $|\alpha_{\rm
A}\rangle_{\rm R}$ are eigenfunctions of the energy operator
(\ref{label8.13}) for the eigenvalue zero
\begin{eqnarray}\label{label8.14}
\hat{H}(t)|\alpha_{\rm A}\rangle_{\rm R} = E(\alpha_{\rm A})
|\alpha_{\rm A}\rangle_{\rm R} = 0.
\end{eqnarray}
This evidences that the energy of the vacuum state is infinitely
degenerated, and the vacuum state is not invariant under the field
translations (\ref{label8.3}). The wave functions of the vacuum state
$|\alpha_{\rm A}\rangle_{\rm R}$ and $|\alpha'_{\rm A}\rangle_{\rm R}$
are not orthogonal to each other for $\alpha'_{\rm A}\not= \alpha_{\rm
A}$ and the scalar product ${_{\rm R}\langle}\alpha'_{\rm
A}|\alpha_{\rm A}\rangle_{\rm R}$ amounts to \cite{[40]}
\begin{eqnarray}\label{label8.15}
{_{\rm R}\langle}\alpha'_{\rm A}|\alpha_{\rm A}\rangle_{\rm R} =
e^{\textstyle - (\alpha'_{\rm A} - \alpha_{\rm A})^2}.
\end{eqnarray}
However, since the eigenvalue of the wave functions $|\alpha_{\rm
A}\rangle$ is zero, they can be orthogonalized by any appropriate
orthogonalization procedure as used in molecular and nuclear physics.

We would like to emphasize that the results expounded above are not
related to the impossibility to determine the two--point Wightman
function (\ref{label8.6}) in the infrared region of $\varphi$--field
fluctuations.  In fact, the analysis of the non--invariance of the
vacuum wave function under the symmetry transformations
(\ref{label8.3}) treats the massless scalar field at the tree
level. This is an appropriate description, since the massless scalar
field $\varphi$ is free, no vacuum fluctuations are entangled and the
quanta of the massless $\varphi$--field are on--mass shell.

Thus, following the Itzykson--Zuber analysis of the 1+1--dimensional
massless scalar field theory of the $\varphi$--field described by the
Lagrangian (\ref{label8.1}) we have shown that (i) the translation
symmetry (\ref{label8.3}) is spontaneously broken, (ii) Goldstone
bosons appear and they are quanta of the $\varphi$--field, (iii) the
ground state is not invariant under the field--translation symmetry
and (iv) the energy of the ground state is infinitely
degenerated. Hence, all requirements for a continuous symmetry to be
spontaneously broken are available in the 1+1--dimensional quantum
field theory of a massless scalar field described by the Lagrangian
(\ref{label8.1}).

Now let us show that chiral symmetry in our approach to the massless
Thirring model is spontaneously broken, i.e. the wave function of the
ground state is not invariant under chiral rotations and a Goldstone
boson exists. The fact of the non--invariance of the ground state of
the massless Thirring model under chiral rotations has been
demonstrated in Eq.(\ref{labelE.21}). The Goldstone bosons are the
quanta the $\vartheta$--field and the effective Lagrangian of the
fermion system, quantized around the minimum of the effective
potential (\ref{label3.21}), is invariant under chiral rotations. In
order to show this we suggest to rewrite the partition function
(\ref{label3.29}) as follows
\begin{eqnarray}\label{label8.16}
Z_{\rm Th} =\int {\cal D}\vartheta {\cal D}\psi{\cal D}\bar{\psi}\,
\exp\,i\int d^2x\,{\cal L}_{\rm eff}[\bar{\psi},\psi;\vartheta],
\end{eqnarray}
where the effective Lagrangian ${\cal L}_{\rm
eff}[\bar{\psi},\psi;\vartheta]$ is determined by
\begin{eqnarray}\label{label8.17}
{\cal L}_{\rm eff}[\bar{\psi},\psi; \vartheta] =
\bar{\psi}(x)\Big(i\gamma^{\mu}\partial_{\mu} - M\,e^{\textstyle
i\,\gamma^5 \beta\vartheta(x)}\,\Big)\psi(x).
\end{eqnarray}
For convenience we have renormalized the $\vartheta$-- field and the
parameter $\beta$ is given by (\ref{label4.5}).  The Lagrangian
(\ref{label8.17}) is invariant under chiral rotations
(\ref{labelE.19})
\begin{eqnarray}\label{label8.18}
{\cal L}_{\rm eff}[\bar{\psi},\psi;\vartheta] \to {\cal L}_{\rm
eff}[\bar{\psi}\,',\psi\,';\vartheta\,'] =
\bar{\psi}\,'(x)\Big(i\gamma^{\mu}\partial_{\mu} - M\,e^{\textstyle
i\,\gamma^5 \beta \vartheta\,'(x)}\,\Big)\psi\,'(x),
\end{eqnarray}
where the field $\vartheta\,'(x)$ is defined by
\begin{eqnarray}\label{label8.19}
\vartheta\,'(x) = \vartheta(x) - 2\,\alpha_{\rm A}/\beta.
\end{eqnarray}
Such an invariance under chiral rotations is trivially explained by the
Mexican--hat shape of the effective potential (\ref{label3.21})
depicted in Fig.2 allowing arbitrary changes of the $\vartheta$--field
around the hill at fixed $\rho$.

In the bosonic description the effective Lagrangian (\ref{label8.17})
 is equivalent to the effective Lagrangian depending only on the
 $\vartheta$--field (\ref{label3.35})
\begin{eqnarray}\label{label8.20}
{\cal L}_{\rm eff}[\bar{\psi},\psi;\vartheta] \to {\cal L}_{\rm
eff}[\vartheta] =
\frac{1}{2}\,\partial_{\mu}\vartheta(x)\partial^{\mu}\vartheta(x),
\end{eqnarray}
which is invariant under field translations (\ref{label8.19}) caused
by chiral rotations.

Thus, the fermionic system described by the massless Thirring model
and quantized around the minimum of the effective potential
(\ref{label3.21}) satisfies all requirements for a spontaneously
broken chiral symmetry as has been discussed above
(\ref{label8.7})--\ref{label8.15}). This evidences that the
bosonization of this fermionic system runs via the chirally broken
phase and is finally described by the Goldstone boson field
$\vartheta$.

For the calculation of the effective Lagrangian (\ref{label8.20}) we
break chiral symmetry explicitly by a local chiral rotation
\begin{eqnarray}\label{label8.21}
 \psi(x) \to e^{\textstyle - i\gamma^5\beta \vartheta(x)/2}\psi(x).
\end{eqnarray}
This reduces the effective Lagrangian (\ref{label8.17}) to the form
\begin{eqnarray}\label{label8.22}
{\cal L}_{\rm eff}[\bar{\psi},\psi; \vartheta] =
\bar{\psi}(x)\Big(i\gamma^{\mu}\partial_{\mu} +
\frac{1}{2}\,\beta\gamma^{\mu}\varepsilon_{\mu\nu}
\partial^{\nu}\vartheta(x) - M\Big)\psi(x).
\end{eqnarray}
The term proportional to $M$ breaks chiral symmetry explicitly
\begin{eqnarray}\label{label8.23}
{\cal L}_{\rm eff}[\bar{\psi},\psi; \vartheta] &\to& {\cal L}_{\rm
eff}[\bar{\psi}\,',\psi\,';\vartheta\,']  =\nonumber\\
&=&\bar{\psi}\,'(x)\Big(i\gamma^{\mu}\partial_{\mu} +
\frac{1}{2}\,\beta\gamma^{\mu}\varepsilon_{\mu\nu}
\partial^{\nu}\vartheta(x) - M\,e^{\textstyle -2\,i\,\gamma^5
\alpha_{\rm A}}\,\Big)\psi\,'(x).
\end{eqnarray}
Such a violation of chiral invariance is caused by a gauge fixing
specifying the starting point at the Mexican--hat for counting of the
chiral phase of a fermion field during a travel around the hill. Due
to the Abelian symmetry Faddeev--Popov ghosts do not appear and the
Faddeev--Popov determinant is equal to unity.

Let us show that such a gauge fixing does not affect the result. The
evaluation of the effective Lagrangian of the $\vartheta$--field does
not depend on $\alpha_{\rm A}$ and coincides with (\ref{label8.20}).

Indeed, the effective Lagrangian of the $\vartheta$--field is defined
by the two--vertex fermion diagram
\begin{eqnarray}\label{label8.24}
\hspace{-0.3in}&&{\cal L}_{\rm eff}(x) = -
\frac{1}{8\pi}\int\frac{d^2x_1d^2k_1}{(2\pi)^2}\,e^{\textstyle
-ik_1\cdot(x_1 - x)}A_{\mu}(x)A_{\nu}(x_1)\nonumber\\
\hspace{-0.3in}&&\times\int \frac{d^2k}{\pi i}\,{\rm
tr}\Bigg\{\frac{1}{\displaystyle M\,e^{\textstyle -2\,i\,\gamma^5
\alpha_{\rm A}} - \hat{k}}\gamma^{\mu}\frac{1}{\displaystyle
M\,e^{\textstyle -2\,i\,\gamma^5 \alpha_{\rm A}} - \hat{k} -
\hat{k}_1}\gamma^{\nu}\Bigg\}=\nonumber\\
\hspace{-0.3in}&&= -
\frac{1}{8\pi}\int\frac{d^2x_1d^2k_1}{(2\pi)^2}\,e^{\textstyle
-ik_1\cdot(x_1 - x)}A_{\mu}(x)A_{\nu}(x_1)\nonumber\\
\hspace{-0.3in}&&\times\int \frac{d^2k}{\pi i}\,{\rm
tr}\Bigg\{\frac{\displaystyle M\,e^{\textstyle -2\,i\,\gamma^5
\alpha_{\rm A}} + \hat{k}}{\displaystyle M^2 -
k^2}\gamma^{\mu}\frac{\displaystyle M\,e^{\textstyle -2\,i\,\gamma^5
\alpha_{\rm A}} + \hat{k} + \hat{k}_1}{M^2 - (k +
k_1)^2}\gamma^{\nu}\Bigg\}=\nonumber\\
\hspace{-0.3in}&&= -
\frac{1}{8\pi}\int\frac{d^2x_1d^2k_1}{(2\pi)^2}\,e^{\textstyle
-ik_1\cdot(x_1 - x)}A_{\mu}(x)A_{\nu}(x_1)\nonumber\\
\hspace{-0.3in}&&\times\int \frac{d^2k}{\pi i}\,{\rm
tr}\Bigg\{\frac{1}{\displaystyle M^2 - k^2}\frac{1}{M^2 - (k +
k_1)^2}\Bigg[M\,e^{\textstyle -2\,i\,\gamma^5 \alpha_{\rm
A}}\gamma^{\mu}M\,e^{\textstyle -2\,i\,\gamma^5 \alpha_{\rm
A}}\gamma^{\nu}\nonumber\\
\hspace{-0.3in}&& + \hat{k}\gamma^{\mu}M\,e^{\textstyle
-2\,i\,\gamma^5 \alpha_{\rm A}}\gamma^{\nu} +M\,e^{\textstyle
-2\,i\,\gamma^5 \alpha_{\rm A}}\gamma^{\mu}(\hat{k} +
\hat{k}_1)\gamma^{\nu} + \hat{k}\gamma^{\mu}(\hat{k} +
\hat{k}_1)\gamma^{\nu}\Bigg] \Bigg\}.
\end{eqnarray}
The second and the third terms vanish due to the trace over Dirac
matrices. Therefore, the r.h.s. of (\ref{label8.24}) does not depend
on $\alpha_{\rm A}$ and the result of the evaluation of the momentum
integral leads to the effective Lagrangian (\ref{label8.20}). This
Lagrangian of the $\vartheta$--field is invariant under
$\vartheta$--field translations (\ref{label8.19}) caused by chiral
rotations. As the vacuum of the $\vartheta$--field is not invariant
under (\ref{label8.19}), the symmetry becomes spontaneously broken in
the way described above and the Goldstone boson is the quantum of the
$\vartheta$--field \cite{[40]}.

This result agrees completely with the derivation of Effective Chiral
Lagrangians within the ENJL model with chiral $U(N_F)\times U(N_F)$
\cite{[19]}--\cite{[22]}, where $N_F$ is the number of quark
flavours. In fact, the starting Lagrangian of the ENJL model with
massless quark fields is invariant under the chiral group
$U(N_F)\times U(N_F)$. Then, via the chirally broken phase with
dynamical quarks the Lagrangian of the ENJL model after the
integration over quark degrees of freedom acquires the form of the
Effective Chiral Lagrangian containing only local boson fields. This
Effective Chiral Lagrangian is invariant under the chiral group
$U(N_F)\times U(N_F)$ \cite{[19]}--\cite{[22]}. However, the bosonic
system described by this Effective Chiral Lagrangian is not stable
under symmetry breaking, and the phase of spontaneously broken chiral
symmetry is energetically preferable. In this spontaneously broken
phase all vacuum expectation values of local bosonic fields are
defined by parameters of the chirally broken phase of the quark system
described by the ENJL model \cite{[19]}--\cite{[22]}.

Now we are able to discuss the problem of the vanishing of the vacuum
expectation value of $\bar{\psi}(x)\psi(x)$ in
(\ref{label4.19}). Using the Abelian bosonization rules
(\ref{label4.17}) we get
\begin{eqnarray}\label{label8.25}
\hspace{-0.3in}&&\langle 0|\bar{\psi}(x)\psi(x)|0\rangle = \langle
0|[\sigma_+(x) + \sigma_-(x)]|0\rangle = 
\frac{1}{2}\,Z^{-1}\langle\bar{\psi}\psi\rangle\langle 0|[A_-(x) +
A_+(x)]|0\rangle =\nonumber\\
\hspace{-0.3in}&&= \frac{1}{2}\,Z^{-1}\langle\bar{\psi}\psi\rangle\int
{\cal D}\vartheta\,e^{\textstyle -i\,\frac{1}{2}\int
d^2z\,\vartheta(z)\,(\Box + \mu^2)\,\vartheta(z)}\Big(e^{\textstyle
i\beta\vartheta(x)} + e^{\textstyle - i\beta\vartheta(x)}\Big) =
\nonumber\\
\hspace{-0.3in}&&=Z^{-1}\langle\bar{\psi}\psi\rangle\,e^{\textstyle
\frac{1}{2}\beta^2\,i\Delta(0)} =Z^{-1}\langle\bar{\psi}\psi\rangle\,
\Bigg(\frac{\mu}{M}\Bigg)^{\beta^2/4\pi}.
\end{eqnarray}
The cut--off $\mu$ has been introduced by Coleman \cite{[3]} in order
to regularized the two--point Green function of the $\vartheta$--field
in the infrared region. Therefore, a regularized correlation function
should be obtained in the limit $\mu \to 0$.  Setting $\mu = 0$ we get
\begin{eqnarray}\label{label8.26}
\langle 0|\bar{\psi}(x)\psi(x)|0\rangle_{\rm R} = 0.
\end{eqnarray}
We would like to emphasize that in our approach the vanishing of the
fermion condensate (\ref{label8.26}) is not due to the triviality of
the vacuum. Our vacuum is essentially different from the vacua which
were used in \cite{[3],[5]}--\cite{[16]} as we have explained in
Sections 4--6.

As has been stated by Itzykson and Zuber \cite{[27]} such a
vanishing of the correlation function $\langle
0|\bar{\psi}(x)\psi(x)|0\rangle$ is caused by the sorrowful fact that
{\it a massless scalar field theory is undefined in a 1+1--dimensional
world due to severe infrared divergences.} \cite{[27]}. The former
leads to a randomization of the $\vartheta$--field in the infrared
region that averages $\langle 0|[A_-(x) + A_+(x)]|0\rangle$ to zero
\cite{[27], [28]}.

Since the problem of the vanishing of the correlation function
(\ref{label8.26}) is related to full extent to the definition of the
massless scalar field in 1+1--dimensional space--time, we suggest to
regularize the correlation function (\ref{label8.25}) within 
dimensional regularization. In dimensional regularization the
two--point Green function of the $\vartheta$--field described by the
effective Lagrangian (\ref{label8.20}) is defined by
\begin{eqnarray}\label{label8.27}
\hspace{-0.3in}&&i\Delta(x-y)  =  \int
\frac{d^2p}{(2\pi)^2i}\,\frac{1}{p^2 + i0}\,e^{\textstyle
-ip\cdot(x-y)}\to \nonumber\\
\hspace{-0.3in}&&\to \int
\frac{d^dp}{(2\pi)^di}\,\frac{\lambda^{2-d}}{p^2 + i0}\,e^{\textstyle
-ip\cdot(x-y)} =
-\frac{1}{4\pi^{d/2}}\,[-\lambda^2(x-y)^2\,]^{\textstyle
\frac{2-d}{2}}\,\Gamma\Big(\frac{d-2}{2}\Big),
\end{eqnarray}
where $\lambda$ is a dimensional parameter making the Green function
dimensionless. We fix this parameter below. In order to obtain the
regularized value $i\Delta(0)_{\rm R}$, we keep $d = 2-\varepsilon$
and set $x-y = 0$. This yields $i\Delta(0)_{\rm R} = 0$.

The same result can be obtained within analytical regularization
\begin{eqnarray}\label{label8.28}
\hspace{-0.7in}&&i\Delta(x-y) = \int
\frac{d^2p}{(2\pi)^2i}\,\frac{1}{p^2 + i0}\,e^{\textstyle
-ip\cdot(x-y)}\to \nonumber\\
\hspace{-0.7in}&&\to -\int
\frac{d^2p}{(2\pi)^2i}\,\frac{\lambda^{2\alpha -2 }}{( -p^2 +
i0)^{\alpha}}\,e^{\textstyle -ip\cdot(x-y)} = -
\frac{1}{4^{\alpha}\pi}\,[-\lambda^2(x-y)^2\,]^{\textstyle \alpha
-1}\,\frac{\Gamma(1-\alpha)}{\Gamma(\alpha)}.
\end{eqnarray}
Keeping $\alpha = 1 + \varepsilon/2$ at $\varepsilon \to +0$ we get
again $i\Delta(0)_{\rm R} = 0$.

Using the regularized Green function $i\Delta(0)_{\rm R} = 0$ the vacuum
expectation value of the fermion condensate (\ref{label8.25}) is equal
to
\begin{eqnarray}\label{label8.29}
\hspace{-0.3in}\langle 0|\bar{\psi}(x)\psi(x)|0\rangle_{\rm R}
=Z^{-1}\langle\bar{\psi}\psi\rangle\,e^{\textstyle
\frac{1}{2}\beta^2\,i\Delta(0)_{\rm R}}
=Z^{-1}\langle\bar{\psi}\psi\rangle.
\end{eqnarray}
Since there are no divergences we should set $Z=1$. This gives the
fermion condensate
\begin{eqnarray}\label{label8.30}
\hspace{-0.3in}\langle 0|\bar{\psi}(x)\psi(x)|0\rangle_{\rm R} =
\langle\bar{\psi}\psi\rangle,
\end{eqnarray}
which is in complete agreement with our result obtain within the BCS
formalism.

Thus, we have shown that the problem of the fermion condensate,
averaged to zero by the $\vartheta$--field fluctuations, can be
avoided by using dimensional or analytical regularization.

The solution of the massless Thirring model in the sense of an
explicit evaluation of any correlation function 
\begin{eqnarray}\label{label8.31}
\Big\langle 0\Big|{\rm
T}\Big(\prod^{p}_{i=1}\prod^{n}_{j=1}\sigma_+(x_i)\sigma_-(y_j)\Big)
\Big|0\Big\rangle
\end{eqnarray}
runs as follows. Using the Abelian bosonization rules
(\ref{label4.17}) the fermion correlation function (\ref{label8.31})
reduces to the $\vartheta$--field correlation function
\begin{eqnarray}\label{label8.32}
\hspace{-0.3in}&&\Big\langle 0\Big|{\rm
T}\Big(\prod^{p}_{i=1}\prod^{n}_{j=1}\sigma_+(x_i)
\sigma_-(y_j)\Big)\Big|0\Big\rangle = \frac{\langle
\bar{\psi}\psi\rangle^{p+n}}{(2Z)^{p+n}} \Big\langle 0\Big|{\rm
T}\Big(\prod^{p}_{i=1}\prod^{n}_{j=1}A_-(x_i)
A_+(y_j)\Big)\Big|0\Big\rangle=\frac{\langle
\bar{\psi}\psi\rangle^{p+n}}{(2Z)^{p+n}}\nonumber\\
\hspace{-0.3in}&&\times\int {\cal D}\vartheta\,e^{\textstyle
-i\,\frac{1}{2}\int
d^2x\,\vartheta(x)\,\Box\,\vartheta(x)}A_-(x_i)A_+(y_j) =\frac{\langle
\bar{\psi}\psi\rangle^{p+n}}{(2Z)^{p+n}} \exp\Big\{
\frac{1}{2}\,\beta^2\,(p+n)\,i\Delta(0)\Big\}\nonumber\\
\hspace{-0.3in}&&\times\,\exp\Big\{ \beta^2\sum^{p}_{j <k}i\Delta(x_j
- x_k) + \beta^2\sum^{n}_{j <k}i\Delta(y_j - y_k) - \beta^2\sum^{p}_{j
= 1}\sum^{n}_{k = 1}i\Delta(x_j - y_k)\Big\}=\nonumber\\
\hspace{-0.3in}&&=\frac{\langle
\bar{\psi}\psi\rangle^{p+n}}{(2Z)^{p+n}}\exp\Big\{ \beta^2\sum^{p}_{j
<k}i\Delta(x_j - x_k) + \beta^2\sum^{n}_{j <k}i\Delta(y_j - y_k) -
\beta^2\sum^{p}_{j = 1}\sum^{n}_{k = 1}i\Delta(x_j - y_k)\Big\},
\nonumber\\
\hspace{-0.3in}&&
\end{eqnarray}
where we have used $i\Delta(0)_{\rm R}=0$ obtained within dimensional
and analytical regularization. Assuming now that all relative
distances do not vanish, we should take the limits $\varepsilon \to +0$
and take away the dimensional or analytical regularization of the
two--point Green functions. In the limit $\varepsilon \to +0$ the
Green function $i\Delta(x-y)$ reads
\begin{eqnarray}\label{label8.33}
i\Delta(x-y) = \frac{1}{2\pi\varepsilon} + \frac{1}{4\pi}\,{\ell
n}[-\lambda^2(x-y)^2].
\end{eqnarray}
Absorbing the divergent parts of the Green functions in the constant
$Z^{p+n}$ we obtain the regularized correlation function
\begin{eqnarray}\label{label8.34}
\hspace{-0.35in}&&\Big\langle 0\Big|{\rm
T}\Big(\prod^{p}_{i=1}\prod^{n}_{j=1}\sigma_+(x_i)
\sigma_-(y_j)\Big)\Big|0\Big\rangle =\frac{\langle
\bar{\psi}\psi\rangle^{p+n}}{2^{p+n}} \frac{\displaystyle \prod^{p}_{j
<k}[-\lambda^2(x_j - x_k)^2]^{\beta^2/4\pi}\prod^{n}_{j
<k}[-\lambda^2(y_j - y_k)^2]^{\beta^2/4\pi}}{\displaystyle\prod^{p}_{j
= 1}\prod^{n}_{k = 1}[-\lambda^2(x_j - y_k)^2]^{\beta^2/4\pi}}
\nonumber\\
\hspace{-0.35in}&&=\frac{\langle
\bar{\psi}\psi\rangle^{p+n}}{2^{p+n}}\lambda^{\textstyle
\frac{\beta^2}{4\pi}[(p-n)^2-(p+n)]}\frac{\displaystyle \prod^{p}_{j
<k}[-(x_j - x_k)^2]^{\beta^2/4\pi}\prod^{n}_{j <k}[-(y_j -
y_k)^2]^{\beta^2/4\pi}}{\displaystyle\prod^{p}_{j = 1}\prod^{n}_{k =
1}[-(x_j - y_k)^2]^{\beta^2/4\pi}}.
\end{eqnarray}
In order to fix a parameter $\lambda$ we suggest to compare our
expression for the correlation function of self--coupled fermion
fields (\ref{label8.34}) with the correlation function of free fermion
fields calculated by Klaiber \cite{[5]} for $p=n$ and space--like
distances:
\begin{eqnarray}\label{label8.35}
\hspace{-0.3in}\Big\langle 0\Big|{\rm
T}\Big(\prod^{n}_{i=1}\sigma_+(x_i)
\sigma_-(y_i)\Big)\Big|0\Big\rangle = \frac{1}{(2\pi)^{2n}}\,
\frac{\displaystyle \prod^{n}_{j <k}[-(x_j - x_k)^2]\prod^{n}_{j
<k}[-(y_j - y_k)^2]}{\displaystyle\prod^{n}_{j = 1}\prod^{n}_{k =
1}[-(x_j - y_k)^2]}.
\end{eqnarray}
Taking the {\it mathematical} limit $\beta^2/4\pi \to 1$ and setting
$p=n$ we obtain from the comparison of (\ref{label8.34}) and
(\ref{label8.35}) that $\lambda =\pm\,\pi\,\langle
\bar{\psi}\psi\rangle$. It is reasonable to choose $\lambda$ positive,
$\lambda = - \pi\,\langle \bar{\psi}\psi\rangle$. Using this
expression for $\lambda$ we recast the correlation function
(\ref{label8.34}) into the form
\begin{eqnarray}\label{label8.36}
\hspace{-0.35in}&&\Big\langle 0\Big|{\rm
T}\Big(\prod^{p}_{i=1}\prod^{n}_{j=1}\sigma_+(x_i)
\sigma_-(y_j)\Big)\Big|0\Big\rangle =\frac{\langle
\bar{\psi}\psi\rangle^{p+n}}{2^{p+n}}[-\pi\langle
\bar{\psi}\psi\rangle]^{\textstyle
\frac{\beta^2}{4\pi}[(p-n)^2-(p+n)]}\nonumber\\
\hspace{-0.35in}&&\times\,\frac{\displaystyle \prod^{p}_{j <k}[-(x_j -
x_k)^2]^{\beta^2/4\pi}\prod^{n}_{j <k}[-(y_j -
y_k)^2]^{\beta^2/4\pi}}{\displaystyle\prod^{p}_{j = 1}\prod^{n}_{k =
1}[-(x_j - y_k)^2]^{\beta^2/4\pi}}.
\end{eqnarray}
Thus, dimensional (or analytical) regularization of the theory for the
massless scalar $\vartheta$--field leads to the expressions for
correlation functions which agree fully with the BCS formalism.  We
should emphasize that the r.h.s. of (\ref{label8.34}) does not vanish
even if $p\not= n$. Recall, that the vacuum expectation value
(\ref{label8.30}) calculated for the trivial chiral invariant vacuum
vanishes for $p \not= n$ \cite{[3],[5]}. The explicit evaluation of
the correlation function (\ref{label8.31}) in the form
(\ref{label8.36}) implies the solution of the massless Thirring model
in our approach.

The obtained results show that a massless scalar field theory in
1+1--dimensional space--time is ill--defined in agreement with
Coleman's statements. Therefore, in the infrared region there are no
single--particle Goldstone states \cite{[26]}. The Goldstone bosons
being the quanta of the $\vartheta$--field exist in the infrared
region in the form of randomized ensemble. The fermion condensate,
averaged over the $\vartheta$--field, vanishes due to the contribution
of the randomized ensemble of infrared Goldstone bosons. Since it is
fully a dimensional problem the application of dimensional (or
analytical) regularization allows to escape the problem of the
randomization of low--frequency quanta of the massless scalar field
$\vartheta$ and calculate the non--vanishing value of the fermion
condensate averaged over the $\vartheta$--field fluctuations in
agreement with the result obtained within the BCS formalism
(\ref{labelE.13}) and (\ref{label1.14}).

\section{Conclusion}
\setcounter{equation}{0}

\hspace{0.2in} We investigated the problem of the solution of the
massless Thirring model and the equivalence between the massive
Thirring model and the SG model in the chirally broken phase of the
fermion system. We found that the fermion system described by the
massless Thirring model, invariant under the chiral group $U_{\rm
V}(1)\times U_{\rm A}(1)$, possesses a chiral symmetric phase with a
trivial perturbative vacuum and a phase of spontaneously broken chiral
symmetry with a non--perturbative vacuum. We have shown that the
ground state of the massless Thirring model in the chirally broken
phase coincides with the ground of the BCS theory of
superconductivity. Using the wave--function of the ground state in the
BCS theory of superconductivity we have calculated the energy density
of the non--perturbative vacuum ${\cal E}(M)$ in the massless Thirring
model.  We have shown that the energy density of the non--perturbative
vacuum ${\cal E}(M)$ coincides with the effective potential $V[M]$
(\ref{label3.21}), which is defined by the integration over fermion
field fluctuations, and acquires a minimum when the dynamical mass $M$
of fermions satisfies the gap--equation (\ref{label1.14}). The mass
spectrum of vacuum fluctuations of fermions is restricted from
below. In fact, when $M$ is kept constant at $\Lambda \to \infty$ the
energy density tends to the limit ${\cal E}(M) = V[M] \to -M^2/4\pi$.

The chiral symmetric phase corresponds to a system with massless
fermions, $M = 0$, and vanishing fermion condensate. The chirally
broken phase is characterized by (i) a non--zero value of a fermion
condensate, (ii) the appearance of dynamical fermions with a dynamical
mass $M\not= 0$ and (iii) fermion--antifermion pairing
\cite{[18]}. The energy density of the ground state of the fermion
system ${\cal E}(M)$ reaches a maximum, ${\cal E}(0)= 0$, in the
chiral symmetric phase and it is negative, ${\cal E}(M) < 0$, in the
chirally broken one. Hence, the chirally broken phase is energetically
preferable and the Thirring model should be bosonized in the chirally
broken phase accompanied by fermion--antifermion pairing.

Using the path integral technique we bosonized explicitly the massless
Thirring model. We have shown that in the bosonic description the
massless Thirring model is a quantum field theory of a massless free
scalar field $\vartheta(x)$. The generating functional of Green
functions in the massless Thirring model can be expressed in terms of
a path integral over the massless scalar field $\vartheta(x)$ coupled
to external sources of fermion fields via $A_{\pm}(x) = e^{\textstyle
\pm\,i\,\beta\,\vartheta(x)}$ couplings.  This allows to represent any
Green function in the massless Thirring model in the fermionic
description by a Gaussian path integral of products of $A_{\pm}(x) =
e^{\textstyle \pm\,i\,\beta\,\vartheta(x)}$ couplings in the bosonic
formulation.  Since these Gaussian path integrals can be calculated
explicitly, this provides a solution of the massless Thirring model.

The evaluation of correlation functions of massless Thirring fermions
by means of the integration over massless $\vartheta$--field
fluctuations is related to the Mermin--Wagner theorem \cite{[25]},
Hohenberg's \cite{[37]} and Coleman's \cite{[26]} proofs concerning
the vanishing of long--range order for systems with a continuous
symmetry described by quantum field theories in 2--dimensional space
\cite{[25],[38]} and 1+1--dimensional space--time \cite{[26]}. The
vanishing of the long--range order parameter implies that there is no
spontaneously broken continuous symmetry in quantum field theories
defined in 2--dimensional space and a 1+1--dimensional space--time.

Coleman's proof of this statement has been focused upon the
impossibility to define a free massless scalar field theory in a
1+1--dimensional space--time. Coleman found that a free massless
scalar field theory is ill--defined due to meaningless infrared
divergences that screen fully one--particle Goldstone boson
states. This screening of a pole--singularity in the Fourier transform
of the two--point Wightman function is formulated by Coleman as the
absence of the Goldstone bosons in a free massless scalar field theory
in a 1+1--dimensional space--time. This statement has been extended by
Coleman onto any quantum system with a continuous symmetry embedded in
a 1+1--dimensional space--time \cite{[26]}.

The relation of Coleman's statement to the Mermin--Wagner--Hohenberg
theorem \cite{[25],[38]} runs in the way explained, for example, in
\cite{[27],[28]}. In fact, the infrared divergences of a free massless
scalar fields lead to the appearance of a randomized ensemble of very
low--frequency quanta of a massless scalar field. Due to this
randomization the fermion condensate, proportional to $\cos\beta
\vartheta(x)$, averages over the $\vartheta$--field fluctuations to
zero.

Using Itzykson--Zuber's analysis of a free massless scalar field
theory \cite{[40]} and adjusting it to 1+1--dimensional space--time we
have shown that (i) a continuous symmetry related to global scalar
field translations is spontaneously broken, (ii) the vacuum wave
function is not invariant under symmetry transformations and the
vacuum energy level is infinitely degenerated. Following Itzykson and
Zuber \cite{[40]} we argue that Goldstone bosons appear as quanta of a
free massless scalar field.

Accepting this point that a continuous symmetry of a quantum field
theory of a free massless scalar field can be spontaneously broken, we
have suggested that the problem of the vanishing of the fermion
condensate in the massless Thirring model, averaged over the
randomized ensemble of low--frequency quanta of the $\theta$--field,
can be solved within an appropriate regularization. We have applied
dimensional and analytical regularizations. By virtue of these
regularization procedures we have succeeded in smoothing the infrared
behaviour of the $\vartheta$--field and get the fermion condensate
averaged over the $\vartheta$--field fluctuations to a non--zero
value, in complete agreement with our results obtained within the
Nambu--Jona--Lasinio prescription (\ref{label1.13})--(\ref{label1.16})
and the BCS formalism (\ref{labelE.13}).

The bosonization of the massive Thirring model runs parallel the
bosonization of the massless one. Starting with the fermion system in
the phase of spontaneously broken chiral symmetry we arrive at the
bosonized version described by the SG model. The parameters of the SG
model can be expressed in terms of the parameters of the massive
Thirring model and read
\begin{eqnarray*}
\alpha = - \beta^2\,m\,\langle \bar{\psi}\psi\rangle +
\frac{m^2}{g}\,\beta^2,
\end{eqnarray*}
where the fermion condensate is defined by Eq.(\ref{label1.16}) and
the coupling constant $\beta$ depends on $g$ via relation
(\ref{label4.5})
\begin{eqnarray*}
\frac{8\pi}{\beta^2} = 1 - e^{\textstyle
-2\pi/g}.
\end{eqnarray*}
The new relation between $\beta$ and $g$ leads to the fact that in our
approach the coupling constant $\beta^2$ is always greater than
$8\pi$, $\beta^2 > 8\pi$. This disagreement with Coleman \cite{[3]} is
caused by different initial conditions for the evolution of the
fermion system described by the Thirring model.  In fact, when the
fermion system evolves in the chiral symmetric phase Coleman's
relation between $\beta$ and $g$ is valid. In turn, if the fermion
system starts with the chirally broken phase the bosonized version of
the fermion system is described by the SG model with the relation
between $\beta$ and $g$ given by Eq.(\ref{label4.5}) and $\alpha = -
\beta^2\,m\,\langle \bar{\psi}\psi\rangle + m^2\beta^2/g$.

The evaluation of correlation functions in the massive Thirring model
in terms of the path integral over the SG model field can be carried
by the Abelian bosonization rules given by Eq.(\ref{label5.21})
\begin{eqnarray*}
Z\,m\,\bar{\psi}(x)\Bigg(\frac{1\mp \gamma^5}{2}\Bigg)\psi(x) =
-\frac{\alpha}{2\beta^2}\,e^{\textstyle \pm i \beta\vartheta(x)} +
\frac{m^2}{2g}.
\end{eqnarray*}
At leading order in the $m$ expansion this expression reduces to the
Abelian bosonization rules derived by Coleman (\ref{label1.10})
\cite{[3]}.

The existence of the chirally broken phase in the massless Thirring
model we have also confirmed within the standard operator
formalism. We have shown that starting with the chiral invariant
Lagrangian and normal ordering the fermion operators in the
interaction term at the scale $M$ we arrive at the Lagrangian of the
massive Thirring model for fermions with mass $M$ only if the
gap--equation (\ref{label1.13}) is fulfilled. Using the equations of
motion for the fermion fields in the massless Thirring model we have
shown that the chirally broken phase is stable during the evolution of
the fermion system when it started to evolve from the chirally broken
phase.

The stability of the chirally broken phase with the non--perturbative
vacuum could in principle be destroyed by the contribution of the
fluctuations of the $\rho$--field around the minimum of the effective
potential (\ref{label3.21}), the $\tilde{\rho}$--field
fluctuations\,\footnote{This question has been raised by Valerii
Rubakov.}. In Appendix B we have shown that the $\tilde{\rho}$--field,
rescaled in an appropriate way in order to get the correct kinetic
term, acquires a mass proportional the cut--off $\Lambda$ and in the
limit $\Lambda \to \infty$ becomes fully decoupled from the
system. This result agrees completely with the Appelquist--Carazzone
decoupling theorem \cite{[41]}. This testifies that the chirally
broken phase with the non--perturbative vacuum cannot be ruined by
fluctuations around the minimum of the effective potential and is
fully determined by the effective potential (\ref{label3.21}).

We have revealed that the existence of the chirally broken phase in
the massless Thirring model changes crucially the Schwinger term in
the equal--time commutation relation $[j_0(x,t), j_1(y,t)]$. We have
shown that the Schwinger term calculated for the non--perturbative
vacuum in the chirally broken phase depends explicitly on the coupling
constant $g$. For the chiral symmetric phase and the trivial vacuum
the Schwinger term is equal to the value calculated previously by
Sommerfield \cite{[34]}. In the limit $g \to 0$ our value of the
Schwinger term reduces to that obtained by Sommerfield.

Now let us clarify the physical meaning of the inequality $\beta^2 >
8\pi$ obtained in our approach. For this aim we suggest to rescale the
$\vartheta$--field, $\beta \vartheta(x) \to \vartheta(x)$. Then in
natural units $\hbar = c = 1$ the action $S$ reads
\begin{eqnarray}\label{label9.1}
S = \frac{1}{\beta^2} \int
d^2x\,\Big[\,\frac{1}{2}\,\partial_{\mu}\vartheta(x)
\partial^{\mu}\vartheta(x)
+ \alpha \,(\cos\vartheta(x) - 1)\Big]
\end{eqnarray}
with $0 \le \vartheta(x) \le 2\pi$. This allows to interpret $\beta^2$
in the sense of ``$\hbar$'' distinguishing ``quantum'' and
``classical'' states of the SG model. In the classical limit $\beta^2
\to 0$ we arrive at a system of classical Klein--Gordon--waves and
solitons.  The action $S$
\begin{eqnarray}\label{label9.2}
S = \frac{1}{\beta^2}\int
d^2x\,\Big[\,\frac{1}{2}\,\partial_{\mu}\vartheta(x)
\partial^{\mu}\vartheta(x) + \beta^2 \Big(- m\,\langle
\bar{\psi}\psi\rangle + \frac{m^2}{g}\,\Big)\,(\cos \vartheta(x) - 1)\Big],
\end{eqnarray}
where for simplicity we have kept the leading terms in the $m$
expansion, describes in the $\beta^2 \to 0$ limit a theory of massless
classical $\vartheta$--waves
\begin{eqnarray}\label{label9.3}
\vartheta(x,t) = \vartheta_-(t - x) + \vartheta_+(t + x)
\end{eqnarray}
obeying
\begin{eqnarray}\label{label9.4}
\Box\,\vartheta(x,t) = \Bigg(\frac{\partial^2}{\partial t^2} -
\frac{\partial^2}{\partial x^2}\Bigg)\vartheta(x,t) = 0,
\end{eqnarray}
with arbitrary functions $\vartheta_-(t - x)$ and $\vartheta_+(t +
x)$.  The mass of Goldstone bosons caused completely by quantum
effects, $M_{\textstyle \vartheta} = (-m\,\beta^2\langle
\bar{\psi}\psi\rangle)^{1/2}$, vanishes in the limit $\beta^2 \to
0$. In turn, the soliton mass tends to infinity, $M_{\textstyle \rm
sol} \propto 1/\beta \to \infty$, and solitons decouple from the
system.

For $\beta^2 > 8$ the
mass of the Goldstone boson becomes greater than the mass of a single
soliton:
\begin{eqnarray}\label{label9.5}
\frac{M_{\textstyle \rm sol}}{M_{\textstyle \vartheta}} =
\frac{8}{\beta^2} < 1.
\end{eqnarray}
This implies that in the ``quantum limit'', $\beta^2 \gg 1$, the
creation of non--perturbative soliton configurations is energetically
preferable with respect to the creation of Goldstone bosons. This
yields that at $\beta^2 > 8\pi$, when $M_{\textstyle \vartheta} \gg
M_{\textstyle \rm sol}$, the Goldstone bosons are decoupled from the
system and there exist practically only solitons.  Hence, the
inequality $\beta^2 > 8\pi$ corresponds to the non--perturbative phase
of the SG model populated by soliton states only.

We have shown that the topological current of the SG model coincides
with the Noether current of the massive Thirring model related to the
$U_{\rm V}(1)$ invariance. Since this Noether current is responsible
for the conservation of the fermion number in the massive Thirring
model, the topological charge of the SG model has the meaning of the
fermion number.  Since many--soliton solutions obey Pauli's exclusion
principle, this should prove Skyrme's statement \cite{[4]} that the SG
model solitons can be interpreted as massive fermions. Thus, via
spontaneously broken chiral symmetry the massive Thirring fermions get
converted into extended particles with the properties of fermions and
masses much heavier than their initial mass
\begin{eqnarray*}
M^2_{\textstyle \rm sol} =- \frac{64}{\beta^2}\,m\,\langle \bar{\psi}\psi\rangle
+ O(m^2) \gg m^2.
\end{eqnarray*}
Finally, we would like to mention that recently \cite{[42]} one of
the authors suggested a generalization of the sine--Gordon model to
3+1--dimensions. This model has also stable solitonic excitations
characterized by a winding number defining a chirality for
fermions. The results obtained in the present paper can be of use for
the derivation of the model suggested in Ref.\cite{[42]} as a
bosonized version of the 3+1--dimensional NJL model with chiral
$SU(2)\times SU(2)$.

Numerous applications to hadron physics of the chiral soliton model
based on the linear $\sigma$--model of Gell--Mann and Levy, the
extended linear $\sigma$--model and the Nambu--Jona--Lasinio quark
model with $SU(2)\times SU(2)$ and $SU(3)\times SU(3)$ chiral symmetry
one can find in papers written by Klaus G\"oke with co--workers
starting with 1985 \cite{[43]}.

\section*{Acknowledgement}

\hspace{0.2in} The authors are greatful to Walter Thirring, Wolfgang
Kummer and Valerii Rubakov for fruitful discussions. We thank Jan
Thomassen for critical comments on our approach at the initial stage
of the work. Numerous and fruitful discussions with Max Meinhardt are
greatly appreciated.

This work was supported in part by Fonds zur F\"orderung der
Wissenschaftlichen Forschung P13997-TPH\,\footnote{After this paper
has been completed we have been known about the paper by Vigman and
Larkin [38]. In a 1+1--dimensional chiral invariant model with
four--fermion interactions Vigman and Larkin investigated the problem
of the appearance of a fermion mass. Analysing the infrared asymptotic
behaviour of the one--particle Green function in the approximation of
a large number of fermion fields Vigman and Larkin showed that
fermions become massive as a result of four--fermion interactions. The
vanishing of the fermion condensate has been declared as the absence
of {\it spontaneous symmetry breakdown}.}.

\newpage

\section*{Appendix A. Chiral Jacobian}

\hspace{0.2in} In this Appendix we adduce the calculation of the
Jacobian induced by chiral rotations in Eq.(\ref{label3.30}). We show
that by using an appropriate regularization scheme this Jacobian can
be obtained to be equal to unity. 

We follow the procedure formulated in
Refs.\cite{[12]}--\cite{[17]}. For the calculation of the chiral
Jacobian we start with the Lagrangian implicitly defined in
Eqs.(\ref{label3.29}) and (\ref{label3.30})
$$
{\cal L}_{\psi}(x) = \bar{\psi}(x)\Big(i\gamma^{\mu}\partial_{\mu} -
M\,e^{\textstyle i\gamma^5\vartheta(x)}\Big)\psi(x) =
\bar{\psi}(x)D(x;0)\psi(x),\eqno({\rm A}.1)
$$
where $D(x;0)$ is the Dirac operator given by 
$$
D(x;0) = i\gamma^{\mu}\partial_{\mu} -
M\,e^{\textstyle i\gamma^5\vartheta(x)}.\eqno({\rm A}.2)
$$
By  a chiral rotation 
$$
\psi(x) = e^{\textstyle - i\alpha \gamma^5\vartheta(x)/2}\,\chi(x),
$$
$$
\bar{\psi}(x) = \bar{\chi}(x)\,e^{\textstyle - i\alpha
\gamma^5\vartheta(x)/2},\eqno({\rm A}.3)
$$
where $0\le \alpha \le 1$, we reduce the Lagrangian $({\rm A}.1)$ to
the form
$$
{\cal L}_{\chi}(x) = \bar{\chi}(x\,)D(x,\alpha)\,\chi(x).\eqno({\rm A}.4)
$$
The Dirac operator $D(x;\alpha)$ reads
$$
D(x;\alpha) = i\gamma^{\mu}\partial_{\mu} +
\frac{1}{2}\,\alpha\,\gamma^{\mu}\gamma^5\,\partial_{\mu}\vartheta(x) -
M\,e^{\textstyle i(1-\alpha)\gamma^5\vartheta(x)}.\eqno({\rm A}.5)
$$
At $\alpha = 1$ we obtain the Lagrangian
$$
{\cal L}_{\chi}(x) = \bar{\chi}(x)\,D(x,1)\,\chi(x) =
\bar{\chi}(x)\,(i\gamma^{\mu}\partial_{\mu} +
\frac{1}{2}\,\gamma^{\mu}\gamma^5\,\partial_{\mu}\vartheta(x) -
M)\,\chi(x),\eqno({\rm A}.6)
$$
where the term $-M\,\bar{\chi}(x)\chi(x)$ has the meaning of a
mass term of the $\chi(x)$ field (see Eq.(\ref{label3.30})).

Due to the chiral rotation (\ref{label3.30}) the fermionic  measure
changes as follows
$$
{\cal D}\psi\,{\cal D}\bar{\psi} = J[\vartheta]\,{\cal D}\chi\,{\cal
D}\bar{\chi},\eqno({\rm A}.7)
$$
For the calculation $J[\vartheta]$ we follow Fujikawa's procedure
\cite{[12]}--\cite{[17]} and introduce eigenfunctions $\varphi_n(x;\alpha)$ and
eigenvalues $\lambda_n(\alpha)$ of the Dirac operator $D(x;\alpha)$:
$$
D(x;\alpha)\,\varphi_n(x;\alpha) =
\lambda_n(\alpha)\,\varphi_n(x;\alpha).\eqno({\rm A}.8)
$$
In terms of the eigenfunctions and eigenvalues of the Dirac operator
the Jacobian $J[\vartheta]$ is defined by \cite{[13],[14]}
$$
J[\vartheta] =
\exp\,2\,i\!\!\int^1_0d\alpha\,w[\vartheta;\alpha],\eqno({\rm A}.9)
$$
where the functional $w[\vartheta;\alpha]$ is given by \cite{[13],[14]}
$$
w[\vartheta;\alpha] = \lim_{\textstyle \Lambda_{\rm F} \to
\infty}\sum_n\varphi^{\dagger}_n(x;\alpha)\,\frac{1}{2}\,
\gamma^5\,\vartheta(x)\,e^{\textstyle i\lambda^2_n/\Lambda^2_{\rm F}}
\varphi_n(x;\alpha)=
$$
$$
=\lim_{\textstyle \Lambda_{\rm F} \to \infty} \frac{1}{2}\,\int
d^2x\,\vartheta(x)\int\frac{d^2k}{(2\pi)^2}\,{\rm
tr}\Big\{\gamma^5\Big\langle x\Big|e^{\textstyle
iD^2(x;\alpha)/\Lambda^2_{\rm F}}\Big|x\Big\rangle\Big\}.
\eqno({\rm A}.10)
$$
For the calculation of the matrix element $\langle x|\ldots|x\rangle$
we use plane waves \cite{[12]}--\cite{[17]} and get
$$
\Big\langle x\Big|e^{\textstyle
iD^2(x;\alpha)/\Lambda^2_{\rm F}}\Big|x\Big\rangle =
$$
$$
= \exp\Big\{\frac{i}{\Lambda^2_{\rm F}}\,[k^2 -
2\,M\,\cos((1-\alpha)\vartheta(x))\,\gamma^{\mu}k_{\mu} +
\frac{1}{2}\,i\,\alpha\,\gamma^{\mu}\gamma^{\nu}\gamma^5\,
\partial_{\mu}\partial_{\nu}\vartheta(x)
$$
$$
+ M^2\,e^{\textstyle 2\,i\,(1-\alpha)\,\gamma^5\,\vartheta(x)} +
(1-2\alpha)\,M\,\gamma^{\mu}\gamma^5\,\partial_{\mu}\,
\vartheta(x)\,\cos((1-\alpha)\vartheta(x))
$$
$$
+ i\,(1-\alpha)\,M\,\gamma^{\mu}\partial_{\mu}\vartheta(x)\,
\sin((1-\alpha)\vartheta(x)) - \frac{1}{4}\,\alpha^2\,
\partial_{\mu}\vartheta(x)\,\partial^{\mu}\vartheta(x)]\Big\}.
\eqno({\rm A}.11)
$$
Substituting $({\rm A}.11)$ in $({\rm A}.10)$ we obtain
$$
\lim_{\textstyle \Lambda_{\rm F} \to \infty}
\int\frac{d^2k}{(2\pi)^2}\,{\rm tr}\Big\{\gamma^5\Big\langle
x\Big|e^{\textstyle iD^2(x;\alpha)/\Lambda^2_{\rm
F}}\Big|x\Big\rangle\Big\} =
$$
$$
= \frac{1}{4\pi}\,{\rm tr}\{- \frac{1}{2}\,\alpha\,
\gamma^{\mu}\gamma^{\nu}\partial_{\mu}
\partial_{\nu}\vartheta(x) - M^2\,\sin(2(1-\alpha)\vartheta(x))\}=
$$
$$
=\frac{1}{4\pi}\,[- \alpha\,\partial^{\mu}\partial_{\mu}\vartheta(x) -
2\,M^2\,\sin(2(1-\alpha)\vartheta(x))].\eqno({\rm A}.12)
$$
The functional $w[\vartheta,\alpha]$ is given then by
$$
w[\vartheta,\alpha] = \frac{1}{8\pi}\int d^2x\,\vartheta(x)\,
[- \alpha\,\partial^{\mu}\partial_{\mu}\vartheta(x) -
2\,M^2\,\sin(2(1-\alpha)\vartheta(x))].\eqno({\rm A}.13)
$$
Inserting $w[\vartheta,\alpha]$ into $({\rm A}.9)$ and integrating over
$\alpha$ we get the Jacobian
$$
J[\vartheta] = \exp\,i\,\frac{1}{4\pi}\!\int d^2x\,\Big[\frac{1}{2}\,
\partial^{\mu}\vartheta(x)\partial_{\mu}\vartheta(x) 
+ M^2\,(\cos2\vartheta(x) - 1)\Big].\eqno({\rm A}.14)
$$
For the derivation of the first term we have integrated by parts and
dropped the surface contributions.

Our result agrees well with that obtained by Dorn for the massive
Schwinger model \cite{[16]}. However, Dorn pointed out that the term
proportional to $M^2$ is {\it renormalization--scheme dependent}
and it is {\it unambiguously defined if one insists on vector gauge
invariance} \cite{[13],[15]}.

Since the Thirring model is not vector gauge invariant, the term
proportional to $M^2$ may not be well defined and may, in
principle, be removed by an appropriate regularization. First, let us
give physical reasons for the absence of this term. Indeed, the term
in Eq.(${\rm A}.14$) proportional to $M^2$ does not depend on the
derivatives $\partial_{\mu}\vartheta(x)$. Therefore, it contributes to
the effective potential. However, we have calculated the effective
potential explicitly in Eqs.(\ref{label3.5}) -- (\ref{label3.21}) and
have shown that it does not depend on the
$\vartheta$--field. Therefore, one can conclude that the determinant
(\ref{label3.29})
$$
{\rm Det}\Big(i\gamma^{\mu}\partial_{\mu} -
M\,e^{\textstyle i\,\gamma^5\,\vartheta}\,\Big)\eqno({\rm A}.15)
$$
is invariant under global rotations, $\vartheta(x)\to \vartheta(x) +
\theta$, where $\theta$ is an arbitrary constant. On the other hand,
after the chiral rotation (\ref{label3.30})
$$
{\rm Det}\Big(i\gamma^{\mu}\partial_{\mu} - M\,e^{\textstyle
i\,\gamma^5\,\vartheta}\,\Big)= J[\vartheta]\,{\rm
Det}\Big(i\gamma^{\mu}\partial_{\mu} +
\frac{1}{2}\,\varepsilon^{\mu\nu}\gamma_{\mu}\partial_{\nu}\vartheta -
M\Big),\eqno({\rm A}.16)
$$
the determinant 
$$
{\rm Det}\Big(i\gamma^{\mu}\partial_{\mu} +
\frac{1}{2}\,\varepsilon^{\mu\nu}\gamma_{\mu}\partial_{\nu}\vartheta -
M\Big),\eqno({\rm A}.17)
$$
depending only on the gradient of the $\vartheta$--field, is also
invariant under global rotations, $\vartheta(x)\to \vartheta(x) +
\theta$. This proves that the Jacobian $J[\vartheta]$ should not
violate invariance under global rotations, $\vartheta(x)\to
\vartheta(x) + \theta$, and the presence of the term proportional to
$M^2$ in Eq.(${\rm A}.14$) is a problem of the regularization
procedure.

In order to confirm our statement we do not need to employ any other
regularization procedure, for example, analytical regularization
expounded in Ref.\cite{[44]}.  We show that exactly in our case of the
fermion field quantized in the chirally broken phase relative to the
non--perturbative vacuum the chiral Jacobian is equal to unity, even if
the Fujikawa regularization procedure is applied. This means that not
only the terms proportional to $M^2$, dependent on $\vartheta(x)$ and
violating rotational invariance, but also the terms proportional to
$\partial_{\mu}\vartheta(x)\partial^{\mu}\vartheta(x)$, invariant
under global rotations $\vartheta(x) \to \vartheta(x) + \theta$
vanish.

According to Refs.[49--51] (see also [52]) the exponent in ({\rm
A}.14) is ill--defined and depends on the regularization
procedure. Therefore, the chiral Jacobian can be written as
$$
J[\vartheta] = \exp\,i\,\frac{\alpha_J}{4\pi}\!\int
d^2x\,\Big[\frac{1}{2}\,
\partial^{\mu}\vartheta(x)\partial_{\mu}\vartheta(x) +
M^2\,(\cos2\vartheta(x) - 1)\Big],\eqno({\rm A}.18)
$$
where $\alpha_J$ is an arbitrary parameter.

The term proportional to $M^2$ gives a contribution to the effective
potential breaking its symmetry under global rotations
$\vartheta(x)\to \vartheta(x) + \theta$, where $\theta$ is an
arbitrary constant. The effective potential (\ref{label3.21}) is
evaluated explicitly without chiral rotations. It is invariant under
global rotations $\vartheta(x)\to \vartheta(x) + \theta$. Therefore,
the requirement of rotational symmetry of all contributions to the
effective potential gives $\alpha_J = 0$. This yields $J[\vartheta] =
1$ [6].

The same result can be obtained taking into account that the
integration over $k$ is nothing more that the integration over fermion
field fluctuations with 2--momenta $k$. In the chirally broken phase
fluctuations of fermion fields are restricted from above. In the
Euclidean momentum space we have $k^2_{\rm E} \le \Lambda^2$. Thereby,
restricting high--energy fluctuations of fermionic fields from above
by the cut--off $\Lambda$ one can recast the r.h.s. of ({\rm A}.12)
into the form
$$
\lim_{\textstyle \Lambda_{\rm F} \to \infty}
\int\frac{d^2k}{(2\pi)^2}\,{\rm tr}\Big\{\gamma^5\Big\langle
x\Big|e^{\textstyle iD^2(x;\alpha)/\Lambda^2_{\rm
F}}\Big|x\Big\rangle\Big\} =
$$
$$
=\lim_{\textstyle \Lambda_{\rm F} \to \infty}\Big(1 - e^{\textstyle
-i\Lambda^2/\Lambda^2_{\rm F}}\,\Big)\,\frac{1}{4\pi}\,[-
\alpha\,\partial^{\mu}\partial_{\mu}\vartheta(x) -
2\,M^2\,\sin(2(1-\alpha)\vartheta(x))] = 0.\eqno({\rm A}.19)
$$
Since the cut--off $\Lambda_{\rm F}$ is independent on $\Lambda$, the
r.h.s. of ({\rm A}.18}) vanishes in the limit $\Lambda_{\rm F} \to
\infty$.  This yields $w[\vartheta,\alpha] = 0$, that entails the unit
value for chiral Jacobian $J[\vartheta] = 1$.  

Thus, local chiral rotations in the Thirring model with fermion fields
quantized in the chirally broken phase do not produce non--trivial
chiral Jacobians.

\section*{Appendix B. Stability of chirally broken phase under 
$\tilde{\rho}$--field fluctuations}

\hspace{0.2in} Here we discuss the stability of the chirally broken
phase under $\tilde{\rho}$--field fluctuations. For this aim we
calculate the effective Lagrangian of the $\tilde{\rho}$--field and
demonstrate the decoupling of the $\tilde{\rho}$--field.

The evaluation of the effective Lagrangian of the
$\tilde{\rho}$--field runs in the following way. First, we rewrite the
functional determinant (\ref{label3.5})
$$
{\rm Det}(i\gamma^{\mu}\partial_{\mu} - \sigma - i\gamma^5\varphi) =
{\rm Det}\Big(i\gamma^{\mu}\partial_{\mu} - \rho\,e^{\textstyle
i\,\vartheta}\Big) =
$$
$$
= {\rm Det}\Big(e^{\textstyle
i\,\gamma^5\,\vartheta/2}\,(i\gamma^{\mu}\partial_{\mu} +
\gamma^{\mu}A_{\mu} - \rho)\,e^{\textstyle
i\,\gamma^5\,\vartheta/2}\,\Big) = {\rm
Det}(i\gamma^{\mu}\partial_{\mu} + \gamma^{\mu}A_{\mu} -
\rho),\eqno({\rm B}.1)
$$
where $A_{\mu}$ is given by Eq.(\ref{label3.31})
$$
A_{\mu}(x) =
\frac{1}{2}\,\varepsilon_{\mu\nu}\,\partial^{\nu}\vartheta(x).
\eqno({\rm B}.2)
$$
In Eq.({\rm B}.1) we have used the fact that the Jacobian of the
chiral rotation is equal to unity, see Eq.({\rm A}.26).

Secondly, we make a shift $\rho = M + \tilde{\rho}$ and represent the
functional determinant in the r.h.s. of Eq.({\rm B}.1) as follows 
$$
{\rm Det}(i\gamma^{\mu}\partial_{\mu} + \gamma^{\mu}A_{\mu} - \rho) =
{\rm Det}(i\gamma^{\mu}\partial_{\mu} - M + \gamma^{\mu}A_{\mu}){\rm
Det}\Big(1 - \frac{1}{i\gamma^{\mu}\partial_{\mu} - M +
\gamma^{\mu}A_{\mu}}\tilde{\rho}\Big).\eqno({\rm B}.3)
$$
The determinant ${\rm Det}(i\gamma^{\mu}\partial_{\mu} - M +
\gamma^{\mu}A_{\mu})$ describes the effective Lagrangian of the
$\vartheta$--field that is given by (\ref{label3.35}). It is
convenient to recast the determinant containing the
$\tilde{\rho}$--field into the form
$$
{\rm Det}\Big(1 - \frac{1}{i\gamma^{\mu}\partial_{\mu} - M +
\gamma^{\mu}A_{\mu}}\tilde{\rho}\Big) ={\rm Det}\Big(1 -
\frac{1}{i\gamma^{\mu}\partial_{\mu} - M}\tilde{\rho} +
\frac{1}{i\gamma^{\mu}\partial_{\mu} -
M}\gamma^{\nu}A_{\nu}\frac{1}{i\gamma^{\mu}\partial_{\mu} -
M}\tilde{\rho}
$$
$$
 -\frac{1}{i\gamma^{\mu}\partial_{\mu} -
 M}\gamma^{\alpha}A_{\alpha}\frac{1}{i\gamma^{\mu}\partial_{\mu} -
 M}\gamma^{\beta}A_{\beta}\frac{1}{i\gamma^{\mu}\partial_{\mu} -
 M}\tilde{\rho} + \ldots\Big).\eqno({\rm B}.4)
$$
It is obvious that even if the scale of the $\tilde{\rho}$--field is
of order $O(M)$, the contribution of the $\vartheta$--field should be
of order $O(\partial_{\mu}\vartheta/M)$. This implies that the
$\vartheta$--field is decoupled from the $\tilde{\rho}$--field and
Eq.({\rm B}.4). This allows to consider the approximate form for the
determinant ({\rm B}.4)
$$
{\rm Det}\Big(1 - \frac{1}{i\gamma^{\mu}\partial_{\mu} - M +
\gamma^{\mu}A_{\mu}}\tilde{\rho}\Big) \simeq {\rm Det}\Big(1 -
\frac{1}{i\gamma^{\mu}\partial_{\mu} - M}\tilde{\rho}\Big).\eqno({\rm
B}.5)
$$
The effective Lagrangian of the $\tilde{\rho}$--field is then given by
$$
{\cal L}_{\rm eff}[\tilde{\rho}(x)] = - i{\rm tr}\,\Big\langle
x\Big|{\ell n}\Big(1 - \frac{1}{i\gamma^{\mu}\partial_{\mu} -
M}\tilde{\rho}\Big)\Big|x\Big\rangle - \frac{M}{2g}\,\tilde{\rho}(x) -
\frac{1}{2g}\,\tilde{\rho}^2(x) - i\,\delta^{(2)}(0)\,{\ell
n}\,\Bigg(1 + \frac{\tilde{\rho}(x)}{M}\Bigg).\eqno({\rm B}.5)
$$
The last term comes from the measure of the path integral in the polar
representation \cite{[45]}: ${\cal D}\sigma {\cal D}\varphi = {\rm
Det}\,\rho\,{\cal D}\rho{\cal D}\vartheta = \exp\{\delta^{(2)}(0)\int
d^2x\,{\ell n}\rho(x)\}{\cal D}\rho{\cal D}\vartheta =
\exp\{\delta^{(2)}(0)\int d^2x\,{\ell n}(M + \tilde{\rho}(x))\}{\cal
D}\tilde{\rho}{\cal D}\vartheta$. This adds a contact term
$-\,i\,\delta^{(2)}(0)\,{\ell n}\,(1 + \tilde{\rho}(x)/M)$, which
serves to cancel divergences appearing from the loop contributions of
the $\tilde{\rho}$--field \cite{[45]}, \cite{[46]}.

It is obvious that the effective potential of the
$\tilde{\rho}$--field independent of gradients
$\partial_{\alpha}\tilde{\rho}(x)$ is completely defined by the
effective potential (\ref{label3.21}) and can be written as
$$
V[\tilde{\rho}(x)] = \frac{1}{4\pi}\Bigg[(M^2 + 2M\tilde{\rho}(x) +
\tilde{\rho}^2(x)){\ell n}\Big(1 + 2\frac{\tilde{\rho}(x)}{M} +
\frac{\tilde{\rho}^2(x)}{M^2}\Big)
$$
$$
 - (\Lambda ^2 + M^2 +
2M\tilde{\rho}(x) + \tilde{\rho}^2(x)){\ell n}\Big(1 +
2\frac{M\tilde{\rho}(x)}{\Lambda^2 + M^2} +
\frac{\tilde{\rho}^2(x)}{\Lambda^2 + M^2}\Big)\Bigg],\eqno({\rm B}.6)
$$
where we have used Eq.(\ref{label3.22}).

In order to understand what kind of rescaling of the
$\tilde{\rho}$--field should be carried out it is sufficient to
calculate a two--vertex diagram contribution keeping only the
contribution of the gradient $\partial_{\mu}\tilde{\rho}(x)$. It
reads
$$
{\cal L}^{(2)}_{\rm eff}[\partial_{\alpha}\tilde{\rho}(x)] =
\frac{1}{8\pi}\,\partial_{\alpha}\tilde{\rho}(x)\int\limits^1_0
d\alpha\,\Bigg[\frac{1}{M^2 + \alpha(1-\alpha)\,\Box} -
\frac{1}{\Lambda^2 + M^2 +
\alpha(1-\alpha)\,\Box}\Bigg]\,\partial^{\alpha}\tilde{\rho}(x).
\eqno({\rm B}.7)
$$
Expanding the integrand in powers of $1/M^2$ and $1/(\Lambda^2 + M^2)$
we obtain
$$
{\cal L}^{(2)}_{\rm eff}[\partial_{\alpha}\tilde{\rho}(x)] =
\frac{1}{8\pi}\,\frac{\Lambda^2}{M^2(\Lambda^2 +
M^2)}\partial_{\alpha}\tilde{\rho}(x)\partial^{\alpha}\tilde{\rho}(x)
$$
$$
+ \frac{1}{48\pi}\,\frac{\Lambda^2(\Lambda^2 + 2M^2)}{M^4(\Lambda^2 +
M^2)^2}\partial_{\alpha}\partial_{\beta}\tilde{\rho}(x)
\partial^{\alpha}\partial^{\beta}\tilde{\rho}(x) + \ldots.  \eqno({\rm
B}.8)
$$
In order to get a correct kinetic term we have to rescale the
$\tilde{\rho}$--field
$$
\tilde{\rho}(x) = \sqrt{4\pi M^2\Big(1 +
\frac{M^2}{\Lambda^2}\Big)}\,v(x).\eqno({\rm B}.9)
$$
In terms of the $v$--field the Lagrangian ({\rm B}.8) reads
$$
{\cal L}^{(2)}_{\rm eff}[\partial_{\alpha}v(x)] =
\frac{1}{2}\,\partial_{\alpha}v(x)\partial^{\alpha}v(x) +
\frac{1}{3}\,\frac{\displaystyle 1 +
\frac{\Lambda^2}{2M^2}}{\displaystyle 1 +
\frac{\Lambda^2}{2M^2}}\frac{1}{M^2}
\partial_{\alpha}\partial_{\beta}v(x)
\partial^{\alpha}\partial^{\beta}v(x) + \ldots.  \eqno({\rm B}.10)
$$
Hence, higher gradients of the $v$--field would enter in the form of
the ratios $O(\partial_{\alpha}/M)$ and can be dropped at leading
order in the $1/M$ expansion.

Thus, the effective Lagrangian of the rescaled $\tilde{\rho}$--field,
the $v$--field, is defined by 
$$
{\cal L}_{\rm eff}[v(x)] =
\frac{1}{2}\,\partial_{\alpha}v(x)\partial^{\alpha}v(x) -
\frac{1}{2}\,M^2_v N[v(x)],\eqno({\rm B}.11)
$$
where $M_v = 2M$ is the mass of the $v$--field. This agrees
with the classical Nambu--Jona--Lasinio model \cite{[19]}--\cite{[22]},
where the mass of the $\sigma$--meson is twice the mass of the
dynamical fermions. Then, the functional  $N[v(x)]$ is equal to
$$
N[v(x)]= \frac{1}{8\pi}\Bigg\{\Bigg[1 + \sqrt{16\pi\Big(1 +
\frac{M^2}{\Lambda^2}\Big)}\,v(x) + 4\pi\Big(1 +
\frac{M^2}{\Lambda^2}\Big)\,v^2(x)\Bigg]
$$
$$
\times\,{\ell n}\Bigg[1 +
\sqrt{16\pi\Big(1 + \frac{M^2}{\Lambda^2}\Big)}\,v(x) + 4\pi\Big(1 +
\frac{M^2}{\Lambda^2}\Big)\,v^2(x)\Bigg]
$$
$$
-\Bigg(1 + \frac{\Lambda^2}{M^2}\Bigg)\Bigg[1 +
 \frac{M^2}{\Lambda^2}\sqrt{\frac{16\pi\Lambda^2}{\Lambda^2 +
 M^2}}\,v(x) + 4\pi\frac{M^2}{\Lambda^2}\,v^2(x)\Bigg]
$$
$$
\times\,{\ell n}\Bigg[1 +
 \frac{M^2}{\Lambda^2}\sqrt{\frac{16\pi\Lambda^2}{\Lambda^2 +
 M^2}}\,v(x) +
 4\pi\frac{M^2}{\Lambda^2}\,v^2(x)\Bigg]\Bigg\}. \eqno({\rm B}.12)
$$
Expanding the functional $N[v(x)]$ in powers of $v(x)$ around $v(x) =
0$ we obtain
$$
N[v(x)] = v^2(x) + O(v^3(x)).\eqno({\rm B}.13)
$$
Since for the derivation of effective Lagrangians the ratio
$M^2/\Lambda^2$ is fixed at $\Lambda \to \infty$, the functional
$\exp\{ -\frac{i}{2}\,M^2_v\int d^2x\,N[v(x)]\}$ reduces in the
$\Lambda \to \infty$ limit to the functional $\delta$--function
$\delta[v(x)]$:
$$
\exp\Big\{-\frac{i}{2}\,M^2_v\int d^2x\,N[v(x)]\Big\}
\stackrel{\Lambda\to\infty}{\longrightarrow}
\prod_{x}\delta[v(x)]. \eqno({\rm B}.14)
$$
Thus, the generating functional of Green functions of the $v$--field
$$
Z[q] = \int{\cal D}v\,\exp\Big\{\frac{i}{2}\int
d^2x\,\Big[\partial_{\alpha}v(x)\partial^{\alpha}v(x) - M^2_vN[v(x)]
$$
$$
 - i\,\delta^{(2)}(0)\,{\ell n}\,\Big(1 + \sqrt{4\pi \Big(1 +
\frac{M^2}{\Lambda^2}\Big)}\,v(x)\Big)\Big] + i\int
d^2x\,q(x)v(x)\Big\} \eqno({\rm B}.15)
$$
reduces in the $\Lambda \to \infty$ limit to the from
$$
Z[q] = \int{\cal D}v\,\delta[v]\exp\Big\{\frac{i}{2}\int
d^2x\,\partial_{\alpha}v(x)\partial^{\alpha}v(x)\Big\}.\eqno({\rm B}.16)
$$
The appearance of the $\delta$--functional, $\delta[v]$, evidences that
the classical value of the $v$--field is zero, $v_{cl}(x) = 0$.

The generating functional ({\rm B}.16) does not depend on the external
source $q(x)$. This confirms the decoupling of the $v$--field that
corresponds to the decoupling of the $\tilde{\rho}$--field. Thereby,
no contributions can appear due to fluctuations of the $\rho$--field
around the minimum of the effective potential (\ref{label3.21}). This
implies the stability of the chirally broken phase and the
non--perturbative vacuum described by the effective potential
(\ref{label3.21}) under fluctuations of the $\rho$--field. We would
like to accentuate that the decoupling of the $\tilde{\rho}$--field
agrees fully with the decoupling theorem derived by Appelquist and
Carazzone \cite{[41]}.

\section*{Appendix C. Solutions of equations 
of motion (\ref{labelH.20}) and (\ref{labelH.21}) for the ansatz
(\ref{labelH.30})}

\hspace{0.2in} It is easy to show that for the ansatz
(\ref{labelH.30}) the equations of motion (\ref{labelH.20}) reduce to
the form
$$
-\Bigg(\frac{\partial }{\partial t} 
+ \frac{\partial }{\partial x}\Bigg)\xi(x,t)  =
 M\,e^{\textstyle - \omega},
$$
$$
\Bigg(\frac{\partial }{\partial t} - \frac{\partial }{\partial
x}\Bigg)\eta(x,t) = M\,e^{\textstyle + \omega},\eqno({\rm C}.1)
$$
whereas the equations of motion become splitted into a set of
first--order differential equations 
$$
-\frac{\partial }{\partial t}\xi(x,t) = 
\frac{M}{g}\,\Bigg(+\frac{a-b}{2}\,e^{\textstyle +\omega} 
+ \frac{a+b}{2}\,e^{\textstyle -\omega}\Bigg),
$$
$$
-\frac{\partial }{\partial x}\xi(x,t) = 
\frac{M}{g}\,\Bigg(-\frac{a-b}{2}\,e^{\textstyle +\omega} 
+ \frac{a+b}{2}\,e^{\textstyle -\omega}\Bigg),
$$
$$
\frac{\partial }{\partial t}\eta(x,t) = 
\frac{M}{g}\,\Bigg(+\frac{a+b}{2}\,e^{\textstyle +\omega} 
+ \frac{a-b}{2}\,e^{\textstyle -\omega}\Bigg),
$$
$$
\frac{\partial }{\partial x}\eta(x,t) = \frac{M}{g}\,
\Bigg(-\frac{a+b}{2}\,e^{\textstyle +\omega} +
\frac{a-b}{2}\,e^{\textstyle -\omega}\Bigg).\eqno({\rm C}.2)
$$
Due to the relation $a + b = g$ Eqs.~({\rm C}.1) are
consistent with ({\rm C}.2). Using the relations $a + b = g$ and $a -
b = 1/c$ Eqs.~({\rm C}.1) and ({\rm C}.2) can be rewritten in the
equivalent form
$$
\Bigg(\frac{\partial }{\partial t} 
- \frac{\partial }{\partial x}\Bigg)\xi(x,t)  =-
 \frac{M}{gc}\,e^{\textstyle + \omega},
$$
$$
\Bigg(\frac{\partial }{\partial t} 
+ \frac{\partial }{\partial x}\Bigg)\xi(x,t)  = 
- M\,e^{\textstyle - \omega}\eqno({\rm C}.3)
$$
and 
$$
\Bigg(\frac{\partial }{\partial t} 
- \frac{\partial }{\partial x}\Bigg)\eta(x,t)  = +
M\,e^{\textstyle + \omega},\eqno({\rm C}.4a)
$$
$$
\Bigg(\frac{\partial }{\partial t} 
+ \frac{\partial }{\partial x}\Bigg)\eta(x,t)  =
 -\frac{M}{gc}\,e^{\textstyle - \omega}\eqno({\rm C}.4b)
$$
The solutions of the differential equations ({\rm C}.4) read
$$
\xi(x,t) = \xi_0 - \frac{M}{gc}\,e^{\textstyle + \omega}\,(t-x) -
M\,e^{\textstyle - \omega}\,(t+x) ,
$$
$$
\eta(x,t) = \eta_0 + M\,e^{\textstyle + \omega}\,(t-x) - \frac{M}{gc}
\,e^{\textstyle - \omega}\,(t+x),\eqno({\rm C}.5)
$$
where $\xi_0$ and $\eta_0$ are integration constants and $c$ is the
Schwinger term (\ref{labelG.10}). 

For $\vartheta(x,t)$ we obtain
$$
\vartheta(x,t) = \frac{1}{\beta}\,[\xi(x,t) + \eta(x,t)] =
$$
$$
=\frac{\xi_0 + \eta_0}{\beta} + \frac{M}{\beta}\,\Bigg(1 - \frac{1}{gc}\Bigg)\,
e^{\textstyle + \omega}\,(t-x) - \frac{M}{\beta}\, \Bigg(1 +
\frac{1}{gc}\Bigg)\,e^{\textstyle - \omega}\, (t+x) ,\eqno({\rm C}.6)
$$
where $\beta$ is defined by (\ref{label4.5}).

Thus we have confirmed the consistency of the equations of motion
(\ref{labelH.20}) and (\ref{labelH.21}) and their consistency with the
ansatz (\ref{labelH.30}).

Notice that the factors $e^{\textstyle \pm\,\omega}$ can be removed by
an appropriate Lorentz boost. This yields
$$
\xi(x,t) =\xi_0 - \frac{M}{gc}\,(t-x) -
M\,(t+x) =\xi_0 - M\Big(1 + \frac{1}{gc}\Big)\,t -  
M\Big(1 - \frac{1}{gc}\Big)\,x,
$$
$$
\eta(x,t) =\eta_0 + M\,(t-x) - \frac{M}{gc} \,(t+x)
=\eta_0  + M\Big(1 - \frac{1}{gc}\Big)\,t - M\Big(1 +
\frac{1}{gc}\Big)\,x ,
$$
$$
\vartheta(x,t) = \frac{\xi_0 + \eta_0}{\beta} +
\frac{M}{\beta}\,\Bigg(1 - \frac{1}{gc}\Bigg)\,(t-x) -
\frac{M}{\beta}\, \Bigg(1 + \frac{1}{gc}\Bigg)\,(t+x) = \frac{\xi_0 +
\eta_0}{\beta} -\frac{2M}{\beta gc}\,t -
\frac{2M}{\beta}\,x.\eqno({\rm C}.7)
$$
This simplifies the solutions following from the ansatz
(\ref{labelH.30}) describing a helical wave discussed in \cite{[1]}.

\section*{Appendix D. Quantum field theory of free massive and massless 
fermion fields in 1+1--dimensional space--time}

\hspace{0.2in} The main aim of this Appendix is to specify the
definitions of massive and massless fermion fields in 1+1--dimensional
space--time.

Let $\psi(x)$ be a free massive fermion field with mass $m$ obeying
the Dirac equation of motion
$$
(i\gamma_{\mu}\partial^{\mu} - m)\psi(x)=0.\eqno({\rm D}.1)
$$
The quantization of field $\psi(x)$ goes via a solution of Eq.({\rm
D}.1) in terms of plane--waves
$$
\psi(x) = \int^{\infty}_{-\infty}\frac{dp^1}{\sqrt{2\pi}}\,
\frac{1}{\sqrt{ 2p^0}}\,\Big[u(p^0,p^1)a(p^1)\,e^{\textstyle -ip\cdot
x} + v(p^0,p^1)b^{\dagger}(p^1)\,e^{\textstyle ip\cdot x}\Big],
$$
$$
\bar{\psi}(x) =\psi^{\dagger}(x)\gamma^0 =
\int^{\infty}_{-\infty}\frac{dp^1}{\sqrt{2\pi}}\,
\frac{1}{\sqrt{2p^0}}\Big[\bar{u}(p^0,p^1)a^{\dagger}(p^1)
\,e^{\textstyle ip\cdot x} + \bar{v}(p^0,p^1)b(p^1)\,e^{\textstyle
-ip\cdot x}\Big],\eqno({\rm D}.2)
$$
where $p\cdot x = p^0x^0 - p^1x^1$. The creation
$a^{\dagger}(p^1)\,(b^{\dagger}(p^1))$ and annihilation
$a(p^1)\,(b(p^1))$ operators of fermions (antifermions) with momentum
$p^1$ and energy $p^0 = \sqrt{(p^1)^2 + m^2}$ obey the anticommutation
relations
$$
\{a(p^1), a^{\dagger}(q^1)\} = \{b(p^1), b^{\dagger}(q^1)\} =
\delta(p^1-q^1),
$$
$$
\{a(p^1), a(q^1)\} = \{a^{\dagger}(p^1), a^{\dagger}(q^1)\} = 
\{b(p^1), b(q^1)\} =
\{b^{\dagger}(p^1), b^{\dagger}(q^1)\} = 0.\eqno({\rm D}.3)
$$
The wave functions $u(p^0,p^1)$ and $v(p^0,p^1)=u(-p^0,-p^1)$ are the
solutions of the Dirac equation in the momentum representation for
positive and negative energies, respectively. They are defined by
$$
u(p^0,p^1) = {\displaystyle
\left(\begin{array}{c}\sqrt{p^0 + p^1} \\ \sqrt{p^0 - p^1}
\end{array}\right)}\;,\;\bar{u}(p^0,p^1) = (\sqrt{p^0 - p^1},
\sqrt{p^0 + p^1})
$$
$$
v(p^0,p^1) = {\displaystyle
\left(\begin{array}{c}\sqrt{p^0 + p^1} \\ -\sqrt{p^0 - p^1}
\end{array}\right)}\;,\;
\bar{v}(p^0,p^1) = (- \sqrt{p^0 - p^1}, \sqrt{p^0 + p^1})\eqno({\rm
D}.4)
$$
at $p^0 = \sqrt{(p^1)^2 + m^2}$ and normalized to 
$$
u^{\dagger}(p^0,p^1)u(p^0,p^1) = v^{\dagger}(p^0,p^1)v(p^0,p^1) = 2p^0,
$$
$$
\bar{u}(p^0,p^1)u(p^0,p^1) = - \bar{v}(p^0,p^1)v(p^0,p^1) = 2m,
$$
$$
\bar{u}(p^0,p^1)v(p^0,p^1) = \bar{v}(p^0,p^1)u(p^0,p^1) = 0.\eqno({\rm
D}.5)
$$
The functions $u(p^0,p^1)$ and $v(p^0,p^1)$ satisfy the following matrix
relations
$$
u(p^0,p^1)\bar{u}(p^0,p^1) = {\displaystyle {\sqrt{p^0 + p^1}\choose
\sqrt{p^0 - p^1}}}(\sqrt{p^0 - p^1}, \sqrt{p^0 + p^1}) =
$$
$$
= \left(\begin{array}{cc}\sqrt{(p^0)^2 - (p^1)^2} & p^0 + p^1 \\ p^0 -
p^1 & \sqrt{(p^0)^2 - (p^1)^2}
\end{array} \right) =  \left(\begin{array}{cc} m  & p^0 + p^1 \\
p^0 - p^1 &  m 
\end{array} \right) = 
$$
$$
= \gamma^0p^0 - \gamma^1p^1 + m = \hat{p} + m,
$$
$$
v(p^0,p^1)\bar{v}(p^0,p^1) = {\displaystyle
{\sqrt{p^0 + p^1}\choose -\sqrt{p^0 - p^1}}}(-\sqrt{p^0 - p^1}, 
\sqrt{p^0 +
p^1}) = 
$$
$$
= \left(\begin{array}{cc}-\sqrt{(p^0)^2 - (p^1)^2} & p^0 + p^1 \\ 
p^0 - p^1 &
-\sqrt{(p^0)^2 - (p^1)^2}
\end{array} \right) =  \left(\begin{array}{cc} - m  & p^0 + p^1 \\
p^0 - p^1 & - m
\end{array} \right) = 
$$
$$
= \gamma^0p^0 - \gamma^1p^1 - m = \hat{p} - m.\eqno({\rm D}.6)
$$
The causal Green function of a free massive fermion field $S_F(x-y)$
is defined by
$$
S_F(x-y) = i\langle 0|{\rm T}(\psi(x)\bar{\psi}(y))|0\rangle =
$$
$$
= i\theta(x^0 - y^0)\int^{\infty}_{-\infty}\frac{dp^1}{2\pi}\,
\frac{\gamma^0p^0 - 
\gamma^1p^1 + m}{2p^0}\,e^{\textstyle -ip^0(x^0-y^0) + ip^1(x^1-y^1)}
$$
$$
- i\theta(y^0 -
x^0)\int^{\infty}_{-\infty}\frac{dp^1}{2\pi}\,\frac{\gamma^0p^0 -
\gamma^1p^1 - m}{2p^0}\,e^{\textstyle - ip^0(y^0 -x^0) + ip^1(y^1 -
x^1)},\eqno({\rm
D}.7)
$$
where $\theta(z^0)$ is the Heaviside function.

Using the integral representation for the Heaviside function
\cite{[47]}
$$
\theta(z^0) = \int^{\infty}_{-\infty}\frac{dq^0}{2\pi
i}\,\frac{\displaystyle e^{\textstyle i\,q^0z^0}}{q^0
-i\,0}
\eqno({\rm D}.8)
$$
we recast the r.h.s. of ({\rm D}.7) into the form
$$
S_F(x-y) = i\langle 0|{\rm T}(\psi(x)\bar{\psi}(y))|0\rangle =
$$
$$
=\int^{\infty}_{-\infty}\frac{dp^1}{2\pi}\int^{\infty}_{-\infty}
\frac{dq^0}{2\pi}\,\frac{\gamma^0p^0 - \gamma^1p^1 + m}{2p^0(q^0 -
i\,0)}\,e^{\textstyle i(q^0 - p^0)(x^0-y^0) + ip^1(x^1-y^1)}
$$
$$
-\int^{\infty}_{-\infty}\frac{dp^1}{2\pi}\int^{\infty}_{-\infty}
\frac{dq^0}{2\pi}\,\frac{\gamma^0p^0 - \gamma^1p^1 - m}{2p^0(q^0 -
i\,0)}\,e^{\textstyle - i(q^0 -p^0)(x^0 -y^0) - ip^1(x^1 - y^1)}=
$$
$$
=\int^{\infty}_{-\infty}\frac{dp^1}{2\pi}\int^{\infty}_{-\infty}
\frac{dq^0}{2\pi}\,\frac{\gamma^0p^0 - \gamma^1p^1 + m}{2p^0(p^0 - q^0
- i\,0)}\,e^{\textstyle - iq^0(x^0-y^0) + ip^1(x^1-y^1)}
$$
$$
-\int^{\infty}_{-\infty}\frac{dp^1}{2\pi}\int^{\infty}_{-\infty}
\frac{dq^0}{2\pi}\,\frac{\gamma^0p^0 - \gamma^1p^1 - m}{2p^0(p^0 + q^0
- i\,0)}\,e^{\textstyle - iq^0(x^0 -y^0) - ip^1(x^1 - y^1)}=
$$
$$
=\int^{\infty}_{-\infty}\frac{d^2p}{(2\pi)^2}\,\frac{m + \hat{p}} {m^2
- p^2 - i\,0}\,e^{\textstyle - i p\cdot(x-y)}.\eqno({\rm D}.9)
$$
A direct calculation of the integral over $p$ yields
$$
S_F(x-y) = \frac{m}{2\pi}\,\frac{- (\hat{x}
-\hat{y})}{\sqrt{-(x-y)^2}}\,K_1\Big(m \sqrt{-(x-y)^2 + i\,0}\,\Big) +
i\,\frac{m}{2\pi}\,K_0\Big(m \sqrt{-(x-y)^2 + i\,0}\,\Big).
\eqno({\rm D}.10)
$$
For the product $(-i)\gamma_{\mu}S_F(x-y)\gamma^{\mu}$ we get
$$
(-i)\gamma_{\mu}S_F(x-y)\gamma^{\mu} = \frac{m}{2\pi}\,K_0\Big(m
\sqrt{-(x-y)^2 + i\,0 }\,\Big).\eqno({\rm D}.11)
$$
At $(x-y)\to 0$ this agrees with our calculation of
$\gamma_{\mu}\langle 0|\psi(x,t)\bar{\psi}(x,t)|0\rangle \gamma^{\mu}$
given by (\ref{labelH.14}) and (\ref{labelH.15}).

The solution of a massless fermion field can be obtained from the
solution ({\rm D}.2) in the limit $m \to 0$. The functions
$u(p^0,p^1)$ and $v(p^0,p^1)$ are defined by ({\rm D}.4) at $p^0 =
|p^1|$.

We would like to emphasize that our solution for a free massless
fermion field has a different phase convention from that used by
Thirring \cite{[3]} and Klaiber \cite{[5]} who set
$$
u(p^0,p^1) = v(p^0,p^1) = {\displaystyle
\left(\begin{array}{c}\sqrt{p^0 + p^1} \\ \sqrt{p^0 - p^1}
\end{array}\right)} = \sqrt{2p^0}\,{\displaystyle
\left(\begin{array}{c}\theta(+p^1) \\ \theta(-p^1)
\end{array}\right)},\eqno({\rm
D}.12)
$$
where $\theta(\pm p^1)$ are Heaviside functions.

The causal Green function $S_F(x-y)$ of a free massless fermion field
is defined by
$$
S_F(x-y) = i\langle 0|{\rm
T}(\psi(x)\bar{\psi}(y)|0\rangle = 
$$
$$
= i\,\theta(x^0-y^0)\,\langle 0|\psi(x)\bar{\psi}(y)|0\rangle -
i\,\theta(y^0 - x^0)\,\,\langle 0|\bar{\psi}^T(y)\psi^T(x)|0\rangle=
$$
$$
=
i\,\theta(x^0-y^0)\int^{\infty}_{-\infty}\frac{dp^1}{2\pi}\,
\frac{1}{2p^0}\,
[u(p^0,p^1)\bar{u}(p^0,p^1)]\,e^{\textstyle -ip^0(x^0 - y^0) +
ip^1(x^1-y^1)}
$$
$$
- i\,\theta(y^0 - x^0)\int^{\infty}_{-\infty}\frac{dp^1}{2\pi}\,
\frac{1}{2p^0}\, [v(p^0,p^1)\bar{v}(p^0,p^1)]\,
e^{\textstyle - ip^0(x^0 -y^0)
+ ip^1(x^1 - y^1)}=
$$
$$
=i\,\theta(x^0 -
y^0)\int^{\infty}_{-\infty}\frac{dp^1}{2\pi}\,
\frac{\gamma^0p^0 - \gamma^1p^1}{2p^0}\,e^{\textstyle -ip^0(x^0 - y^0) +
ip^1(x^1 - y^1)}
$$
$$
- i\,\theta(y^0 - x^0)\int^{\infty}_{-\infty}\frac{dp^1}{2\pi}\,
\frac{\gamma^0p^0
- \gamma^1p^1}{2p^0}\,e^{\textstyle - ip^0(y^0 - x^0) + ip^1(y^1 - x^1)}
$$
$$
=i\,(\hat{x} - \hat{y})\,\delta((x-y)^2) +
\frac{1}{2\pi}\,\frac{\hat{x} - \hat{y}}{(x-y)^2} =
\frac{1}{2\pi}\,\frac{\hat{x} - \hat{y}}{(x-y)^2 - i\,0}.  \eqno({\rm
D}.13)
$$
This is a well--known expression for the causal Green function of a
free massless fermion field in 1+1--dimensional space--time
\cite{[48]} (see also \cite{[47]}).


\newpage

\begin{thebibliography}{9}
\bibitem{[1]} 
M. Remoissenet, 
in {\it WAVES CALLED SOLITONS, Concepts
and Experiments}, Springer--Verlag, Heidelberg, 1994.
\bibitem{[2]}
W. Thirring,
Ann. Phys. (N.Y.) {\bf 3}, 91 (1958).
\bibitem{[3]}
S. Coleman,
Phys. Rev. D {\bf 11}, 2088 (1975).
\bibitem{[4]} 
T. H. R. Skyrme, Proc. Roy. Soc. A {\bf 247}, 260
(1958); {\it ibid.} A {\bf 262}, 237 (1961).
\bibitem{[5]}
B. Klaiber, 
in {\it LECTURES IN THEORETICAL PHYSICS},
Lectures delivered at the Summer Institute for Theoretical Physics,
University of Colorado, Boulder, 1967, edited by A. Barut and
W. Brittin, Gordon and Breach, New York, 1968, Vol. X, part A, pp.141--176.
\bibitem{[6]}
K. Furuya, Re. E. Gamboa Saravi and F. A. Schaposnik,
Nucl. Phys. B {\bf 208}, 159 (1982).
\bibitem{[7]}
J. Fr\"ohlich and P. Marchetti,
Commun. Math. Phys. {\bf 116}, 127 (1988).
\bibitem{[8]} 
S. Mandelstam, 
Phys. Rev. D {\bf 11}, 3026 (1975).
\bibitem{[9]}
P. H. Damgaard, H. B. Nielsen and R. Sollacher,
Nucl. Phys. B {\bf 385}, 227 (1992) and references therein.
\bibitem{[10]}
J. Thomassen,
PhD, TUWien, 15 January 2000.
\bibitem{[11]}
J. Alfaro and P. H. Damgaard,
Ann. of Phys. (N.Y.) {\bf 202}, 398 (1990).
\bibitem{[12]} 
K. Fujikawa, Phys. Rev. Lett. {\bf 44}, 1195 (1979);
Phys. Rev. D {\bf 21}, 2848 (1982); Phys. Rev. D {\bf 22}, 1499(E)
(1980).
\bibitem{[13]}
R. Roskies and F. A. Schaposnik,
Phys. Rev. D {\bf 23}, 558 (1981);
R. E. Gamboa Saravi, F. A. Schaposnik and J. E. Solomin,
Nucl. Phys. B {\bf 153}, 112 (1979);
R. E. Gamboa Savari, M. A. Muschetti, F. A. Schaposnik 
and J. E. Solomin,
Ann. of Phys. (N.Y.) {\bf 157}, 360 (1984).
\bibitem{[14]}
O. Alvarez,
Nucl. Phys. B {\bf 238}, 61 (1984).
\bibitem{[15]}
S.K. Hu, B. L. Young and D. W. McKay,
Phys. Rev. D {\bf 30}, 836 (1984); 
C. M. Na${\acute{\rm o}}$n,
Phys. Rev. D {\bf 31}, 2035 (1985).
\bibitem{[16]}
H. Dorn,
Phys. Lett. B {\bf 167}, 86 (1986).
\bibitem{[17]}
R. A. Bertlmann,
in {\it ANOMALIES IN QUANTUM FIELD THEORY}, Oxford Science 
Publications, Clarendon Press $\bullet$ Oxford, 1996, pp.248--286.
\bibitem{[18]}
Y. Nambu and G. Jona--Lasinio,
Phys. Rev. {\bf 122}, 345 (1961); {\it ibid.} {\bf 124}, 246 (1961).
\bibitem{[19]} 
K. Kikkawa, Progr. Theor. Phys. {\bf 56}, 947 (1976);
H. Kleinert, Proc. of Int. Summer School of Subnuclear Physics, Erice
1976, Ed. A. Zichichi, p.289; A. Dhar, R. Shankar and S. R. Wadia,
Phys. Rev. D {\bf 31}, 3256 (1985); D. Ebert and H. Reinhardt,
Nucl. Phys. B {\bf 271}, 188 (1986); M. Wakamatsu, Ann. of
Phys. (N.Y.) {\bf 193}, 287 (1989).
\bibitem{[20]}
S. P. Klevansky,
Rev. Mod. Phys. {\bf 64}, 649 (1992);
T. Hatsuda and T. Kunihiro,
Phys. Rep. {\bf 247}, 221 (1994).
\bibitem{[21]}
A. N. Ivanov, M. Nagy and N. I. Troitskaya,
Int. J. Mod. Phys. A {\bf 7}, 7305 (1992) 7305;
A. N. Ivanov,
Int. J. Mod. Phys. A {\bf 8}, 853 (1993).
\bibitem{[22]}
M. Faber, A. N. Ivanov, W. Kainz and N. I. Troitskaya,
Z. Phys. C {\bf 74}, 721 (1997);
Phys. Lett. B {\bf 386}, 198 (1996);
M. Faber, A. N. Ivanov, A. M\"uller, N. I. Troitskaya and M. Zach,
Eur. Phys. J. C {\bf 7}, 685 (1999); {\it ibid.} C {\bf 10}, 537 (1999).
\bibitem{[23]} 
J. Bardeen, L. N. Cooper and J. R. Schrieffer,
Phys. Rev. {\bf 106}, 162 (1957); Phys. Rev. {\bf 108}, 1175 (1957).
\bibitem{[24]} 
A. N. Ivanov, 
{\it NONPERTURBATIVE PHENOMENA IN QUANTUM CHROMODYNAMICS}, 
Lectures delivered at Institute of Nuclear Physics in
Vienna University of Technology in Winter semester in 1995, Part 2.
\bibitem{[25]}
N. D. Mermin and H. Wagner,
Phys. Rev. Lett. {\bf 17}, 1133 (1966).
\bibitem{[26]}
S. Coleman,
Comm. Math. Phys. {\bf 31}, 259 (1973).
\bibitem{[27]}
C. Itzykson and J.--B. Zuber,
in {\it QUANTUM FIELD THEORY}, McGraw--Hill Book Company, 
New York, 1980, p.525.
\bibitem{[28]}
K. Huang,
in {\it QUANTUM FIELD THEORY, From Operators to Path Integrals}, 
John Willey $\&$ Sons, Inc., New York, 1998, pp.363--367.
\bibitem{[29]} 
E. M. Lifshitz and L. P. Pitaevski${\check{\rm i}}$, 
in {\it RELATIVISTIC QUANTUM THEORY}, part 2, Pergamon Press, New York,
1974, p.503;
(see also Ref.[17], p.136).
\bibitem{[30]} 
C. G. Callan, R. F. Dashen, and D. H. Sharp,
Phys. Rev. {\bf 165}, 1883 (1968).
\bibitem{[31]}
H. Sugawara,
Phys. Rev. {\bf 170}, 1659 (1968).
\bibitem{[32]}
C. M. Sommerfield,
Phys. Rev. {\bf 176}, 2019 (1968);
H. Georgi,
Phys. Rev. D {\bf 2}, 2908 (1970).
\bibitem{[33]} 
J. Schwinger,
Phys. Rev. Lett. {\bf 3}, 296 (1959).
\bibitem{[34]} 
C. M. Sommerfield,
Ann. of Phys. (N.Y.) {\bf 26}, 1 (1963).
\bibitem{[35]}
D. Finkelstein,
J. Math. Phys. {\bf 7}, 1218 (1966);
L. D. Faddeev and V. E. Korepin,
Phys. Rep. C {\bf 42}, 1 (1978); 
(see Ref.\cite{[28]}, pp.380--388).
\bibitem{[36]} 
M. Gell--Mann, R. J. Oakes and B. Renner,
Phys. Rev. {\bf 175}, 2195 (1968).
\bibitem{[37]} 
P. C. Hohenberg,
Phys. Rev. {\bf 158}, 383 (1967).
\bibitem{[38]} 
P. B. Vigman and A. I. Larkin,
Zh. Exper. Theor. Fiz. {\bf 72}, 857 (1977) (in Russian).
\bibitem{[39]}
J. Goldstone,
Nuovo Cimento {\bf 19}, 154 (1961);
J. Goldstone, A. Salam, and S. Weinberg,
Phys. Rev. {\bf 127}, 965 (1962). 
\bibitem{[40]}
(see [27], pp.519--521)
\bibitem{[41]} 
T. Appelquist and J. Carazzone,
Phys. Ref. D {\bf 11}, 2856 (1975).
\bibitem{[42]} 
M. Faber, 
Few Body System {\bf 30}, 149 (2001).
\bibitem{[43]} 
K. G\"oke, J. N. Urbano, M. Fiolhais, and M. Harvey,
Phys. Lett. B {\bf 164}, 249 (1985).
\bibitem{[44]} 
(see Ref.[17] pp.284--286);
L. Bonora, M. Bregola, and P. Pasti,
Phys. Rev. D {\bf 31}, 2665 (1985).
\bibitem{[45]}
D. J. Gross and R. Jackiw,
Phys. Rev. D {\bf 6}, 477 (1972);
L. Dolan and R. Jackiw,
Phys. Rev. D {\bf 9}, 2904 (1974).
\bibitem{[46]}
A. Salam and J. Strathdee,
Phys. Rev. D {\bf 2}, 2869 (1970);
J. Honerkamp and K. Meetz,
Phys. Rev. D {\bf 3}, 1996 (1971);
J. M. Charap,
Phys. Rev. D {\bf 3}, 1998 (1971).
\bibitem{[47]} 
N. N. Bogoliubov and D. V. Shirkov, 
in {\it INTRODUCTION
TO THE THEORY OF QUANTIZED FIELDS}, Interscience Publishers, 
Inc., New York, 1959,pp.136--153; p.648.  
\bibitem{[48]}
B. Klaiber,
Helv. Phys. Acta 37 (1964) 554.
\bibitem{[49]}
G. V. Christos,
Z. Phys. C {\bf 18}, 155 (1983).
\bibitem{[50]}
R. Banerjee,
Z. Phys. C {\bf 25}, 251 (1984);
T. Ikehashi,
Phys. Lett. B {\bf 313}, 103 (1993).
\bibitem{[51]}
R. Jackiw and R. Rajaraman,
Phys. Rev. Lett. {\bf 54}, 1219 (1985);
K. Harada, H. Kubota, and I. Tsutsui,
Phys. Lett. B {\bf 173}, 77 (1986).
\bibitem{[52]}
M. Faber and A. N. Ivanov,
{\it On the solution of the massless
Thirring model  with fermion
fields quantized in the chiral symmetric phase}, hep--th/0112183.
\end{thebibliography}
\end{document}